# A Consensus Privacy Metrics Framework for Synthetic Data

Lisa Pilgram[1,2,3], Fida K. Dankar[2], Jörg Drechsler[4,5], Mark Elliot[6], Josep Domingo-Ferrer[7], Paul Francis[8], Murat Kantarcioglu[9], Linglong Kong[10], Bradley Malin[11, 12, 13], Krishnamurty Muralidhar[14], Puja Myles[15], Fabian Prasser[16], Jean Louis Raisaro[17], Chao Yan[11], Khaled El Emam[1,2]

[1]School of Epidemiology and Public Health, University of Ottawa, Ontario, Canada;

[2]CHEO Research Institute, Ontario, Canada;

[3]Department of Nephrology and Medical Intensive Care, Charité – Universitätsmedizin Berlin, Berlin, Germany;

[4]Department for Statistical Methods, Institute for Employment Research, Nuernberg, Germany;

[5]Institute for Statistics, Ludwig-Maximilians-Universität, Munich, Germany; Joint Program in Survey Methodology, University of Maryland, USA

[6]The Cathie Marsh Institute Research, School of Social Sciences, University of Manchester, Manchester, United Kingdom;

[7]Department of Computer Engineering and Mathematics, Universitat Rovira i Virgili, Tarragona, Catalonia;

[8]Max Planck Institute for Software Systems, Germany;

[9]Department of Computer Science, Virginia Tech, USA;

[10]Department of Mathematical and Statistical Sciences, University of Alberta, Alberta, Canada;

[11]Department of Biomedical Informatics, Vanderbilt University Medical Center, Nashville, Tennessee, USA

[12]Department of Biostatistics, Vanderbilt University Medical Center, Nashville, Tennessee, USA

[13]Department of Computer Science, Vanderbilt University, Nashville, Tennessee, USA

[14]Department of Marketing and Supply Chain Management, University of Oklahoma, Oklahoma, USA

[15]Medicines and Healthcare products Regulatory Agency, London, UK

[16]Berlin Institute of Health at Charité – Universitätsmedizin Berlin, Medical Informatics Group, Berlin, Germany

[17]Biomedical Data Science Center, University Hospital Lausanne, Lausanne, Switzerland

# Summary

Synthetic data generation is one approach for sharing individual-level data. However, to meet legislative requirements, it is necessary to demonstrate that the individuals' privacy is adequately protected. There is no consolidated standard for measuring privacy in synthetic data. Through an expert panel and consensus process, we developed a framework for evaluating privacy in synthetic data. Our findings indicate that current similarity metrics fail to measure identity disclosure, and their use is discouraged. For differentially private synthetic data, a privacy budget other than close to zero was not considered interpretable. There was consensus on the importance of membership and attribute disclosure, both of which involve inferring personal information about an individual without necessarily revealing their identity. The resultant framework provides precise recommendations for metrics that address these types of disclosures effectively. Our findings further present specific opportunities for future research that can help with widespread adoption of synthetic data.

# 1.   Introduction

Data access for secondary analysis remains a challenge [1], sometimes taking many months to get data [2,3]. An analysis of the success of getting individual-level data for meta-analysis projects from research authors found that the percentage of the time these efforts were successful ranged from 0% to 58% [3–8].

To address data access challenges, there is growing interest in using synthetic data generation (SDG) techniques to enable broader sharing of data for research and analysis [9–20]. In health research, for example, many synthetic datasets have been made available for research including the National COVID Cohort Collaborative (N3C) by the National Institutes of Health of the United States [21], the CMS Data Entrepreneur's Synthetic Public Use files [22], synthetic cardiovascular and COVID-19 datasets available from the CPRD in the UK [23,24], cancer data from Public Health England [25], synthetic variants of the French public health system claims and hospital dataset (SNDS) [26], and synthetic microdata from Israel's National Registry of Live Births [27]. Furthermore, recently study authors have been making synthetic variants of data used in their research papers publicly available to enable open science [28–31].

Such broad sharing of synthetic datasets requires strong assurances that the privacy of data subjects is protected. Unlike statistical disclosure control methods that create protected data by perturbing original data or reducing their detail [32–35], (fully) synthetic data is generated by sampling records from a distribution learned in model training. It is thereby grounded in the original data but should not preserve a one-to-one mapping between the synthetic records and real individuals. For this reason, one might naively conclude that synthetic data has a low privacy vulnerability. However, if the SDG model overfits the original data, the resulting synthetic data may be arbitrarily close to the original data, and privacy breaches can occur, such that an adversary using the synthetic data to conduct their attack may be able to learn some sensitive information about data subjects. Therefore, a privacy assessment is still needed to demonstrate that the generated synthetic datasets do indeed have low privacy vulnerability.

In certain domains, such as the healthcare sector, a privacy assessment of synthetic data is of particular importance considering the sensitive nature of the data and the more significant potential harms that would arise from privacy breaches. At the same time, however, a recent review showed that many published studies where SDG was used as a privacy enhancing technology (PET) did not include a privacy assessment of synthetic data [36]. The authors posit that the lack of consensus on how to measure privacy vulnerabilities in synthetic data may have contributed to those vulnerabilities not being evaluated.

Similarly, an urgent call for developing privacy frameworks for synthetic health data has just recently been published [37].

More generally, multiple authors note that the large number of metrics, as well as taxonomies for privacy metrics for PETs, makes application and interpretation difficult [36,38–41]. For example, Wagner et al. describe more than eighty privacy metrics [38]. The diversity in metrics is exacerbated by the fact that decades of experience in measuring disclosure vulnerabilities and agreed-to thresholds in PETs have been based on assessing the risk of identity disclosure. Identity disclosure is concerned about correctly assigning an identity to a record in a dataset [42,43] which may not be directly applicable to synthetic records.

Therefore, there is a need to assess and consolidate the current work on SDG privacy and its measurement. Such a consolidation will enable the comparison of SDG methods, the development of standardized benchmark datasets and software, support decisions on sharing synthetic data, and make it easier for regulators to establish greater regulatory certainty for synthetic data.

The objective of this study was to develop a consensus framework for how to evaluate privacy vulnerabilities in synthetic data. The approach of the study was to:

a. critically analyze privacy metrics and evaluation practices in synthetic data based on the current body of work, and identify the strengths and weaknesses, and

b. convene a global panel of privacy experts to develop consensus recommendations on how to evaluate privacy in synthetic data.

Note that the objective was not to ameliorate weaknesses or fix problems in existing metrics nor to develop new metrics that improve on existing approaches. Such additional improvements are left for future work.

## 2.    Methods

### 2.1    Study Process

The typical process to achieve expert agreement involves a literature review, followed by a report and a formal consensus method. Such processes have been widely used in health research, e.g., to develop guidelines or to decide on important fields of research [44–49]. As a formal consensus method, we conducted a Delphi process as it aims at free choice while preventing personal bias and dominance ("halo effect"). We applied the classic Delphi design as follows:

1. Round 0: Develop a report on privacy metrics that is reviewed by the panellists; formulating most relevant findings into clear recommendations (i.e., statements).

2. Round 1: Scoring the statements generated in round 0.

3. Rounds 2-n: Re-scoring the statement after controlled feedback (see section 2.4) with a minimum of n = 2 until the stopping criterion is met.

This process was aligned with respective guidelines [44–48,50,51] and is depicted in Figure 1.

The overall study process lasted from February 2024 until November 2024. Three scoring rounds were required to meet the stopping criterion. The participation rate was 100% in all Delphi rounds.

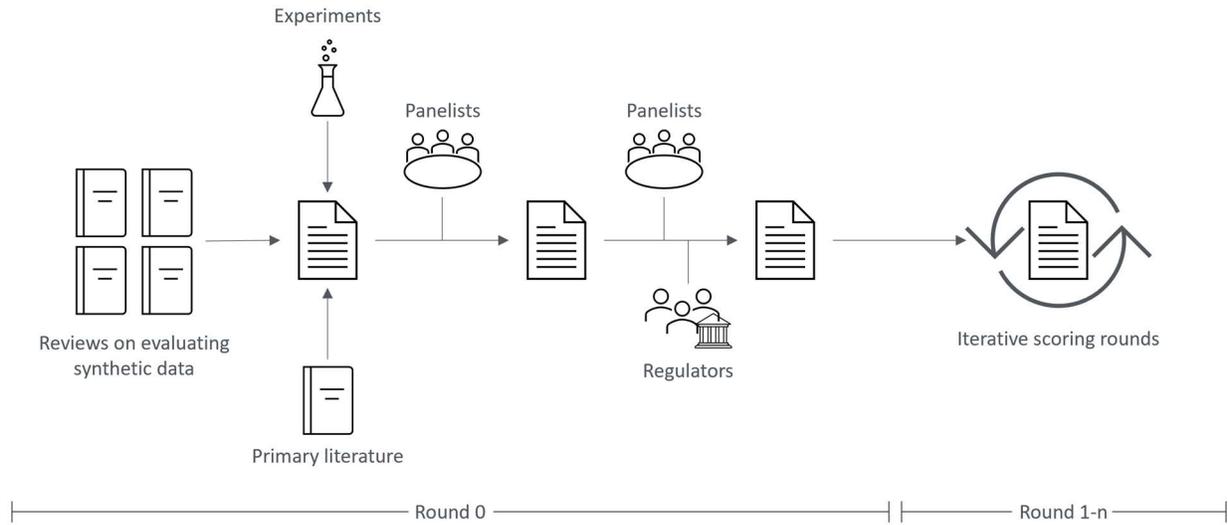

**Figure 1: Study Process.** Four literature reviews on the evaluation of synthetic data served as a starting point to identify commonly used privacy metrics [36,40,41,52]. Their primary literature was reviewed, and various additional simulation experiments conducted to better understand their behaviour. Panelists and regulators were invited to comment on the report (round 0). Statements were then created from the report's conclusions and rated by the panelists in Delphi rounds (round 1-n).

## 2.2   Expert Panel

The choice of the panel is known to heavily impact a consensus study's results [48,53]. For this study, an expert panel of thirteen individuals was set up. It has been argued that panelists should reflect the diversity of the topic [48]. While privacy, indeed, is a multidisciplinary topic, the purpose of this consensus study was to evaluate existing privacy metrics for synthetic data from the technical perspective. Consequently, the panelists were expected to have a high level of technical expertise in privacy metrics to understand, discuss and give opinions from a technical perspective. Consistent with that scope, we did not include patients' representatives nor other professions such as members of ethics committees.

We used a mixed recruitment strategy for the panel [48]. Identification of experts can be done, for example, through objective criteria such as a literature review or subjective criteria such as a colleague recommendation. In this study, the following criteria were used to select experts for the panel: editorial board member in Transaction on Data Privacy during the period 2019-2024 and consistent conference committee membership of Privacy in Statistical Databases during the period 2019-2024. The journal Transaction on Data Privacy was chosen due to its outstanding role in communicating high-quality findings in data privacy technologies [54]. Privacy in Statistical Databases is a key conference attracting a global audience in the field of data privacy. It was sponsored by the UNESCO Chair in Data Privacy [55]. Both resources are very focused on technical privacy topics and their members are representative of the respective research community. We identified eleven experts according to this criterion and invited them by email to participate in the panel. We extended this approach by including recommended experts. These were nominated by the initially identified experts or by the study's coordinators. For these additional nominated experts, we confirmed that they have published scholarly work relevant to our topic in the last five years. In this way, an additional nine experts were identified and invited to participate. Of those twenty experts that were identified thirteen responded positively to the invitation to participate in this study. This is within the range of typical panel sizes reported in literature [48,56].

## 2.3    Literature-Informed Critical Analysis (Round 0)

The literature-informed critical analysis focused on ways to assess privacy vulnerability in synthetic data. By privacy, we mean informational privacy which is concerned with whether personal information is disclosed, rather than the potential harms that may result from such disclosure. Privacy vulnerability is a measure of the likelihood of a privacy violation as a characteristic of the data [57]. This is distinct from "risk" which includes contextual factors such as the likelihood of an attack [58]. By synthetic data, we mean fully synthetic tabular data. Instead of conducting yet another systematic review on privacy metrics in synthetic data in round 0, we used four recently published reviews on the evaluation of synthetic data [36,40,41,52] and conducted a critical analysis report built upon their findings. This report is provided as Appendix A.

The analysis report followed an inductive classification approach where categories of privacy metrics were defined based on common conceptual themes as presented in the literature [36,40,41,52]. This classification grouped privacy metrics according to what they were aiming to measure. This is similar to the "adversary goal" as described in the privacy metrics classification by Wagner et al. [38]. A privacy metric within a category represents a specific implementation of that category. We then analyzed the metrics to identify strengths and weaknesses across, and also within, the defined categories.

The following four categories were inductively formulated: (1) record-level similarity, (2) membership disclosure, (3) attribute disclosure, and (4) differential privacy.

Record-level similarity captures how close the values of the synthetic records are to the values of the original records (i.e., distance to closest record) [59]. Membership disclosure is the ability of an adversary to determine that a target individual was in the training dataset for the SDG model (i.e., a member of the training dataset). Attribute disclosure has been defined as when an adversary can infer sensitive information about a target individual from the dataset's attributes [60–62]. Differential Privacy (DP) is described as a protection "from any […] harm that [individuals] might face due to their data being in the private database x that they would not have faced had their data not been part of x" [63].

While most metrics and privacy concepts that have appeared in the literature could be classified into membership or attribute disclosure, this was not the case for the metrics evaluating record-level similarity and for DP. A number of record-level similarity metrics were trying to capture identity disclosure in synthetic data [64,65]. Identity disclosure is the probability of assigning a correct identity to a record in a dataset [42,43]. DP distinguishes itself from the other categories by being an *a priori* feature of the process (i.e., SDG). Its parameter $\varepsilon$, known as a privacy (loss) budget, is used to characterize the privacy of resulting datasets and additional privacy evaluations beyond the privacy budget are typically not conducted. In this sense, it could be seen as a stand-alone category.

In the report, each category was defined, illustrated through exemplar metrics, and this was followed by a critical appraisal. The exemplar metrics were not intended to provide a comprehensive coverage of all metrics in a category, but to highlight specific ways that metrics that fall into a respective category have been defined.

Round 0 included iterative revisions involving feedback from the panelists. We also invited experts from six privacy and health regulators that have done work on synthetic data privacy to comment on the report although they did not participate in the Delphi rounds. These were from Canada, Italy, Singapore, South Korea, United Kingdom and United States of America. Their views did not represent their agency and did not imply endorsement. We considered their feedback in round 0.

From the report, the most relevant questions on privacy metrics in synthetic data were identified and formulated as statements. The report and the statements were further refined throughout the entire study process based on the panelists' feedback.

## 2.4    Scoring Rounds and Analysis

In the scoring rounds, panelists scored statements while having the report as background information. The level of agreement was indicated on a five-point Likert scale, which is commonly used in Delphi studies [53]. Comments could be provided to give explanations for the indicated level of agreement. The scoring rounds were conducted online using the Welphi software [66] and were pilot tested within the coordinator's research lab beforehand. Throughout all Delphi rounds anonymity was maintained.

After each round, responses were analyzed both quantitively and qualitatively. Relative frequency distributions, median and interquartile range (IQR) were calculated. Panelists received a personalized statistical summary with the relative frequency distribution of responses of the previous round alongside their own previous response [48]. Comments were analyzed with two objectives: First, to refine the statements and the report in between the rounds and second, to identify counterarguments. Details of the qualitative analysis and its results are given in Appendix C. Wherever relevant, key findings are also summarized in the main manuscript.

## 2.5    Stopping Criterion

The stopping criterion was defined *a priori* as group stability. We didn't use consensus as the stopping criterion to avoid forced consensus and to account for scenarios where plurality (i.e., no consensus) might be a stable outcome [47,67,68]. We therefore checked for significant group differences between two successive rounds by the Wilcoxon matched-paired signed-rank test consistent with other Delphi studies [67,69] and multiple testing was corrected for using a  Bonferroni adjustment. Stability was then defined as a p value greater than 0.05. As soon as all statements achieved stability, no further round of re-scoring was initiated [48].

## 2.6    Consensus Measurement

Consensus in the literature has been defined in various ways with criticisms highlighting that multiple definitions fail to distinguish between stability, consensus and agreement [47,53,67,70,71]. In our study, consensus was measured after stability was achieved and included both: determining (1) whether a consensus, and (2) whether agreement was observed. The IQR has been proposed as an objective way to determine whether consensus is achieved [67]. Consensus was defined as a maximum 1.0-point range of the IQR and non-consensus (i.e., stable plurality) as a more than 1.0-point range [67]. In those statements where consensus was achieved, the median was used to determine the type of consensus (i.e., agreement, uncertainty or disagreement) [51]). The agreement definition in [51] was adapted to the five-point Likert scale as shown in Figure 2. All methods were defined *a priori.*

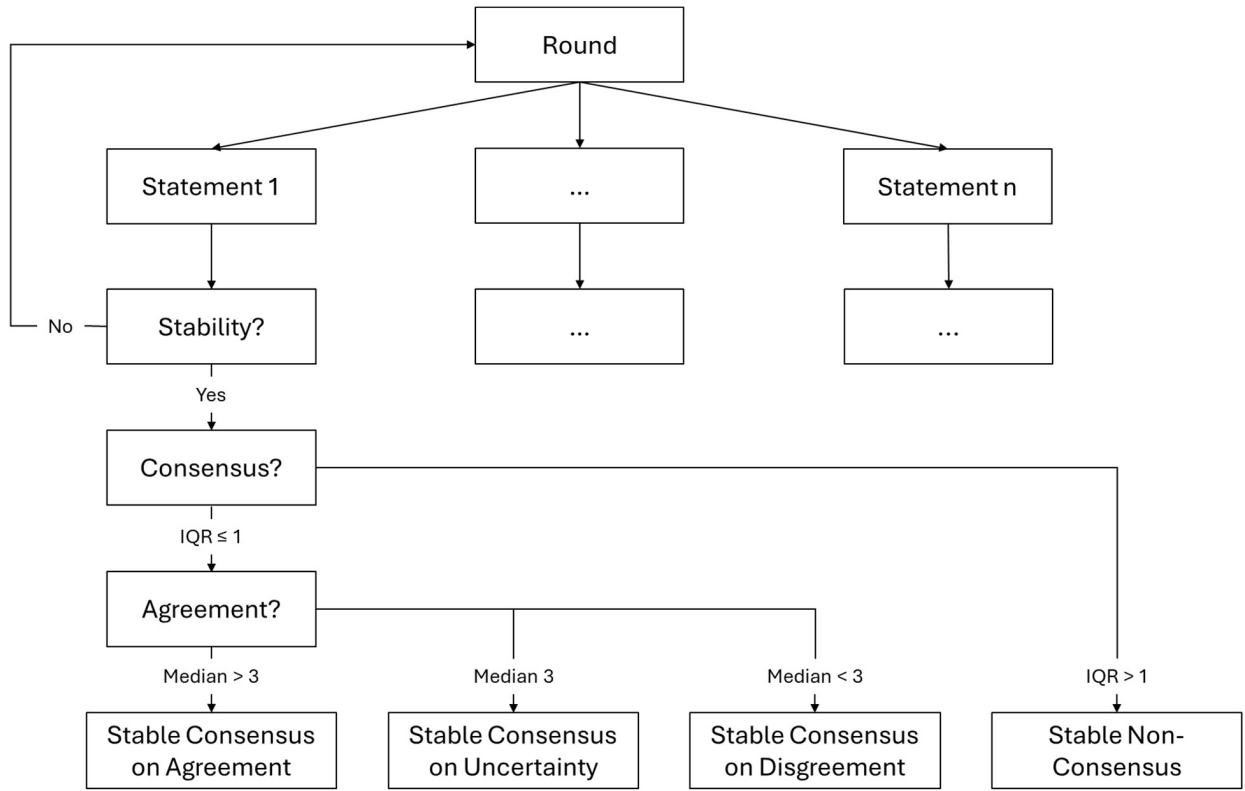

**Figure 2: Consensus Measurement.** Consensus was measured after stability of all statements was achieved. The measurement included two steps: (1) whether consensus and (2) whether agreement is achieved. Four possible outcomes could be expected. IQR: Interquartile range.

## 3.    Results

Stability of responses was achieved after the third Delphi round. Consensus and agreement were then analyzed. In most statements (10/11), the panelists' scorings indicated a stable consensus on agreement. There was 1/11 statement with stable consensus on uncertainty, 0/11 with stable consensus on disagreement and 0/11 with stable non-consensus. The statements are given in Table 1. The results of the consensus analysis are illustrated in Figure 3.

In the following, we present each statement along with a summary of the critical analysis from which it evolved and a qualitative analysis of the panelists' comments wherever it supports the understanding of the findings. The complete critical analysis is attached as Appendix A, all results from the quantitative and qualitative analyses conducted across the rounds are presented in Appendix C.

| Number | Statement |
|---|---|
| R1 | Disclosure vulnerability metrics should be based on quasi-identifiers. These may vary depending on the data context (e.g., can still be all attributes) and are ascertained by the data controller. |
| R2 | When evaluating a specific trained SDG model, disclosure vulnerability metrics need to be reported both for individual and multiple synthetic datasets (e.g. averaged across them and variation). |
| R3 | Disclosure vulnerability metrics should not be calculated on a pre-selected subset of "vulnerable" records but for all of the records. |
| R4 | Stand-alone similarity metrics (i.e., that are not part of attribute or membership disclosure) should not be used to report privacy in synthetic data. |
| R5 | Membership disclosure vulnerability should only be evaluated when the assumptions of the current metrics hold which is that the adversary would learn something new for targets drawn from the same population as the training dataset. |
| R6 | Because the F1 score, which is commonly used in membership disclosure metrics, is prevalence dependent, it needs to be reported relative to an adversary guessing membership. |
| R7 | As an anchor for membership disclosure vulnerability, a relative F1 score vulnerability (i.e., $F_{rel}$ value) of 0.2 is suggested. |
| R8 | Meaningful attribute disclosure vulnerability only applies to individuals that are in the dataset (i.e., members). Penalizing accurate prediction on individuals that have not been part of the dataset (i.e., group privacy) requires a broader ethical framework. |
| R9 | A relative attribute disclosure vulnerability that takes a non-member baseline into account is meaningful. |
| R10 | In attribute disclosure vulnerability, a relative vulnerability higher than its threshold is only considered as unacceptably high when the absolute vulnerability is higher than its threshold. |
| R11 | The privacy budget epsilon is not an adequate metric to report disclosure vulnerability unless it is set to a value close to 0. Even when differential privacy methods are used, disclosure vulnerability would still need to be evaluated using the same metrics as those applied to non-differentially private synthetic data. |

**Table 1: Recommendations in the Third (last) Delphi Round.** These recommendations (or statements) were rated by the panelists in the third round. The report as background information (see Appendix A) and brief explanations within the online tool were also provided (see Appendix C). All statements, except for R7, reached stable consensus on agreement. R7 had a stable consensus on uncertainty.

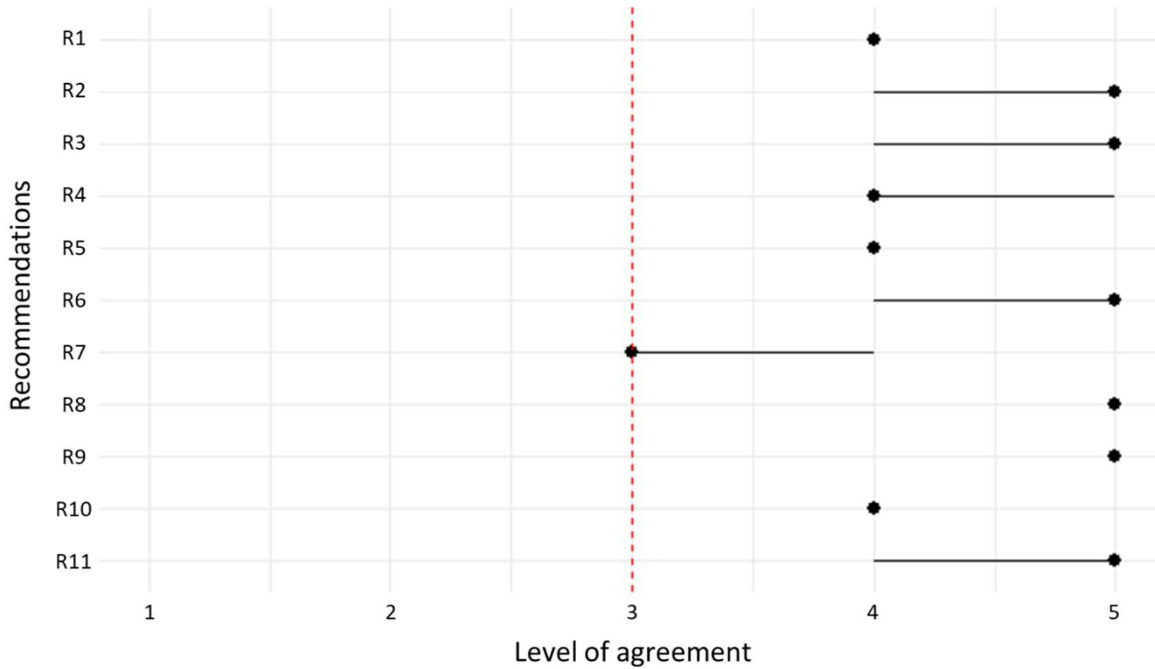

**Figure 3: Consensus Measurement.** The 11 recommendations of the third round were assessed for consensus and agreement. The recommendations are shown in Table 1. The agreement is indicated in 5 levels ranging from strongly disagree (1) to strongly agree (5). When interquartile range (IQR) was equal to or below 1, consensus was assumed, and the agreement level was measured according to the median.

## 3.1 Stable Consensus on Agreement

### 3.1.1 Overarching Considerations

**R1: Disclosure vulnerability metrics should be based on quasi-identifiers. These may vary depending on the data context (e.g., can still be all attributes) and are ascertained by the data controller.**

Quasi-identifies (QIs) represent the background knowledge of an adversary. This can be, for example, the age and gender of an individual. Many of the metrics that are used in the SDG literature use *all* attributes in the dataset rather than determining relevant QIs. Decades of research on identity disclosure in anonymized data and respective guidelines, however, are based on the assumption that the adversary prior knowledge is represented by QIs [35,43,58,62,72–76] and guidance for determining QIs has been published [58,77]. The rationale in the SDG literature for using all attributes is primarily based on three arguments: (1) it is believed to reflect the worst-case scenario (i.e., highest privacy vulnerability); (2) it avoids the subjectivity involved in determining QIs for a certain scenario and, (3) it makes the calculation more amenable to automation.

While a highly knowledgeable adversary clearly holds advantages when attempting an attack, our critical analysis revealed that *using* all attributes does not necessarily result in the highest privacy vulnerability as calculated by the metrics. Therefore, the current practice of using all attributes based on the first argument is not a robust approach. We illustrated this for evaluating membership disclosure where matching is an essential part of calculating vulnerability. Details of this simulation are described in Appendix A and summarized as follows. In our simulation, we used 10 different health datasets (see Appendix B), each of 50,000 records and 20 attributes. These datasets were considered as the population. From each population, an SDG training dataset with 10,000 records was randomly sampled (simple random sampling) based upon which synthetic data was then generated. Another dataset of 10,000 records was randomly sampled from the population. This served as the attack dataset of an adversary who was attempting to determine membership for their targets. Such an adversary would claim membership for a target when a synthetic record matches the target record. We mimicked this attack by matching attack records with synthetic records and considering varying number of attributes. We used subsets of 1 to 20 attributes and considered all combinations of attributes for each subset which gave us a total of 1,048,575 combinations for each dataset. Matching was then performed with each of the combinations. A match could then be true positive in case the matched record was indeed a member or false positive in case they weren't. A mismatch could be true negative in case they were a non-member or false negative case they were a member. From these numbers, we calculated precision and recall, and reported the F1 score which is a typical metric used to report membership disclosure [78]. The average F1 score was calculated by averaging across all combinations for each number of attributes, and the maximum F1 score by choosing the maximum value from them.

We observed that the more attributes were considered the lower the estimated membership vulnerability on average. The maximum vulnerability plateaued and decreased drastically with all attributes considered (see Figure 4). This observation was consistent across all datasets indicating a robust trend.

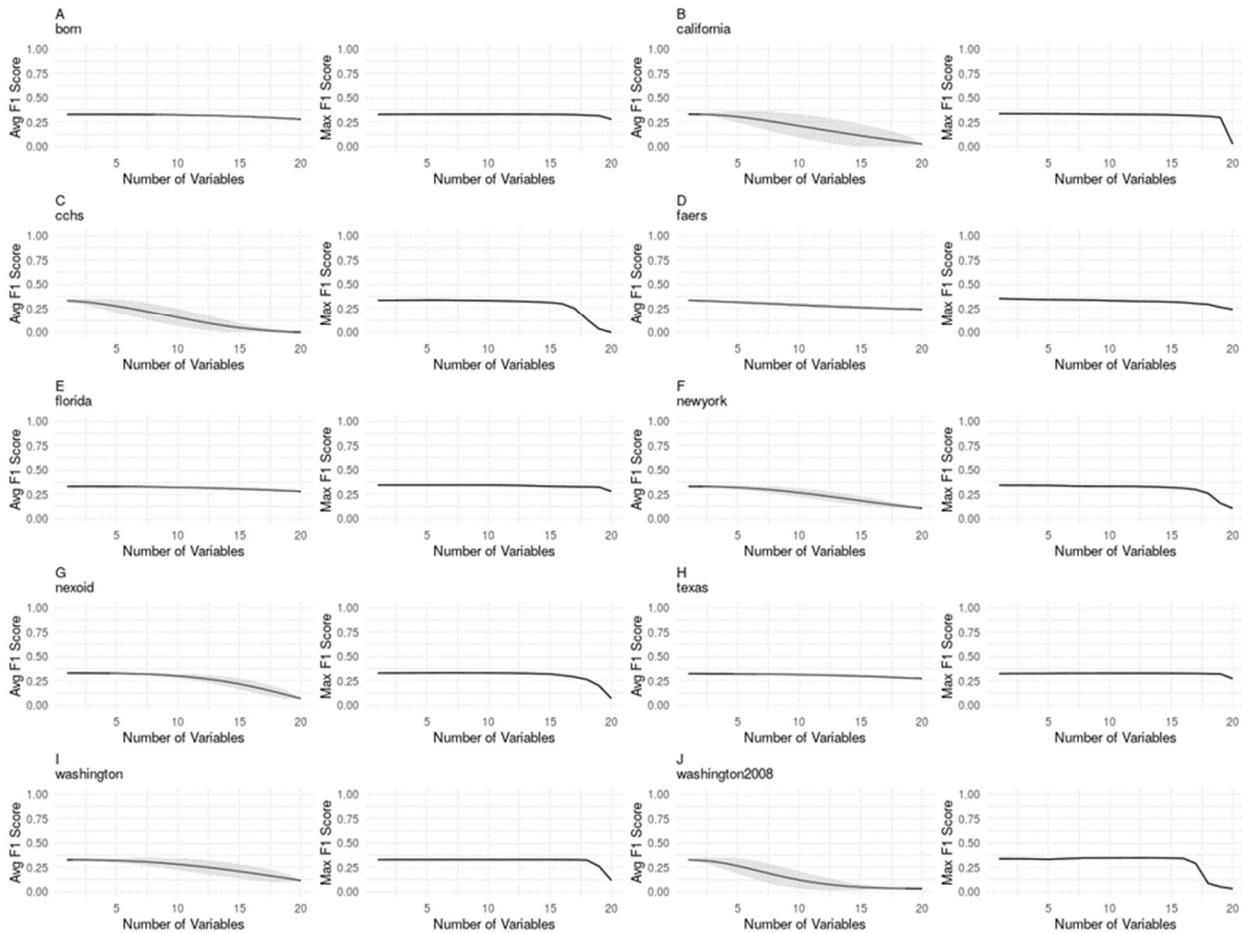

**Figure 4: Membership Disclosure when Varying Number of Attributes for Matching**. Membership disclosure was simulated by drawing an attack dataset from the same population as the training dataset. For each population, the F1 score is reported as the membership metric, calculated when varying the number of variables used in the attack. Maximum F1 score is the maximum value from all combinations for the respective number of attributes; average F1 score their average value. Standard deviation is illustrated in grey. Results for the following 10 datasets are shown: Born (A), California (B), CCHS (C), FAERS (D), Florida (E), New York (F), Nexoid (G), Texas (H), Washington (I) and Washington2008 (J). Dataset details are found in Appendix B.

Consequently, using all attributes in calculating privacy vulnerability is not necessarily a worst-case estimate for synthetic data but could result in underestimating disclosure vulnerability. More generally, for any specific number of known attributes (even if we only consider QIs), any combination of those attributes can be used by the adversary, and the match rate will vary depending on the combination chosen. This will have an impact on the disclosure vulnerability measurement. The data controller will not know *a priori* which combination of QIs the adversary will try. Therefore, it would be prudent to try all combinations for each number of attributes of the QIs. QIs need to be ascertained by the data controller for a given dataset and may vary depending on the assumptions on the adversary, the synthetic data recipient or class of recipients. Out of all combinations, the maximum can serve as the worst-case estimate of vulnerability.

Even though there was stable consensus on agreement for this recommendation, there were 2/13 (15.4%) experts who disagreed with this statement. While our simulation provided compelling evidence against the first argument, subjectivity remained a key topic of disagreement. For example, the diagnosis of obesity or diabetes mellitus can be considered as a QI since its knowability can be quite high (e.g., public awareness, photos, social media) while the diagnosis of interstitial nephropathy may not be a familiar term to the patient themselves. Consequently, it can be challenging to label the attribute *diagnosis* as either QI or non-QI across all diagnoses.

Furthermore, the context of the data may shift over time and an attribute that was once deemed not to be a QI may then be considered one. This contextual interpretation is very likely to result in inter-individual variability. Another key topic argued that any attribute could be a QI depending on the adversary so that entire records should be considered. Contemporary interpretations of data vulnerability consider the background knowledge of the data recipient [79], which may be different depending on the data recipient (and hence the adversary).

While it may be the most prudent approach to treat all attributes as QIs, this could, in practice, become computationally problematic as a prudent approach would also require considering all potential combinations of those QIs in matching.

**R2: When evaluating a specific trained SDG model, disclosure vulnerability metrics need to be reported both for individual and multiple synthetic datasets (e.g., averaged across them and their variation).**

SDG is a generative process with stochastic variability in its output. When evaluating the vulnerability of an SDG model rather than a single synthetic dataset, an aggregate (average and standard deviation) across multiple synthetic datasets from the same model would be appropriate. Stochasticity has also been shown to be relevant in utility evaluation of synthetic data [80], whereby it is recommended to average across ten synthetic datasets to reach a plateau in utility metrics synthetic data [81]. Similarly, a vulnerability across ten synthetic datasets would then reflect the vulnerability of an SDG model tied to a training dataset. However, if the decision-making scenario is data release, then the disclosure vulnerability for the specific synthetic dataset(s) may be the most relevant. Given that it is not always possible to determine *a priori* the exact decision-making scenario, it would be prudent to have both types of results.

**R3: Disclosure vulnerability metrics should not be calculated on a pre-selected subset of "vulnerable" records but for all of the records.**

Privacy vulnerability is often calculated for a pre-selected subset of records only instead of the entire dataset. The selected records are then labeled as "vulnerable" records [82–84]. The idea is to account for the worst-case scenario, assuming that these records are those experiencing the maximum disclosure vulnerability in the dataset. Such an *a priori* assumption is, however, not necessarily true [85,86]. It is essential to clarify the meaning of "vulnerable" within this context. Vulnerable records may be those containing potentially harmful information (e.g., HIV diagnosis), or those that are rare in a univariate or multivariate distribution.

Under the first definition, the main concern is harm. This means that the focus is on individuals who may suffer the most harm from disclosure, rather than on those who are most likely to experience disclosure. Under the second definition, the idea is that an adversary may learn more from an atypical value than from a typical one [87]. Again, this pertains to the *use* of the disclosed information rather than to the vulnerability of the synthetic data. Vulnerability is concerned about the likelihood of disclosure not the potential harm that may arise from it. An assumption about which records have a high attribute disclosure vulnerability cannot be easily made. It depends on the dataset, its correlational structure, and the vulnerability being examined.

Consequently, estimating the maximum disclosure vulnerability across a synthetic dataset always involves calculating vulnerability for each record in the first instance. Basing a vulnerability calculation on a pre-selected subset can result in overlooking records that actually have a higher vulnerability than those selected.

### 3.1.2 Record-Level Similarity
**R4: Stand-alone similarity metrics (i.e., that are not part of attribute or membership disclosure) should not be used to report privacy in synthetic data.**

Record-level similarity metrics are the most prevalent metrics to measure privacy in synthetic data [36]. They assess the distance between the training and synthetic data. This distance can serve as a metric by itself [88,89] meaning that closeness is then considered as high vulnerability. It can also be compared against some sort of baseline [41,90,91].

Similarity between a synthetic and training record, however, must be contingent on the identity disclosure for that very record in the training data. If that vulnerability is very small, then similarity would not necessarily indicate elevated disclosure vulnerability. On the other hand, for a training record with a high identity disclosure vulnerability plus high similarity, then that would be a scenario with unsafe generated data. This is related to the concept of identity disclosure in anonymized data that often encompasses singling out of records [42,43].

In anonymized data, identity disclosure generally aligns with learning something new: when singling out a unique record based on QIs, the adversary correctly learns information that comes from being part of the data *and* from the sensitive attributes. In synthetic data, however, singling out a unique record based on QIs does not necessarily come with learning correct information from the revealed sensitive attributes. If synthetic values of the sensitive attributes do not align with real values of the training record, then the adversary does not learn correct information from the sensitive attributes even if singling out a unique individual. They do, however, learn information that comes from being part of the data for that target. In other cases where both the QIs match and the sensitive values align with real ones the adversary would learn correct information from sensitive attributes.

A strict interpretation of identity disclosure would be that it does not require the adversary gaining information from successfully linking an identity to a record. While that type of disclosure would unlikely bring additional harm in practice, such an interpretation may need to be captured by privacy metrics in synthetic data. There are no current metrics that have successfully defined this concept.

### 3.1.3  Membership Disclosure

**R5: Membership disclosure vulnerability should only be evaluated when the assumptions of the current metrics hold which is that the adversary would learn something new for targets drawn from the same population as the training dataset.**

Membership disclosure metrics mimic an attack of an adversary who tries to determine whether a target was part of the training data used in SDG. Current metrics of partitioning methods (which are the most used in practice) involve splitting the real dataset into a training and holdout dataset. Records are then randomly sampled from these datasets to construct the presumed attack dataset of the adversary. The assumption is that the adversary draws targets from the same population the training dataset is sampled from [82,85,92–94,94–109]. In such implementations, the attack dataset includes some records that are part of the training data and some that are not, but all records are drawn from the same population the training dataset was sampled from.

An often-used narrative for calculating membership disclosure vulnerability is the scenario where population-defining information (e.g., patients diagnosed with HIV) is learned by an adversary through membership disclosure (e.g., in an HIV training dataset) [85,96,98,105]. This narrative is, however, not compatible with the above indicated sampling assumption for the attack dataset. If the population is HIV patients, then the adversary already knows the diagnosis of the target.

An alternative narrative that aligns with the sampling assumptions of current metrics could, for example, be a scenario where the adversary is trying to determine membership in a population-based interventional trial where vaccines are tested, the adversary is a vaccination opponent and harm results from the vaccination opponent inferring the pro-vaccine stance of a target individual based on membership. In this case the target being in the general population does not necessarily mean that they are in the interventional trial (training) dataset. Under such a narrative learning membership is additional information to the adversary prior knowledge. This can be referred to as learning sample-defining information (e.g., application of vaccines). Current membership disclosure metrics are based on the sampling scheme characterized by this narrative.

In our critical analysis, we simulated two attack scenarios and show that their vulnerability estimates can vary by a wide margin. Details of the simulation are given in Appendix A. The first scenario reflected what is currently operationalized in membership disclosure metrics, which is learning sample-defining information, by randomly sampling target records from the same population that the training dataset is sampled from (Figure 5A). In the second scenario, which is the more commonly used narrative where the adversary learns population-defining information, we sampled target records from a super-population of the training dataset (Figure 5B).

The two simulated populations were generated based on sociodemographic and HIV information from Ottawa. The population of interest in the first scenario (Figure 5A) reflected the population of young HIV-positive individuals in Ottawa. The adversary in this scenario already knows the HIV diagnosis but can learn sample-defining information through membership, for example, the drug taken by participants in a study. The population in the second scenario (Figure 5B) mimicked the entire young population of Ottawa. In this scenario, the adversary can learn the HIV diagnosis by learning membership.

We took a random sample from the population of young HIV-positive individuals, and this was used to train a synthetic dataset. A simulated attack was launched against the synthetic dataset. The adversary would consider a target record to be a member if it matched a synthetic record. Membership disclosure was then reported as the precision of correctly assigning membership to a target record.

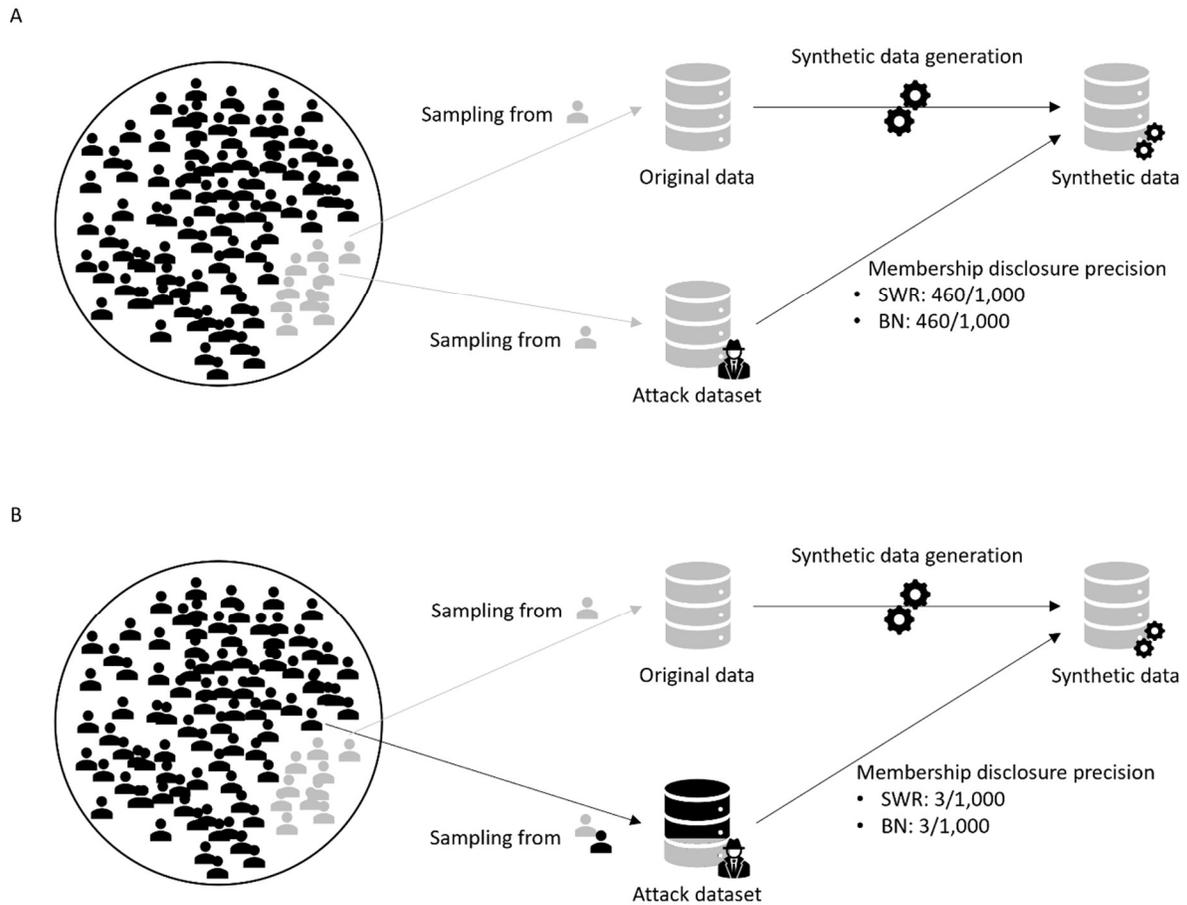

**Figure 5: Different Adversary's Attack Datasets**. The original data consisted of 1,000 people with HIV. It was randomly drawn from the population of young people with HIV in Ottawa (2,172) highlighted in grey. The attack dataset of 1,000 people was drawn from the same population as the original data, so from young people with HIV in Ottawa (A), or from all young people in Ottawa (B). The population of young people in Ottawa was set as 322,835. Membership disclosure is given as the precision of the attack. Synthetic data was generated via sampling with replacement (SWR) and Bayesian Network (BN).

This simple simulation illustrates that precision can be significantly lower when the adversary draws targets from a different population than the training data and thereby learns population-defining information (i.e., scenario B). Current metrics only operationalize scenario A and using these metrics to indicate vulnerability for scenario B gives wrong vulnerability estimates. Current metrics can be used when the assumptions of the metric are in line with the narrative, meaning that the adversary draws targets from the same population as the training data and learns sample-defining information (i.e., scenario A). Other narratives require a different modelling approach for the attack dataset.

**R6: Because the F1 score, which is commonly used in membership disclosure metrics, is prevalence dependent, it needs to be reported relative to an adversary guessing membership.**

It is known that metrics like precision, recall or the F1 score vary with prevalence [110,111]. Therefore, the distribution of the attack dataset (i.e., member prevalence) that is used to determine membership disclosure affects the reported vulnerability [109,112]. Most metrics presented in the literature build an attack dataset that includes 50% members [92,93,96,102,106–108]. Some use another arbitrary fixed value [94,95,113]. If, however, the sampling assumption of current metrics holds and targets are drawn from the same population that the training dataset is sampled from, then the attack dataset needs to account for the presumed member prevalence. In this scenario, this would be the sampling fraction of the training data. The resulting F1 score is, however, not uniformly interpretable across different datasets as the interpretation depends on the actual sampling fraction.

The idea of the naïve baseline as presented in [109] brings the different scores to the same scale. The naïve baseline would be 1.0 when the training dataset is identical with the population (i.e., sampling fraction of 1) and gets close to zero the smaller the sampling fraction. $F_{naive}$ then reflects an adversary's success when guessing membership without actually using the synthetic data. Combining it with the F1 score gives the incremental vulnerability introduced by the synthetic data. This can be accounted for by a relative membership disclosure metric ($F_{rel}$).

### 3.1.4 Attribute Disclosure
**R7: Meaningful attribute disclosure vulnerability only applies to individuals that are in the dataset (i.e., members). Penalizing accurate prediction on individuals that have not been part of the dataset (i.e., group privacy) requires a broader ethical framework.**

Some authors have argued that an individual's privacy can only be violated when that individual is in the dataset (i.e., a member) [57,114]. Therefore, accurately predicting the sensitive value of a target non-member would not be considered a privacy violation and hence not an attribute disclosure. Accurately predicting the sensitive attribute for a member target, in contrast, can be considered an attribute disclosure if and only if the prediction accuracy benefited from being a member. This is based on the concept that the presence or absence of a single individual in a dataset should not affect the outcome [115]. For a data subject, an attribute disclosure consequently occurs from being part of a dataset rather than from being part of the population where the dataset is drawn from. This is not to say that the release of data, its analysis, and knowledge generation cannot be harmful to non-members. It is rather that disclosure vulnerability metrics are not aiming at (and are not capable of) measuring harm that results from data release and this requires a broader ethical framework.

**R9: A relative attribute disclosure vulnerability that takes a non-member baseline into account is meaningful.**

In line with the previous statement, an increasingly accepted principle in technical privacy research is to make sure that being a member in a dataset does not increase the likelihood of an adversary gaining sensitive information about an individual so that everything that can be learned about the individual can also be learned without them being a member of the dataset [57,114,116]. This is sometimes also referred to as population-level (generic) information and contrasts with individual-level (specific) information. Population-level information, or knowledge generation, would not be considered a privacy violation. This would be more consistent with the goals of research where population-level information results from publishing, risk calculators, aggregate statistics and prognostic results (i.e., knowledge generation). Knowledge generation typically involves the identification of general patterns or trends and is not intended to reveal attributes of a specific individual (i.e., individual-level information).

The idea of attribute disclosure vulnerability metrics is therefore to make that distinction either by measuring how being part of the dataset affects the correct inference of attributes about an individual and assigning this value as a measure of disclosure. Alternatively, one can establish a population-level baseline (i.e., non-member baseline), comparing it to the information gained about a member individual and assigning the incremental information as a measure of disclosure.

It is, however, commonly understood that models are generally not capable of performing as well on unseen data as they do on the training data [117–119]. Consequently, it is likely that a difference will always remain [82,86], and an acceptable deviation from the non-member baseline or threshold must be agreed on. An example of deriving such thresholds is provided in Appendix A.

**R10: In attribute disclosure vulnerability, a relative vulnerability higher than its threshold is only considered as unacceptably high when the absolute vulnerability is higher than its threshold.**

The incremental or relative attribute disclosure vulnerability as described above is not sufficient to guide decision-making processes. From the perspective of both the adversary and the data subjects, the accuracy of the learned information (i.e., absolute prediction accuracy) matters. So, it is that not only the difference from the non-member baseline, but also where the difference is within the range of the scale matters. It may, for example, be acceptable that a member is predicted better than a non-member in cases where the accuracy of the learned information is still considered low.

Consider a simple example where the learned attribute is binary (e.g., diagnosis or no diagnosis). If the member prediction is an area under the receiver operating characteristic curve (AUROC) of 0.4 and the non-member prediction is one of 0.1, then the difference between them is arguably large and the incremental disclosure is high. But an AUROC value of 0.4 is worse than chance and therefore the absolute value also matters, which in this case would indicate that this is not an attribute disclosure. On the other hand, if the member AUROC was 0.9 and the non-member AUROC was 0.6, the difference is the same, but the high member AUROC would suggest that an adversary learns the diagnosis with high accuracy, and that fact should be considered.

### 3.1.5   Differential Privacy

**R11: The privacy budget epsilon is not an adequate metric to report disclosure vulnerability unless it is set to a value close to 0. Even when differential privacy methods are used, disclosure vulnerability would still need to be evaluated using the same metrics as those applied to non-differentially private synthetic data.**

A mechanism $M(D)$ that introduces randomness to the dataset $D$ is considered as $\varepsilon$-differentially private if:

$$\Pr\big[M(D) \in S\big] \le e^{\varepsilon}\,\Pr\big[M(D') \in S\big] \tag{1}$$

whereby $D$ and $D'$ differ with one record, and $S$ is the range set of $M$. The parameter $\varepsilon$ is the privacy budget and the $e^{\varepsilon}$ consequently a relative quantification of how much results of the mechanism are allowed to differ. This means that the privacy budget translates exponentially into changes in vulnerability. With a budget close to 0, analytical output from the data hardly changes regardless of whether any particular data subject is in the data. Such a small privacy budget can give the mathematical guarantee that privacy is preserved. If the privacy budget becomes large, however, theoretical privacy presumptions cannot be easily translated into empirical privacy [120]. Then, the interpretation of the privacy budget may differ across different statistical analyses [121] and is likely to depend on the implementation and noise introduced [122]. Given the current lack of interpretability of large epsilon values, privacy still needs to be evaluated empirically, as done, for example, in [116]. For such an evaluation, the same privacy metrics as those used in non-differentially privacy synthetic data should be used to allow for comparisons.

There were 2/13 (15.4%) panelists who indicated uncertainty with this statement. The key topic related to these scorings expressed that "close to 0" should not be demanded but rather small epsilon values. Small epsilon values, however, vary largely in their definition across academia. For example, Muralidhar et al. make use of an epsilon of 1.0 [122], Stadler et al. implement an epsilon of 0.1 [82], Li et al. state that 4 is an empirically reasonable value [94], Rosenblatt et al. consider an epsilon < 3.0 as low [123], and Hayes et al. an epsilon < 10 as acceptable [95]. As illustrated in Appendix A, industrial and government applications also have epsilon valued that vary from 0.1 to >18. This reinforces the need to use of privacy metrics other than the epsilon values as it demonstrates how unclear the definition of small epsilon values is. The question arises whether there exists any epsilon value other than 0 with a clear privacy interpretation.

## 3.2    Stable Consensus on Uncertainty

**R8: As an anchor for membership disclosure vulnerability, a relative F1 score vulnerability (i.e., $F_{rel}$ value) of 0.2 is suggested.**

Whether a privacy vulnerability metric is absolute or relative, having an anchor or benchmark to compare against is necessary to interpret the metric and for decision-making. This is particularly true for a binary decision on the release of synthetic data where thresholds are needed to determine whether a vulnerability metric's value is too high or acceptable.

A widely adopted approach to set thresholds is to rely on precedents. For example, precedents have been used to set thresholds for (absolute) identity disclosure vulnerability with anonymized data [43,73,124]. Consistent with that approach, other disciplines have used empirical benchmarks computed from the literature based on past performance as thresholds for acceptable performance, for example for the defect rate in software development an average industry defect rate can be set as a threshold [125]. In software process assessments, an acceptable interrater agreement has been defined based on 70 process examples [126]. In epidemiology, thresholds for statistical or performance metrics have also been proposed based on literature [127,128].

In line with that general approach, the value of 0.2 as anchor for a relative F1 score in membership vulnerability was suggested. It is a threshold that has been used by several authors [14,109,129]. The idea is to establish an anchor or starting point in practice that can be adjusted up and down depending on context. This context is characterized by the sensitivity of the data, potential harm and appropriateness of consent and notice.

The panel's ratings reflected a consensus on uncertainty regarding this statement. In the qualitative analysis, two key topics were identified: First, any threshold, anchor or default value for disclosure vulnerability is considered as inadequate. One reason that was given for this perspective is the contextual interpretation of privacy vulnerability as described above. Another reason is that vulnerability can be reported very differently, for example, based on an average value as well as a maximum value. A concern is that when giving an anchor value people may rely on this value without considering relevant aspects that may trigger adjustments to the value. The second key topic stressed that there is not enough evidence for the very anchor value proposed. Implications of this are discussed below.

# 4.    Discussion

## 4.1    Summary

The aim of our study was to foster standardized privacy evaluation practices in synthetic data through recommendations based on a critical analysis of privacy metrics used in the literature and agreed upon by experts in a formal consensus process. The critical analysis raised clear concerns on two points:

1.    concerns about using record-level similarity, and

2.    concerns about the privacy budget epsilon for privacy evaluation under differential privacy.

Membership disclosure and attribute disclosure vulnerability were identified to be the most suitable for evaluating privacy in synthetic data. Most published metrics in these categories, however, either rely on assumptions that must be validated for the specific use case, are not interpretable by themselves or make use of an inappropriate baseline. Experts in our study agreed to discourage the use of record-level similarity and the privacy budget, except when it is close to 0. They also agreed on the need to evaluate privacy using membership and attribute disclosure for both non-differentially privacy synthetic datasets and differentially private ones (if the epsilon value was not close to zero). Recommendations on a reasonable way to evaluate membership and attribute disclosure vulnerability with currently available metrics were made, as discussed below.

### 4.1.1    Recommendations for Membership Disclosure

For membership disclosure, the two consensus recommendations address major pitfalls when calculating vulnerability. These can be attributed to an uncritical and/or uncontextualized application of contemporary metrics. In practice, the evaluation of membership disclosure should start with the following question: are the adversarial assumptions in the SDG scenario, as expressed in the membership disclosure narrative, in line with the sampling assumptions of the metric ? The assumption of currently used metrics is that the adversary draws targets from the same population as the training dataset and thereby would learn sample - but not population - defining information?

Only if this question can be answered positively, current metrics are able to give a realistic estimate for vulnerability. Considering further results of the critical analysis, the steps in Figure 6 would then be used to guide and document the calculation.

Introduce an adversarial assumption for the SDG scenario that is in line with the metrics assumptions

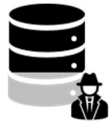
Explicitly model the member prevalence in the attack dataset

$F_\beta$
Choose a reasonable weight of precision and recall depending on the adversarial assumption

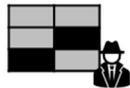
Consider all possible subsets of the known attributes (QIs) when matching to calculate the absolute $F_\beta$ score

$$F_{rel} = \frac{F_\beta - F_{naive}}{1 - F_{naive}}$$
Calculate the relative $F_\beta$ score

**Figure 6: Practical Guidance to Calculate Membership Disclosure Vulnerability.** This guidance assumes that partitioning methods are used.

To take a release-decision, the calculated membership disclosure vulnerability $F_{rel}$ must be compared against a threshold. This value depends on the context and needs to be informed by the sensitivity of the data, potential harm and appropriateness of consent and notice. In our consensus study, there was uncertainty in determining an anchor value to inform the threshold choice. In general, an anchor value can provide a valuable resource to inform thresholds and thereby facilitate the wide adoption of SDG. This requires that the metric it is meant for is precisely defined and the context is carefully considered. The actual value of such an anchor may be determined in future research considering the growing body of evidence and experience with synthetic data.

### 4.1.2 Recommendations for Attribute Disclosure

Attribute disclosure metrics come with the main challenge that they must be able to discriminate privacy violation from knowledge generation. Our consensus recommendations highlight this requirement and gives recommendations on how to operationalize it. As shown in Figure 7, a holdout dataset can be used to calculate the portion of the prediction accuracy attributable to knowledge generation. Interpretation of the prediction accuracy will depend on the performance measurement. For example, an AUROC of 0.5 is no better than random guessing regardless of the class distribution. Other measurements such as precision or recall must, however, account for class distribution to provide meaningful interpretation.

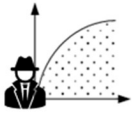

Choose a performance measurement depending on the adversarial assumption (e.g. AUROC)

$$A_{rel} = A_{members} - A_{non-members}$$

Use a holdout dataset as proxy for the non-member (i.e., knowledge generation) baseline

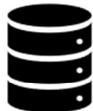

Calculate attribute disclosure vulnerability for the entire dataset, not only for a pre-selected subset

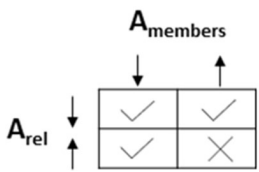

Use both, absolute (member) and relative vulnerability to guide decision-making

**Figure 7: Practical Guidance to Calculate Attribute Disclosure Vulnerability.**

## 4.2   Limitations

Below we highlight some considerations for this study:

a)   While our Delphi design is justified and based on respective guidance, we cannot ultimately exclude that there was group bias in our study or forced consensus (due to a misconception that the stopping criterion was related to consensus).

b) Qualitative results may vary depending on the researcher who carries out the analysis [130]. In this sense, we cannot exclude that there may have been further key topics that have not been identified but could have prompted refinement of the statements or the report.

## 4.3    Future Work

The recommendations presented in this paper were developed through a systematic consensus process built upon a detailed critical analysis reflecting the current body of work. This analysis has also identified certain weaknesses in the measurement of disclosure vulnerability for synthetic data that present specific opportunities for further research. Crucial points that need to be addressed are:

1. **Identity disclosure metrics:** SDG generates data that reflect the statistical properties of the training dataset without preserving any direct link between a synthetic and real record. Our analysis revealed that the concept of identity disclosure lacks precise interpretation, which makes it difficult to apply to synthetic data. Identity disclosure can be strictly interpreted as establishing a link between a record and a real identity without assessing the truthfulness of this link or of any learned information. There is, however, currently no metric available that provides a solution for such an interpretation.

2. **Membership disclosure metrics:** There is a gap in current membership disclosure metrics in terms of providing correct vulnerability estimates in different adversarial sampling scenarios. The common narrative of learning population-defining information is not captured by current metrics that are implemented with the assumption that targets are sampled from the same population that the training data is sampled from.

3. **Attribute disclosure metrics:** Attribute disclosure vulnerability is a prediction task where the sensitive target variable can be categorical (i.e., classification) or continuous (i.e., regression). There are many ways to model both and to quantify their prediction performance. Standardizing on prediction models would provide results that can be comparable across studies.

4. **Adversarial matching strategies:** Matching is an essential part of calculating vulnerability in multiple disclosure metrics. An adversary is motivated to consider subsets of attributes rather than the complete set when matching. The reasoning is that in synthetic data, the more attributes an adversary uses for matching, the higher the chance that a synthetic value differs from the real one, so that a mismatch occurs. If the adversary generalizes the attributes, then success may be further optimized. A comprehensive analysis of different matching strategies is needed to account for worse case scenarios in matching based metrics.

5. **Anchor values and thresholds:** Whether a privacy vulnerability metric is absolute or relative, having a threshold to compare against is necessary to interpret the metric and for decision-making. This is particularly true for a binary decision on the release of synthetic data where thresholds are needed to determine whether a vulnerability metric's value is too high or acceptable. Thresholds also need to be precisely defined in relation to their vulnerability, including the underlying model and specific measurement they are meant for. Further effort is needed to establish thresholds based on precedents or some other criterion.

## Ethics

This project has been approved by the CHEO Research Institute Research Ethics Board protocol 23/65X.

## Resource availability

### Lead Contact

Requests for further information and resources should be directed to and will be fulfilled by the lead contact:

Khaled El Emam
Children's Hospital of Eastern Ontario Research Institute
401 Smyth Road
Ottawa, Ontario K1H 8L1
Canada
kelemam@ehealthinformation.ca

### Material Availability

This study did not generate new unique reagents.

### Data and Code Availability

We used publicly available as well as confidential medical data for our simulations. The data sources are listed alongside the simulations in Appendix B and access can be requested directly at the source. The lead contact can be contacted for further information or support to gain access. All original code for our simulations has been deposited under [131] as of the date of publication.

## Acknowledgements

This work is being funded by CIFAR and the Bill & Melinda Gates Foundation. KEE is funded by a Discovery Grant RGPIN-2022-04811 from the Natural Sciences and Engineering Research Council of Canada, and the Canada Research Chairs program from the Canadian Institutes of Health Research. LP is funded by the Deutsche Forschungsgemeinschaft (DFG, German Research Foundation) – 530282197. JD receives funding from the U.S. Census Bureau for conducting research on formal privacy methods for survey data (corporate agreement CB20ADR016000). JDF is funded by the Government of Catalonia (ICREA Acadèmia Prize) and the EU's NextGenerationEU/PRTR via INCIBE (project "HERMES" and INCIBE-URV cybersecurity chair). BM receives funding from the NIH - RM1HG009034). ME receives funding from the UKRI grant ES/Z502984/1. LK was partially supported by grants from the Canada CIFAR AI Chairs program, the Alberta Machine Intelligence Institute (AMII), and Natural Sciences and Engineering Council of Canada (NSERC), and the Canada Research Chair program from NSERC. JLR is funded by the Innovative Health Initiative Joint Undertaking (IHI JU) under grant agreement No 101172872.

## Author contributions

Conceptualization, design, and analysis: LP and KEE; drafting the manuscript: LP and KEE; review and editing: FD, JD, ME, JDF, PF, MK, LK, BM, KM, PM, FP, JLR and CY.

## Competing Interests Statement

KEE is the scholar-in-residence at the Office of the Information and Privacy Commissioner of Ontario. PM has led on research and development of synthetic data that are made available for researchers for a licence fee by the Clinical Practice Research Datalink (CPRD), which is the Medicines and Healthcare products Regulatory Agency's (MHRA) real world data research service. The views expressed by PM are her own and do not represent the Medicines and Healthcare products Regulatory Agency's (MHRA) policy position. All other authors declare no competing interests.

**A Consensus Privacy Metrics Framework for Synthetic Data**

**Appendix A: Critical Analysis of Disclosure Metrics (Report)**

**Table of Contents**

# 1. General Considerations

## 1.1 Terminology and Scope

### 1.1.1 Synthetic Data Generation

In contemporary discourse, the term "generative models" is often associated with Large Language Models (LLMs). LLMs are trained to generate human-like text. The current report, however, focuses on generated data in the form of fully synthetic structured tabular individual-level data where one row corresponds to an individual not an event. Its generation is referred to as SDG. To avoid overloading terms, we will use the term "SDG models" henceforth instead of the term "generative models".

### 1.1.2 Scope of Synthetic Data

By default, we assume unless otherwise stated, that SDG involves a training dataset. However, SDG in general does not necessarily require training data; it can be based on distributions known a-priori and informed by background knowledge, published summary statistics, and published risk calculations [1–4]. Synthetic data that has been generated that way is less likely to have elevated privacy risks [1], but some privacy considerations may still apply to it.

We exclude partially synthetic and hybrid synthetic data. In both cases, parts of the original dataset are retained (attributes or full records respectively) which makes their privacy evaluation closer to that of perturbation under statistical disclosure control methods [5].

Consequently, in this report, we will use the term synthetic data to mean fully synthetic structured tabular data only.

### 1.1.3 Privacy and Information Disclosure

Privacy has been broadly defined as "the claim of individuals […] to determine for themselves when, how and to what extent information about them is communicated" [6]. This report approaches privacy from a more technical perspective by focusing on data protection by SDG and how to measure it. In this context, we use the term disclosure for "revealing confidential or personally identifiable information from a dataset based on a vulnerability that is found or exploited" as defined by the International Organization of Standardization (ISO) [7]. Traditionally, identity disclosure has been the primary focus when considering privacy vulnerabilities in the context of anonymization. Beyond identity disclosure, however, personal information can also be revealed through other mechanisms [8], which are reflected in the privacy metrics categories in this report.

### 1.1.4 Adversary

We use the term adversary to refer to an "individual or unit that can, whether intentionally or not, exploit potential vulnerabilities" in line with the ISO definition [7]. The goal is to correctly learn new sensitive information about an individual. Sensitive information is meant broadly including any information related to a real person [9,10].

### 1.1.5 Disclosure Vulnerability

The overall disclosure risk can be treated as the combination of the probability of a disclosure given an attempt by an adversary [11].

$$pr(disclosure, attempt) = pr(disclosure \mid attempt) \times pr(attempt) \qquad (1)$$

where $pr(disclosure, attempt)$ is the overall disclosure risk and $pr(attempt)$ the probability that an adversary attempts a disclosure attack. The probability of an attempt is influenced by non-data factors such as the security and privacy controls in place. For example, if there are strong access controls then the adversary may not be able to access the synthetic data easily to attempt an attack. Or if there are contractual restrictions on attack attempts, that would create a disincentive for the adversary.

We consider the first term $pr(disclosure \,|attempt)$ to be the *disclosure vulnerability*, which is based only on the synthetic data [12]. Vulnerability and attempt are not independent concepts, meaning that, for example, a higher disclosure vulnerability may increase the probability of attempt (since it provides incentives for the adversary), and hence higher overall risk. In this report, we assume that privacy metrics are meant to quantify *vulnerability* as a characteristic of the synthetic dataset and thereby cover only one component of the overall risk. This is relevant when discussing specific privacy metrics categories.

Some calculations proposed in the literature factor the probability of attempt rather than just disclosure vulnerability. We will highlight those as we review them.

The idea of harm needs to be separated from disclosure vulnerability. Although they may coincide, they do not necessarily align. Harm encompasses a broader ethical framework and can ultimately only be evaluated from the perspective of the individual experiencing it or subjectively analyzed in the context of an ethics review. The broader scope of harm is exemplified by Abowd et al. who state that "harm is outside the scope of the legal requirements governing the Census Bureau" [13].

The same is true for perceived vulnerability or incorrect claims of disclosure. Claims of disclosure that turn out to be incorrect can have harmful consequences for individuals. Perceived vulnerability by an individual or the public is still relevant for the overall disclosure risk as it may increase the probability of attempt [11].

To be clear, the above concepts are still important and should be considered in an overall risk management process. However, they are orthogonal to the measurement of disclosure vulnerability. The recently published guideline on SDG by the Personal Data and Protection Commission of Singapore is an example for such an overall assessment and management process [14].

### 1.1.6 Quasi-Identifiers

Quasi-identifiers are the attributes that are used to mount privacy attacks. They are knowable by an adversary using public sources, because the adversary has access to non-public sources of information, or the adversary is an acquaintance of the target individual and has private background information about the individual. Decades of research on privacy in anonymized data and respective guidelines are based on quasi-identifiers being a reasonable representation of the adversary's prior knowledge [7,15–17]. Quasi-identifiers may vary and are ascertained by the data controller for a given dataset and a given scenario.

### 1.1.7 Population, Original, Training, Holdout and Attack Dataset

In this report, we assume that there isn't any directly identifying attribute such as unique identifying numbers (e.g., social security numbers) or personal names present in the dataset and that consequently datasets consist of two types of attributes: quasi-identifiers and attributes containing sensitive information (i.e., sensitive attributes).

The dataset that is available to the data controller is referred to as the *original* dataset or the *real* dataset. The *training* dataset is the one used to train the SDG model and may be the same as the original, or a subset of the original dataset (for example, if the original dataset is partitioned into a training dataset and a *holdout* dataset). However, note that the data controller would then apply SDG modeling to the full original dataset and generate the actual released synthetic data from that; the partitioning is only for the purpose of computing privacy metrics.

The record that is available to the adversary is a target record. Privacy metrics, however, are calculated from the data controller perspective that tries to mimic the adversary and their target records. In this context, from the data controller's perspective the terms "attack record" and "attack dataset" are used.

An important assumption is that the original dataset is a simple random sample from the (typically larger) population. In this sense, the original dataset is intended so serve as a representative reflection of the population.

## 1.2 Categories of Disclosure Vulnerability Metrics

This report follows a conceptual classification approach to group privacy metrics according to what they are aiming at. This is similar to the "adversary goal" as described in the privacy metrics classification by Wagner et al. [18].

While writing this report, we have taken into account the recently published reviews on the evaluation of synthetic data and have inductively formed corresponding categories from the listed metrics [19–22]. Consequently, this report groups the privacy metrics into four categories: (1) record-level similarity, (2) membership disclosure, (3) attribute disclosure, and (4) differential privacy. While most metrics and privacy concepts that have appeared in the literature could be classified into membership or attribute disclosure, this was not the case for the metrics evaluating record-level similarity and for differential privacy. This observation is in line with the review in [22] that has chosen the identical categorization. We will see that a number of record-level similarity metrics are very likely trying to capture identity disclosure in synthetic data.

In contrast, differential privacy is agnostic to a specific disclosure concept and further distinguishes itself from the other categories by being an *a priori* feature of the process and not a feature of the dataset. Its parameter $\varepsilon$, known as a privacy (loss) budget, is crucial to characterize the privacy of resulting datasets and additional privacy evaluations beyond the privacy budget are not typically conducted. In this sense, it can be seen as a stand-alone category.

Each category is defined, illustrated through exemplar metrics, and this is followed by a critical appraisal. The example metrics covered in this report are not intended to provide a comprehensive coverage of all metrics in a category, but to highlight specific ways that metrics that fall into a respective category have been defined. Based on the critical appraisal, recommendations are made to be rated by the expert panel.

## 1.3 General Considerations and Principles for Evaluating Disclosure Vulnerability

The following items are important to highlight as they are consistent themes that need to be addressed for each disclosure vulnerability category. While these can be discussed in a general sense, our analysis of privacy metrics suggests that they need to be discussed within the context of the disclosure vulnerability category.

### 1.3.1 Perspectives on Disclosure Vulnerability

When measuring disclosure vulnerability, there are three primary perspectives that one can take: the data subject, the data controller, and the adversary. Each perspective will impose certain constraints on how metrics are defined. For example, a data controller will not have the exact target individuals that the adversary would want to conduct, say, a membership attack on. Therefore, the controller can define a metric that uses holdout samples from the original dataset as well as samples from the training dataset to mimic data that an adversary would have. Where relevant, we will highlight how different perspectives are considered in metric definitions.

The calculation of privacy metrics typically tries to mimic the adversary's perspective utilizing the information that are available to the data controller (e.g., the original and synthetic dataset). The data subject perspective typically informs privacy metrics on a more conceptual level, particularly in determining acceptable reference points or baselines (see 1.3.3).

### 1.3.2   Types of Decisions Based on Disclosure Vulnerability Metrics

The evaluation of privacy in synthetic data can serve different purposes and a disclosure vulnerability metric can trigger different types of decisions. The following three decisions can be distinguished:

1. Benchmarking of SDG models. The vulnerability metric is used to compare different SDG models.

2. Optimizing an SDG model. During training an SDG model is optimized to minimize the disclosure vulnerability (and maximize utility).

3. Releasing synthetic data. A data controller needs to make a binary release / no-release decision. Here we assume that there is no SDG model release but one (or multiple) synthetic datasets are being released. Releasing does not imply that the synthetic data is then publicly available. It could be a limited non-public access or sharing scenario.

Considering these decision-making processes allows for a more focused discussion on the privacy metrics and an easier choice of the "best" metrics. Metrics that inform the binary decision on the release of synthetic data may have more requirements than those for benchmarking or optimizing. We will take this into account whenever reasonable.

Another dimension to the type of decisions is what we are evaluating. There are three ways to aggregate disclosure vulnerability metrics:

a) Disclosure vulnerability of a *particular synthetic dataset*.

b) Disclosure vulnerability of a *specific trained SDG model*. This can be the aggregated vulnerability value from (a) across multiple synthetic datasets generated from the same model.

c) Disclosure vulnerability of a *class of SDG models*. This can be the aggregated vulnerability value from (b) across multiple trained SDG models on different datasets.

Evaluating classes of SDG models as in (c) is challenging because disclosure vulnerability can differ significantly among and also within classes. For example, one study evaluated 6 different generative adversarial network (GAN) models on two datasets and found that privacy depended not only on the GAN model but also on the dataset utilized for evaluation [23]. A similar conclusion can be drawn from a study that evaluated 8 different SDG techniques on twelve datasets and showed that the best performing model varied considerably depending on the dataset [24]. While this example focused only on utility, it is very likely that variation can also be observed in privacy given the experiences in the first example. Therefore, one can question whether aggregation across different datasets as in (c), with expected high variation, would provide actionable results. The disclosure vulnerability of an SDG model will be tied to the characteristics of the dataset. This is the reason why we consider situation (a) and (b) but not (c) in our analysis.

### 1.3.3   Reporting Disclosure Vulnerability: Absolute and Relative Metrics

We make a distinction between two types of disclosure metrics: absolute and relative. Disclosure can be reported, for example, as a simple measure such as prediction model accuracy in attribute disclosure [25–28]. This would be an absolute metric. Relative metrics, in contrast, are metrics that are compared to another value. For example, in attribute disclosure the prediction model accuracy of a model trained on synthetic data (absolute metric) can be compared to the accuracy when making predictions based on the univariate distribution of the original data (i.e., naïve guess). Such a value, in this example the accuracy when making predictions for a sensitive attribute based on the univariate distribution of the original data, provides a reference point and we refer to that reference point as a *baseline*. The combination of absolute metric and baseline is then the relative metric. In this sense, relative metrics give the incremental vulnerability of synthetic data relative to a baseline. For example, if $V_{synthetic}$ is the absolute vulnerability of the synthetic data and $V_{naive}$ is considered as a baseline, then the relative metric $V_{rel}$ can be reported as:

$$V_{rel} = V_{synthetic} - V_{naive} \qquad (2)$$

or

$$V_{rel} = \frac{V_{synthetic} - V_{naive}}{V_{naive}} \qquad (3)$$

Another way to describe absolute metrics is to say that there is a main attack that exploits synthetic data to learn information about a target (i.e., absolute measure, which in this case would be $V_{synthetic}$), and further attacks (i.e., baselines, which in this case would be $V_{naive}$) that assume different sources of information that we compare against. The relative metric can then be a subtraction (as above in equation (2)) or a ratio (as above in equation (3)) taking both, the main and the baseline attack, into account. Selecting a meaningful baseline is a non-trivial and error-prone task [12]. Some baselines solve some problems and create others, as will be demonstrated.

### 1.3.4   Selecting Threshold(s)

Whether a disclosure vulnerability metric is absolute or relative, having a threshold to compare against is necessary to interpret the metric and for decision-making. This is particularly true for a binary decision on the release of synthetic data where thresholds are needed to determine whether a vulnerability metric's value is too high or acceptable, but also can be for optimizing a model.

A common way to define thresholds is to rely on precedents. For example, precedents have been used to set thresholds for (absolute) identity disclosure vulnerability with anonymized data [7,17,29]. Also, precedents for relative thresholds have been used, such as the 0.2 threshold for a relative synthetic data vulnerability metric [30–32].

Consistent with that approach, other disciplines have used empirical benchmarks computed from the literature based on past performance as thresholds for acceptable performance, for example for the defect rate in software development an average industry defect rate can be set as a threshold [33]. In software process assessments, an acceptable interrater agreement has been defined based on 70 process examples [34]. In epidemiology, thresholds for statistical or performance metrics have also been proposed based on literature [35,36]. For disclosure vulnerability, one could also then look at current benchmarks to set thresholds [37].

Therefore, using precedents in the form of commonly established practices or through empirical benchmarks is a widely adopted approach to set thresholds to interpret values on quantitative scales. Consistent with that general approach, we will also use relevant precedents, where available, to recommend thresholds in our analysis.

It should also be noted that, in practice, thresholds that are recommended based on precedents may be adjusted up or down based on the context. Such adjustments need to be considered carefully and be justified. This context has been characterized as the Invasion of Privacy construct [15], and was characterized by three dimensions:

1. The sensitivity of the data. If data are particularly sensitive then the recommended threshold may be adjusted to be more conservative.

2. Potential harm. If the potential harm to the data subjects from an inappropriate disclosure is high, then then the recommended threshold may be adjusted to be more conservative. This may or may not be correlated with sensitivity.

3. Appropriateness of consent and notice. To the extent that the data subjects have consented, been consulted, or been notified to / about the disclosure, even if the data has been through an SDG process, then the recommended threshold may be adjusted to be more permissive.

And we add two more dimensions as part of these criteria:

4. Benefits to data subjects and society. If the disclosure would be highly beneficial to the data subjects (e.g., in the case of a natural disaster) or to broader society, then the recommended threshold may be adjusted to be more permissive to increase its utility and hence its potential benefit.

5. Court or regulator order. If a court orders a particular threshold to be used or there is a regulator order, and that is inconsistent with a recommended threshold, then the order threshold would be used, whether it is more permissive or conservative.

This means that the recommended thresholds are used as anchors and adjustments can be implemented when they are applied in real world situations. These adjustments would need to be justified and documented on a case-by-case basis, while still allowing for the recommended values to be the default. Another important consideration is that all the recommended privacy metrics must be below the threshold for a synthetic dataset or specific trained SDG model to be deemed to have low disclosure vulnerability. This means that if, for example, a synthetic dataset has a low membership disclosure vulnerability but a high attribute disclosure vulnerability, it is considered to have a high disclosure vulnerability: for data custodians and data subjects, all applicable disclosure vulnerabilities would need to be addressed simultaneously.

## 1.4 Overarching Recommendations

There are some weaknesses that have been encountered in multiple metrics across all categories. Their consequences are consistent and not dependent on the type of disclosure vulnerability. These are presented below so as not to repeat the same points throughout the category sections. For some weaknesses, there are metrics that present solutions and we briefly discuss why their solutions can be general recommendations for all privacy metrics in synthetic data. In other cases, the question is more open and different approaches are presented and analyzed.

### 1.4.1 Using Available Information to the Adversary: Quasi-Identifiers

Many of the metrics that are used in the SDG literature assume that the adversary knows all of the attributes in the dataset rather than knowing just the quasi-identifiers. Decades of research on identity disclosure in anonymized data and respective guidelines, however, are based on the adversary only knowing the quasi-identifiers [7,15–17,38–42], and therefore would only be able to match target records on these quasi-identifiers. Furthermore, the concept of adversary power in this literature is sometimes used [43–47], which stipulates that even if there are many quasi-identifiers in a dataset, a specific adversary will likely only know a subset of them as background knowledge about a target individual.

In practice, almost all known identity disclosure attacks have been performed using the quasi-identifiers [48–51]. Guidance for determining quasi-identifiers has been published [15,52]. The External Guidance on the Implementation of the European Medicines Agency policy on the publication of clinical data for medicinal products for human use, for example, provides three conditions that need to be considered in the determination of a quasi-identifier: replicability (the stability of an attribute value over time), distinguishability (the variability of an attribute among individuals) and knowability (the availability of the attribute value to an adversary) [17]. All of these conditions must be met for the attribute to qualify as a quasi-identifier. The US Subcommittee on Disclosure Limitation Methodology refers to similar criteria to assess the risk of information [53]. In [54], these are summarized as replication, resource availability and distinguishability. Quasi-identifiers can also be informed by the types of information that is used in published disclosure attacks (for example, see [50,55]).

While this is the de facto assumption and addressed within the International Organization for Standardization (ISO) standard on Information security, cybersecurity and privacy protection – Privacy enhancing data de-identification framework (ISO/IEC 27559) [7] in the context of anonymization, the identification of quasi-identifiers is context-dependent and thereby involves a certain degree of subjectivity. Therefore, the question is whether this should also be taken as a reasonable assumption under SDG. For example, the diagnosis of obesity or diabetes mellitus can be considered as a quasi-identifier since its knowability can be quite high (e.g., public awareness, photos, discussions on social media) while the diagnosis of interstitial nephropathy may not be a familiar term to the patient themselves. Consequently, it can be challenging to label the attribute *diagnosis* as either quasi-identifying or non-quasi-identifying. Also, the context of the data may shift over time and an attribute that was once deemed not to be a quasi-identifier may then be considered one.

Some may argue that considering all attributes in the dataset as being knowable by an adversary accounts for the worst-case scenario (i.e., the highest vulnerability value). Unlike with anonymization, however, depending on the calculation of the metric, using all of the attributes may not always translate into calculating the highest disclosure vulnerability. The reason is that SDG should disrupt the one-to-one mapping between synthetic and real training records. This has two relevant implications: First, when matching as part of the vulnerability calculation is done on the synthetic data, the more values required to match, the higher the likelihood that a synthetic value may differ from the real one decreasing the matching rate. And second, even when there is a match, the synthetic sensitive value may still differ so that no correct sensitive information is learned. This interplay is illustrated in simplified form in Figure 1.

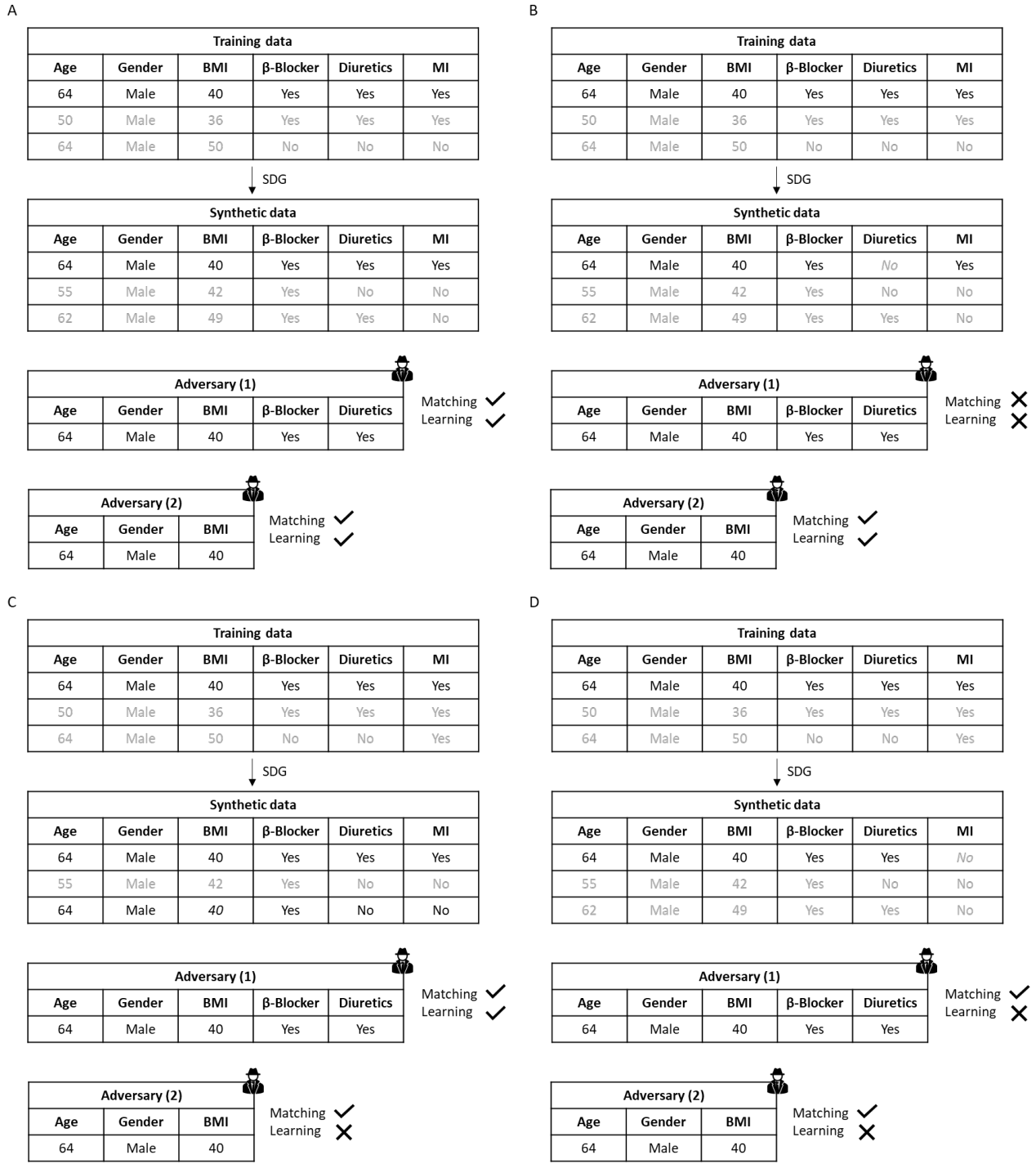

**Figure 1: Learning the sensitive attribute when knowing all attributes or quasi-identifiers only**. Two adversaries are illustrated: (1) an adversary that knows all attributes for a target that is part of the training dataset and (2) one that knows quasi-identifiers only. Their success in learning the sensitive attribute myocardial infarction (MI) is exemplified in four different synthetic datasets: (A) with a replicate of the target record, (B) with a similar record except for one (non-quasi-identifying) attribute, (C) with a replicate of the target record but another record that matches the quasi-identifiers and (D) with a similar record except for the sensitive attribute. BMI: Body-Mass Index (in kg/m$^2$), MI: Myocardial Infarction.

For example, if an adversary knows all except the sensitive attributes for a target record and this record was used and fully replicated in SDG, then they can achieve a 100% accuracy in learning the sensitive attribute by matching the target with its replicate in the synthetic data (Figure 1 , scenario A). They achieve the 100% accuracy if there is only one match, but they can also achieve it if there are multiple matches, provided they have the same value of the sensitive attribute. In this scenario, the assumption of knowing all attributes reflects the highest vulnerability and thereby accounts for a worst-case scenario. Note that, in this scenario, matching on less attributes could also result in a 100% accuracy if either the matching still results in one matched synthetic record or if all synthetic matches share the same sensitive value.

Changing the scenario to replicate the record except for one attribute during SDG causes the matching process with all attributes to fail (Figure 1, scenario B). Hence, more could be learned when matching on less attributes, provided that they are good predictors of the sensitive attribute. In such a case, considering all attributes to calculate the metric does not give the highest disclosure vulnerability possible and thereby not the worst-case scenario.

Scenarios C and D illustrate other situations where having more attributes can be beneficial (Figure 1, scenario C) or would make no difference (Figure 1, scenario D). These examples illustrate that the impact of assuming background information by the adversary will depend on the characteristics of the dataset and the SDG model. The impact of variable selection was, for example, observed in [76] where an increase in the number of attributes used to calculate attribute disclosure did not indirectly or directly correlate with the vulnerability estimate.

This is not to say that an adversary who knows more attributes cannot be a worst-case scenario but rather, using all attributes in their attack is not necessarily a worst-case scenario. More generally, for any specific number of known attributes, any combination of those attributes can be used by the adversary, and the match rate will vary depending on the combination chosen. This will have an impact on the disclosure vulnerability measurement, and we illustrate that below with membership disclosure. For example, if 20 attributes are considered knowable by an adversary, the adversary can potentially achieve a higher disclosure vulnerability by considering a subset of ten attributes, and some combinations of ten out of 20 will have higher vulnerability than other combinations. This further complicates the analysis by the controller about which set of attributes represent the worst case.

We illustrate these points below through a simulation with a membership disclosure metric where matching is an essential part of the overall calculation (see Section 3.1). We used the Washington state 2008 dataset that contains the patients' hospital discharge data from the State Inpatient Database (SID).[1] From this dataset, we randomly chose 50,000 records and selected 20 attributes. The 50,000 records were considered as the population. An SDG training dataset with 10,000 records was then randomly sampled from the population as well as an attack dataset of 10,000 attack records. A synthetic dataset of the same size was created using a Bayesian Network SDG model as implemented by Synthcity [56]. After discretizing continuous attributes[2], we matched the attack records with the synthetic dataset. An attack record could then be:

- true positive: matches the synthetic dataset and is a member of the SDG training dataset;

- false positive: matches the synthetic dataset but is a non-member of the SDG training dataset;

---

[1] The Healthcare Cost and Utilization Project (HCUP), from the Agency for Healthcare Research and Quality. This dataset is available for purchase at https://hcup-us.ahrq.gov/tech_assist/centdist.jsp

[2] Discretizing means converting continuous values into categories. This allows for some deviation from the real numerical value while still being considered as a match. In our simulation, we discretized the value space of a continuous attribute into 20 bins ranging from the minimum to the maximum value.

- true negative: does not match the synthetic dataset and is a non-member of the SDG training dataset; and

- false negative: does not match the synthetic dataset but is a member of the SDG training dataset.

From these numbers, we calculated precision, recall and the F1 score which are the typical metrics used to report membership disclosure (see Figure 2, J-L). For illustrative purposes, we also matched the attack records with the training dataset (see Figure 2, A-F), which reflects a baseline where matching with a non-synthetic dataset is performed.

If the adversary guesses that all of the attack records are in the training dataset, then this naïve guess F1 score will be 0.33 (see [31]). The naïve F1 score gives us the success rate of the membership disclosure attack in our simulation if the adversary did not use the synthetic data.

The number of attributes considered was varied starting from subsets of 1 to 20. For each subset all combinations were considered, which gave us $\sum_{j=1}^{20} \binom{20}{j}$ combinations in total (1,048,575 combinations).

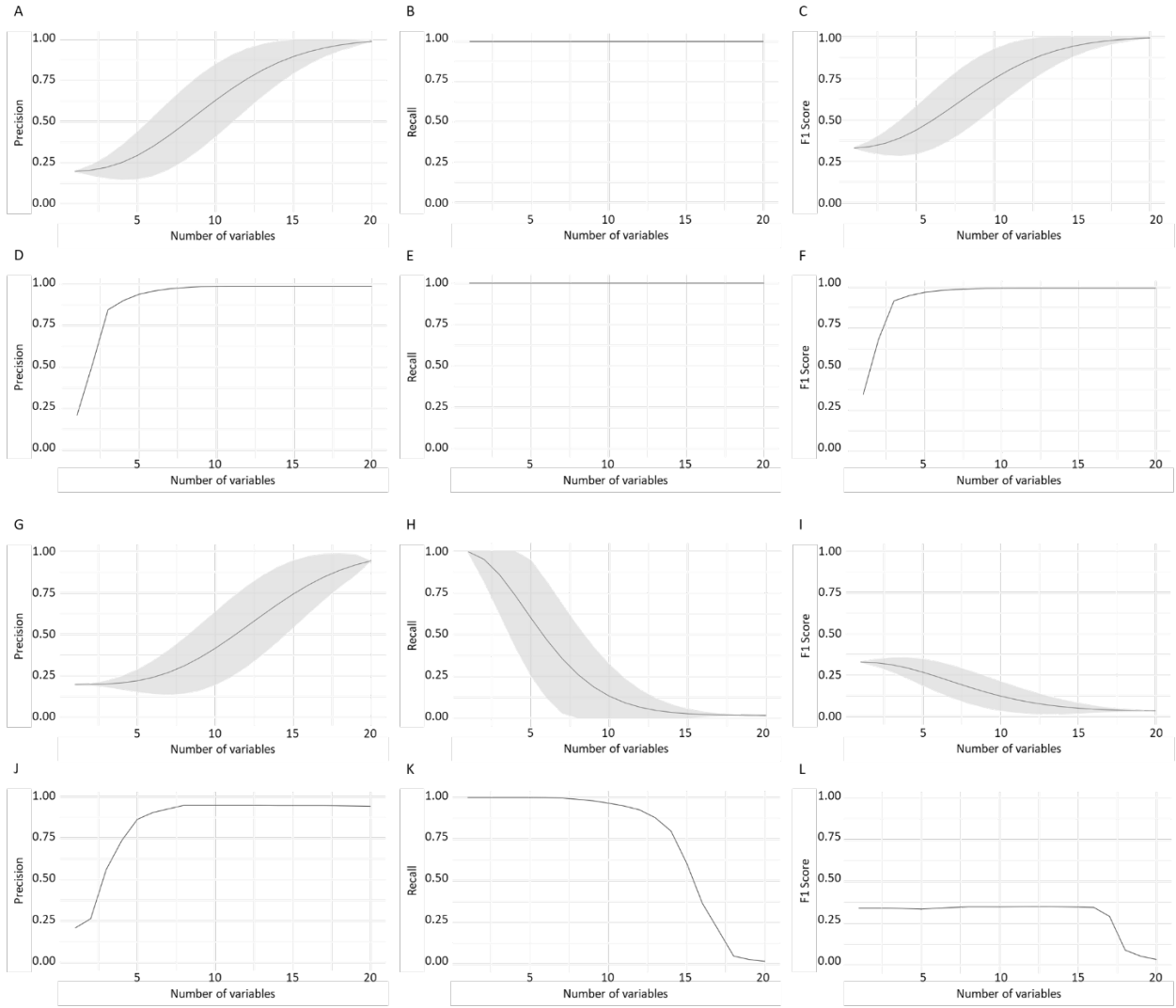

**Figure 2: Membership Disclosure when Varying Number of Attributes for Matching**. Membership disclosure was simulated by drawing an attack dataset from the same population as the training dataset. A Bayesian network SDG model was used to generate the synthetic data. Attack records were then matched with training records (A-H) and synthetic records (G-L). Average and standard deviation of precision, recall and F1 score was calculated from each combination of attributes (A-C, G-I). This means that for each number of all attributes, all combinations out of 20 were evaluated. Maximum precision, recall and F1 score was calculated by choosing the maximum value from all combinations for each number of attributes (D-F, J-L).

A number of observations can be made from these plots:

- In the case of the baseline (plots A-F), the more attributes considered as background knowledge of the adversary, the greater the correct match rate, and hence the greater the disclosure vulnerability. Therefore, in those cases there is no advantage for the adversary to consider subsets and the worst case is the greatest number of attributes.

- In the case of SDG (plots G-L), the more attributes that are considered as part of the background knowledge of the adversary the worse the match rate (see plots H and I), on average. In our particular example, the membership disclosure vulnerability measured as an F1 score decreases with more attributes considered.

- As more attributes are considered, the F1 score decreases below the 0.33 naïve guess score. This means that if the controller was evaluating membership disclosure and was using, say, 17 attributes, the evidence would suggest that the vulnerability is indeed very low and not better than guessing. But this may not be correct.

- Because of this effect in synthetic data, the adversary is motivated to consider subsets of the attributes to maximize the match rate. In fact, there is no compelling reason for the adversary to consider all of the attributes in the matching process.

- When the adversary considers fewer attributes, for every number of attributes there is variation in the membership disclosure vulnerability depending on the combination of attributes that are considered. For example, if the adversary considers ten attributes, some combinations of ten attributes will have better matches and higher membership vulnerability than others.

- Because of this variability, there will be combinations where the F1 score in plot I will be higher than the naïve F1 score, especially at a small number of attributes. This may lead to a different conclusion about whether the vulnerability is acceptable or not.

- The data controller will not know a priori which combination the adversary will try, and therefore it would be prudent for the controller to try all combinations for each number of attributes and take the maximum as the worst case estimate of membership vulnerability.

While it is not shown here, if the adversary generalizes the attributes, then the membership disclosure vulnerability will also change. Therefore, the controller needs to consider all combinations and for each combination all possible generalizations to find the maximum membership disclosure vulnerability (for every combination a generalization lattice can be constructed and a search performed for the highest vulnerability node as in [57]). Depending on the number of attributes and rows in the dataset, this can be a computationally very demanding calculation of disclosure vulnerability. This behavior for SDG is quite different than for anonymization, and therefore the assumptions made in an anonymization context do not necessarily carry forward to SDG.

This analysis, illustrated in both Figure 1 and Figure 2, shows that by considering all of the attributes in the vulnerability calculations, the controller may be significantly underestimating the true vulnerability and potentially releasing datasets that violate the privacy of data subjects. A more prudent approach is to carefully consider which attributes are likely known by an adversary. Once these attributes are determined then evaluate the vulnerability on all combinations of subsets and generalizations and use the maximum across these as the worst case value. Such an approach can decrease the number of combinations that would have to be considered to those that represent realistic scenarios in line with adversarial assumptions. It can thereby avoid computationally demanding situations. If the adversarial knowledge is assumed to be the full record, then the search may become infeasible. While there are ways to address high complexity in search problems, there is currently no such method established. Another way may be to avoid such search by optimizing the matching strategy, e.g. by using dimension-reduction methods or allowing for a certain level of discrepancy when matching on more attributes.

On the other hand, the assumption that the adversary knows all of the attributes makes the analysis more amenable to automation since it avoids a largely subjective and manual step of classifying attributes as quasi-identifiers.

Given these considerations, the subjectivity involved in determining quasi-identifiers cannot easily be avoided. Treating all attributes except the sensitive one as quasi-identifiers is itself an assumption. There may be situations where this is a reasonable consideration but as shown above considering all of them in vulnerability calculations may not necessarily be the worst-case scenario. It is necessary to explicitly take and communicate such an assumption. Otherwise, interpreting the disclosure vulnerability can become challenging. Therefore, quasi-identifiers may vary and are ascertained by the data controller for a given dataset. Furthermore, from a risk management perspective the quasi-identifiers may differ depending on the synthetic data processor or class of processor. For example, in the disclosure control community risk is assessed from the perspective of the "anticipated recipient" under the US HIPAA privacy rule, and a similar interpretation is emerging based European court cases [58]. All of that must be considered when undertaking the critical step of determining quasi-identifiers.

### 1.4.2 Accounting for the Stochasticity of SDG

Another question is whether and how to account for the stochasticity of the SDG process? While some metrics recognize the randomness of an SDG model and account for it [25,26,59], not all follow general rules for combining multiple estimates. Stochasticity has also been shown to be relevant in utility evaluation of synthetic data [60], whereby it is recommended to average across 10 synthetic datasets to reach a plateau in utility metrics synthetic data [61].

For the decision-making scenarios of benchmarking and optimization, it would be more appropriate to aggregate the disclosure vulnerability results across multiple synthetic datasets from the same SDG model. This would provide a general result about the behavior of the model for a specific training dataset. However, if the decision-making scenario is data release, then the disclosure vulnerability for the specific synthetic dataset(s) may be most relevant. Given that it is not always possible to determine a priori the exact decision-making scenario, it would be prudent to have both types of results. These match items (a) and (b) in Section 1.3.2.

### 1.4.3 Disclosure Vulnerability should be calculated on all records

A more open question is how to extrapolate disclosure vulnerability from individual to dataset-level. Similar to the identity disclosure assessment in anonymized data [15], we could think of the maximum probability across all record, the average across all records or the number or proportion of the records that have a probability higher than a certain threshold. Privacy metrics in the SDG literature have used different strategies such as averaging [25,62] or reporting it for selected records only [25,62–64].

Selected records are often labeled as "vulnerable" records [25,59,63]. The idea is to account for the worst-case scenario, assuming that these records are those experiencing the maximum disclosure vulnerability in the dataset. Such an *a priori* assumption is, however, not necessarily true (see 4.3.2) [62,64]. Consequently, determining the maximum disclosure vulnerability of any record always involves evaluating all records in the first instance. Several disclosure vulnerability metrics are computationally very intense which, in practice, makes a pre-selection of records inevitable. Consequently, they cannot be recommended as privacy metrics as it is not feasible to determine disclosure vulnerability for all records of a dataset. This issue has been raised by other authors [62,65].

There may be scenarios where the data context specifically requires evaluating the vulnerability of certain sub-cohorts within the data. For example, understanding the vulnerability of data subjects that have undergone an abortion might be crucial as they could face relevant harm (e.g., legal prosecution under anti-abortion laws) if their data were disclosed. However, it is essential to recognize that such a vulnerability only applies to this sub-cohort and when the question is whether to release data, then the vulnerability of all data subjects must be considered.

## High Level Summary of Key Recommendations

The following are the key high-level recommendations that we make from the analysis in this report:

- A subset of attributes that reflect the background knowledge of the adversary, notwithstanding the subjectivity in their determination, should be used for computing disclosure vulnerability. These avoid the controller making unreasonably permissive assumptions that can elevate privacy violations for data subjects, and make some of the vulnerability metrics more computationally feasible.

- Record-level similarity metrics are not a good way to evaluate the identity disclosure vulnerability in synthetic data.

- There is still some uncertainty on how to apply differential privacy successfully in practice and ensure that the synthetic data has sufficient utility. It remains necessary to measure disclosure vulnerability metrics on differentially private synthetic data when the privacy budget is not close to zero to ensure that the data is safe.

- The types of disclosure vulnerability metrics that are suitable for synthetic data are membership disclosure and attribute disclosure.

- Current membership disclosure vulnerability metrics make assumptions about the sampling and distribution of the attack dataset that require accounting for the sampling fraction of the original data from the population. If the assumptions do not hold for the SDG scenario, then the current metrics may give incorrect vulnerability values.

- The implementation of membership disclosure vulnerability metrics must consider attribute subsets and combinations to find the worst case value.

- Membership disclosure is a binary classification accuracy metric, and due to dependence on prevalence, a relative measure is recommended that accounts for that prevalence.

- Attribute disclosure vulnerability can best be measured by a combination of one absolute metric and one relative metric to address the different decision-making scenarios.

- Attribute disclosure vulnerability requires a clear distinction from knowledge generation and is defined in terms of the difference of learning a sensitive attribute for individuals in the training data versus individuals not in the training data. Learning a sensitive attribute for individuals not in the training data is considered knowledge generation (e.g., research and scientific discovery).

- Thresholds for interpreting these metrics are proposed based on precedents from the literature and current practices. Thresholds enable broader adoption of SDG because they increase certainty for data controllers.

## 2.    Record-level Similarity

### 2.1    Definition

The similarity between two datasets can be measured based on various levels of abstraction. Univariate or multivariate similarity compares distributions and is typically referred to as a utility (fidelity) metric [19,20]. The concept of similarity in synthetic data privacy captures how close the values of the closest synthetic records are to the values of original records (distance to closest record (DCR)) to identify vulnerable synthetic records (VR) [66].

Record-level similarity as a privacy metric comes from the idea that synthetic data should not replicate the training data (i.e., memorization). It is intended to provide an insight into the generalization of the SDG process on the basis that generalization implies better privacy [66].

While similarity metrics can also contribute to calculating membership or attribute disclosure metrics, such usage is typically clearly identified and does not rely solely on similarity. These approaches are therefore discussed within their respective categories.

The reasoning behind using record-level similarity outside the membership or attribute disclosure context is rarely explicitly communicated. There are, however, examples where it has been used to approximate identity disclosure [67,68]. Accordingly, this section will primarily consider this interpretation.

### 2.1.1    Absolute and Relative Metrics

Multiple distance metrics can be used to measure DCR such as Euclidean distance, Hamming distance, cosine similarity or Manhattan distance [21]. DCR can then be assessed between a synthetic and a training record (STD) or the other way around from a training to a synthetic record (TSD). Note that a closest distance measurement does not result in unique matching pairs: one synthetic record may serve as closest record for multiple training records and vice versa.

This distance can serve as a metric by itself [69,70], and the number of exact matches (i.e., STD or TSD equals 0), the number of records that have an STD or TSD lower than a threshold, or the average STD or TSD across the dataset can be reported (i.e., absolute metrics). It can also be compared against some sort of baseline (i.e., relative metrics). Baselines in record-level similarity are calculated as the DCR between a synthetic and a holdout record (SHD), between two training records (TTD) or a training and a holdout record (THD) [21,71,72]. Different example metrics are illustrated in Figure 3.

Relative metrics can define VR as follows [21]:

$$VR_t = \begin{cases} 1, & if \ TSD < THD \\ 0, & \text{otherwise} \end{cases} \tag{4}$$

where $VR_t$ is the comparison of the DCR for one training record $t$ to a synthetic (TSD) and the DCR of the same training record $t$ to a holdout record (THD). Vulnerability is then claimed when the TSD is smaller than the THD. The other way around, it would take the following notation [71]

$$VR_s = \begin{cases} 1, & if \ STD < SHD \\ 0, & \text{otherwise} \end{cases} \tag{5}$$

where $VR_s$ is the comparison for one synthetic record $s$ to training (STD) and to a holdout record (SHD).

It can also take the following formula:

$$VR_t = \begin{cases} 1, & if \ TSD < TTD \\ 0, & \text{otherwise} \end{cases} \tag{6}$$

$VR_t$ is then a training record $t$ that has a TSD that is smaller than the TTD. For equation (4) and equation (6) the metric can then give the number of $VR$ as follows:

$$VR = \sum_{t=1}^{n} VR_t \tag{7}$$

where n is the size of the training dataset, for equation (5) it is the sum of $VR_s$ respectively. The identical or exact match is a special form of DCR where the DCR is zero [71,73].

These baselines create a reference point based on the original data or a holdout dataset that was sampled from the same population as the training data [71].

### 2.1.2 Application of Record-Level Similarity Metrics

When looking at the prevalent privacy metrics used for synthetic data, for example, in medicine, the top privacy metrics stem from different definitions of record-level similarity metrics [19]. Some metrics use similarity indirectly through clustering algorithms [74,75]. Most often Euclidean distance is used to measure DCR [21]. Record-level similarity metrics are, for example, implemented in the SynthGauge Python Library in the form of a simple TSD metric and as the number of identical matches [76]. Record-level similarity metrics are also used by companies as part of the privacy guarantee of their products (e.g. Amazon Web Services [77], MOSTLY AI [78], and Syntegra [30]). When presenting or evaluating SDG methods, some authors rely exclusively on absolute similarity metrics [69,73]. For example, Kotelnikov et al. introduced TabDDPM to model tabular data using diffusion models and evaluated privacy only by STD [69]. Others use relative metrics or use it as one part of a more comprehensive privacy assessment [30,71,79,80].

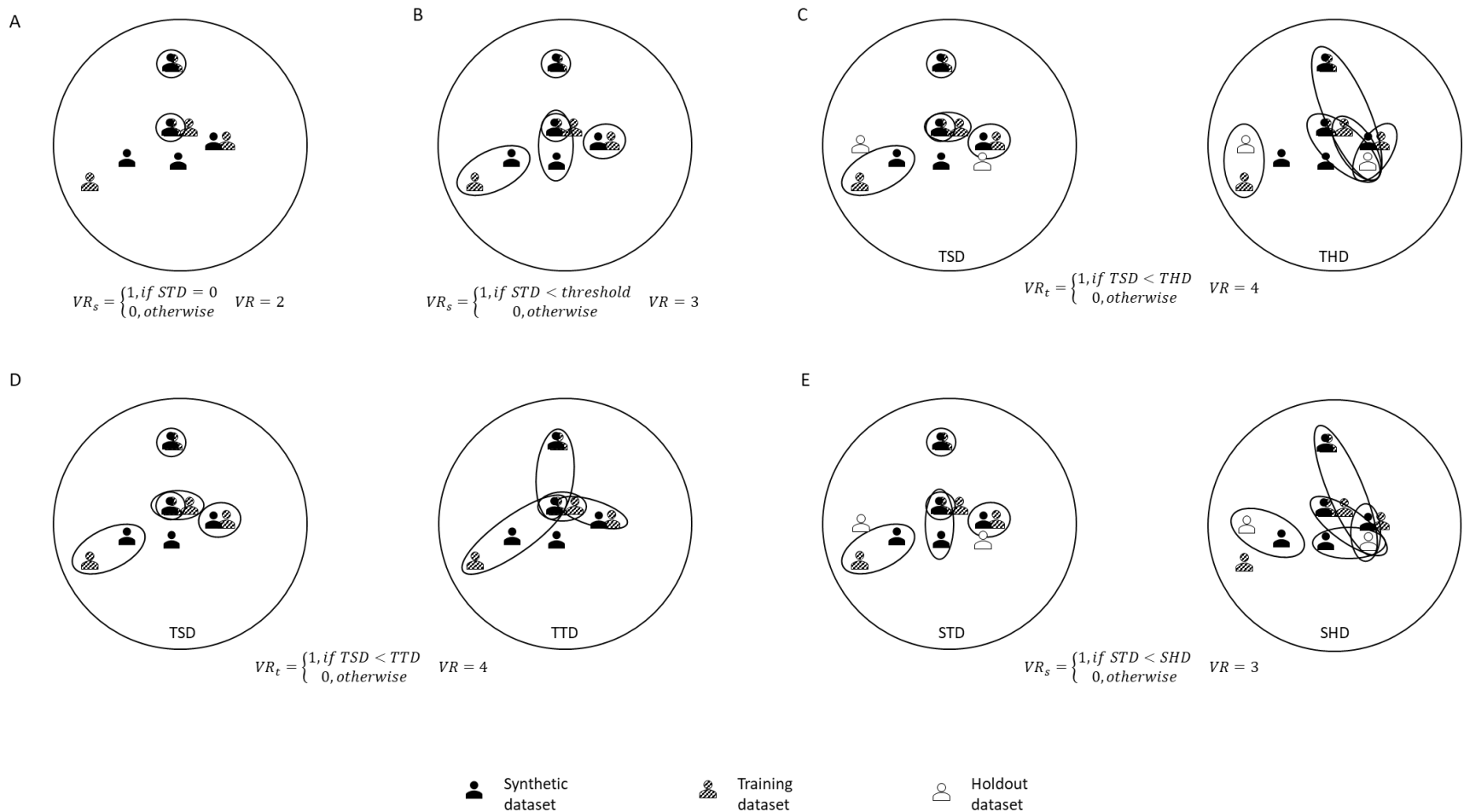

**Figure 3: Similarity Metrics.** Similarity metrics identify vulnerable records (VR). They can be absolute metrics such as the number of exact matches [73] (A), the number of records that have a synthetic-to-training distance (STD) lower than a threshold (B), or the average STD across the dataset [70]. On the other hand, the training-to-synthetic distance (TSD) can be measured and compared to the closest training-to-holdout distance (THD) [21] (C), or to the training-to-training distance (TTD) [72,81] (D). Another relative metric is the comparison of STD to the synthetic-to-holdout distance (SHD) [71] (E). The number of VR is estimated from the distances in this illustration. In (B), an arbitrary threshold was defined as the point where the illustrated individuals intersect. In this illustration, the partition between the training and holdout dataset is an approximate 70:30 split.

## 2.2 Critical Appraisal

One of the challenges of similarity metrics is that they do not necessarily indicate that there is an identity disclosure vulnerability. This is further clarified below. Note that this analysis makes the assumption that similarity is measured between the quasi-identifiers of a record which is commonly considered as the information available to an adversary in identity disclosure (see 1.4.1).

### 2.2.1 Absolute Record-level Similarity Does Not Approximate Identity Disclosure

While record-level similarity indicates a certain closeness of synthetic data to the training data, this may or may not translate into identity disclosure vulnerability. We can consider some examples to illustrate.

If there is a small number of categorical attributes, it is very likely that SDG will replicate records of the original dataset. An extreme example would be the presence of only two binary attributes in a dataset, e.g., gender (female/male) and expert in privacy metrics (yes/no). Having an original dataset of 1000 records with a uniform distribution of the four possible combinations, the synthetic dataset would include multiple replicates because there are only four possible values. Relying on an absolute distance measure, this would be considered as a high-vulnerability situation.

To obtain a more precise estimate of the ground truth, we could assess the identity disclosure vulnerability using k-anonymity [82] (see the sidebar "k-anonymity vs k-map"). K-anonymity (or k-map) is built upon the concept of equivalence classes. An equivalence class is a group of individuals that cannot be distinguished from each other. The equivalence class size (the value k in k-anonymity or k-map) translates into the vulnerability for all individuals of this very class (i.e., $1/k$). In this simple example, there would only be 4 equivalence classes (female privacy expert, male privacy expert, female privacy non-expert, male privacy non-expert). The equivalence class size in each of these classes will, assuming a uniform distribution, be ~250 translating into a small identity disclosure vulnerability that is certainly smaller than any commonly used threshold for acceptable risk [15]. This, in spite of the high similarity (i.e., small STD). This means that high similarity is not equal to high identity disclosure vulnerability. But there are some exceptions which are highlighted below.

The literature on anonymization often talks about k-anonymity. However, it is important to make a distinction between k-anonymity and k-map [82]. The most common interpretation of k-anonymity is when an adversary knows that a particular target individual is in a dataset, and the dataset is a sample from some population. In such a case the identity disclosure vulnerability of that individual is 1/k, where k is calculated based on the equivalence class size in the dataset. The other condition where this is true is when the dataset represents the population, so by definition the target individual is in the dataset.

If the adversary does not know that a target individual is in the dataset because the dataset is a sample from some population, then this is referred to as k-map. The risk is still 1/k but now the value of k refers to the equivalence class size in the population and not in the dataset. If instead the equivalence class size in the dataset is used, we would get a very conservative measure of identity disclosure vulnerability under the assumption that the adversary does not know that a target individual is in the dataset. The reason is that the equivalence class sizes in the population would be the same or larger than in the sample. Because the equivalence class sizes in the population would not be known, they need to be estimated, and there is a large body of work on that estimation problem [83–85].

Because the assumptions and results between k-anonymity and k-map are vastly different, it is important to make a distinction between them.

In practice an adversary is not likely to know that a target individual is in a particular dataset as most datasets are going to be samples. This is consistent with published re-identification attacks and uniqueness studies that are almost always conducted with samples [48]. Assuming k-map is arguably a more reliable default assumption to make.

### 2.2.2   Similarity Metrics Are Evaluating Reconstruction Not Identity Disclosure

More generally, replication of training data records does not mean that there is an identity disclosure because the training record may have had a low identity disclosure vulnerability to start off with (e.g., in a population equivalence class size of 250). This means that similarity between a synthetic and training record must be contingent on the identity disclosure for that very record in the training data. If that risk is very small, then similarity would not necessarily indicate elevated disclosure risks. On the other hand, for a training record with a high identity disclosure vulnerability plus high similarity, then that would be a scenario with unsafe generated data.

Most similarity metrics measure, in fact, the vulnerability of reconstruction which is a prerequisite for but different from identity disclosure where the equivalence class of the record in the population (not in the training dataset) must be taken into account. Taking reconstruction as a proxy for re-identification overestimates the actual risk, which would ultimately lead to poor data utility.

### 2.2.3   Outlier Training Records can be Linked to Synthetic Records By Comparing TSD to TTD

The TTD baseline gives context about the similarity of records in the training dataset (see equation (6)). It assigns a low disclosure vulnerability to replicated records in the synthetic data when there is also a replicate record in the training dataset. A large TTD combined with a small TSD (i.e., $VR_t$) generally appears in training records that are far from other training records (i.e., outlier or unique) and are replicated in the synthetic data (see Figure 4).

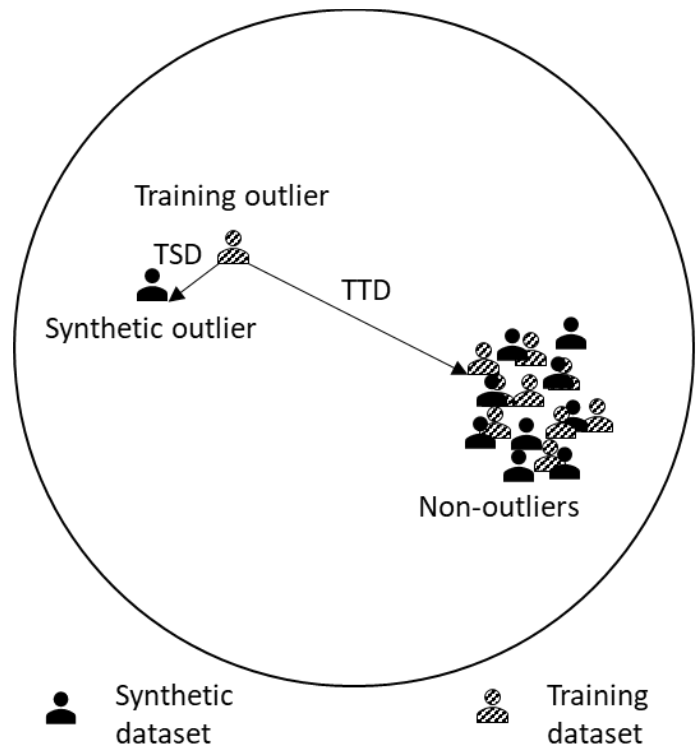

**Figure 4: Outlier detection through TTD.** The training-to-synthetic distance (TSD) can be measured and compared to the closest training-to-training distance (TTD) [81]. A vulnerable record ($VR_t$) occurs when TSD < TTD is true for the training outlier as in this illustration. Based on an image from [21].

Such a metric can successfully establish a link between a synthetic record and a real target if they are

1. an outlier in the training dataset, and

2. known by the adversary to be in the training dataset

For synthetic data, however, such a disclosure may be better described as membership or attribute disclosure (see 2.2.6).

### 2.2.4   Comparing STD to SHD can be An Indicator of Identity Disclosure Vulnerability When the Sample is Large Enough

Comparing STD to SHD is another example where a link between a synthetic record and a real target can be successfully established. In contrast to the assumptions on the target in 2.2.3, a link can be established to a non-outlier target in the population, but only in the case of a holdout sample size large enough. We illustrate this through the following simulation.

The relative vulnerability derived from a holdout baseline is calculated as shown in equation (5) [71]. For simplicity, let's consider exact or identical matches as the similarity metric, which is when the distance is zero. We calculate if there is an exact match between a synthetic record and a training record ($STD = 0$) and/or a holdout record ($SHD = 0$) [71]. The formula for relative vulnerability per record would then be:

$$VR_s = \begin{cases} 1, & if \ STD = 0 \wedge \ SHD \ != 0 \\ 0, & otherwise \end{cases} \qquad (8)$$

The more non-matches (*SHD != 0*) are in the holdout dataset, the more $VR_s$ are present, and vice versa. Consequently, the proportion of exact matches between the synthetic and the holdout dataset should decrease with increasing actual identity disclosure vulnerability to accurately mimic the vulnerability of the original dataset. This would allow relative similarity metrics to accurately gauge the identity disclosure vulnerability of the original dataset.

To demonstrate how SHD can be an indicator of original data identity disclosure vulnerability, we ran a simulation with populations having different identity disclosure risks. The ground truth vulnerability was determined using k-map. Note that we defined k-map vulnerability in the population as we assumed an adversary that does not know that a target is in the dataset [82,86,87]. The actual vulnerabilities ranged from 5% to 100% (i.e., k between 1-20) and were the same for all records within a population. The population size was fixed at 100,000. Original datasets of different sizes (between 10-10,000 by increments of 50) were randomly drawn from the population. Each original dataset was partitioned into a training and holdout dataset with a ratio 80:20 respectively. We then generated a synthetic dataset of the same size as the training dataset via resampling from the training dataset to mimic a poorly generalizing SDG process. Consequently, the proportion of exact matches between synthetic and training records (STD) was 100% independent of the actual identity disclosure vulnerability. Exact matches were then determined between the synthetic and the holdout set (SHD) to analyze its appropriateness to gauge the identity disclosure vulnerability. The simulation results are shown in Figure 5. When having a small original dataset, a low-identity disclosure vulnerability scenario (i.e., 5%) could hardly be distinguished from a high-vulnerability scenario (i.e. 100%), while differences became larger with the dataset size.

This means that the number of exact matches on the hold-out dataset seems to be monotonically related to the identity disclosure vulnerability of the original dataset (as measured using k-map). However, because this relationship is highly dependent on the original dataset size, it is not possible to know whether the disclosure is high or low in an uncalibrated measure of similarity. A link that reflects a disclosure vulnerability can thus only be established if the dataset is sufficiently large though the definition of "sufficiently large" remains unknown.

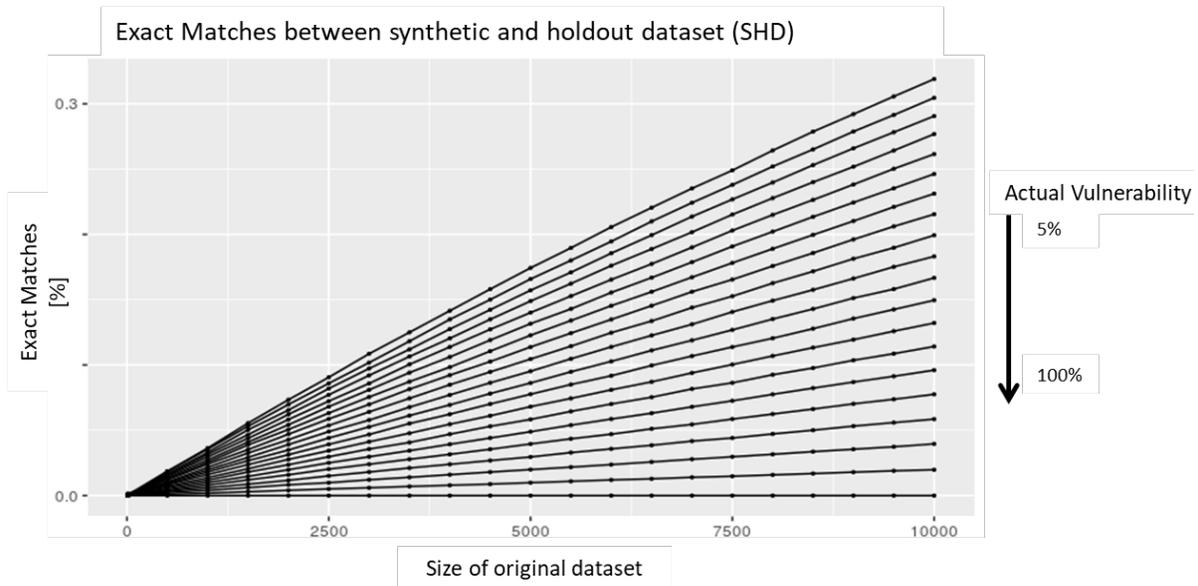

**Figure 5: Dependence of exact matches between synthetic and holdout dataset on dataset size with fixed sampling fraction.** The different actual vulnerability scenarios decrease from 100% to 5% as measured by k-anonymity. The sampling fraction was fixed at 80:20 between training and holdout dataset.

### 2.2.5 Record-level Similarity Must be Contingent on Identity Disclosure Vulnerability on the Record-Level not on the Dataset-Level

It is difficult to interpret similarity metrics when the identity disclosure vulnerability of the original data is low. A valid question then is why would we use SDG at all and have to calculate privacy metrics when the original dataset is of low vulnerability anyway? The application of SDG can reduce the utility of the data that is made available to the data users. Since one of the primary use cases for SDG is to enable responsible sharing of synthetic data, if the real dataset already has low identity disclosure, would it not make sense to just use and disclose that, and eschew the application of additional SDG? Would similarity metrics consequently be meaningful only when applied in scenarios where the real dataset has high identity disclosure?

There are two important caveats to consider. Firstly, SDG may still be applied to reduce other vulnerabilities in the data even if identity disclosure is low (e.g. attribute disclosure and membership disclosure). This means that in principle when the identity disclosure vulnerability is low, there could still be an incremental protective value in applying SDG.

Secondly, similarity metrics need to be contingent on identity disclosure on the record-level not only on the dataset-level. This means that the implications drawn from a dataset-level do not necessarily reflect the actual vulnerability. The four possible combinations of the identity disclosure vulnerability and the similarity metric are illustrated in Figure 6.

**Identity disclosure vulnerability of the training data**

|  |  | Low | High |
|---|---|---|---|
| **Record-level similarity** | **Low** | Low vulnerability | Low vulnerability |
|  | **High** | Low vulnerability | Unclear |

**Figure 6**: **Implications drawn from combinations of identity disclosure vulnerability and similarity metrics on dataset level.** Identity disclosure is measured for the training dataset. Similarity is measured between the synthetic and the training dataset. The vulnerability is then defined for the synthetic dataset.

Intuitively, there is no vulnerability if there is only low similarity. But let's look at the scenarios where similarity is high. Consider the realistic scenario where there are multiple equivalence classes that have very different sizes in the population.

If one equivalence class size is 250 and the identity disclosure vulnerability of the training dataset consequently very low (i.e., 0.004), then any synthetic record that has a high similarity can achieve a vulnerability that is at maximum 0.004. This would still be considered a low vulnerability. Let's assume there is another smaller equivalence class, one of size 2, then the identity disclosure vulnerability of the whole training dataset would be high (i.e., the maximum across all of the records, which would be 0.5).

The vulnerability of the synthetic data could still be low if high similarity is only measured in records of the equivalence class of size 250. It could, however, also be high, if high similarity is measured for a record of the equivalence class of size 2. In this sense, vulnerability depends on two conditions:

1. the synthetic record is similar to a training record, and

2. that the training record has high identity disclosure vulnerability.

The disclosure vulnerability would not depend on whether there is high similarity in general across the dataset.

Different from our simulation, equivalence classes of various sizes (i.e., identity disclosure vulnerabilities) exist side by side in most datasets. Accordingly, similarity metrics must be decided on the record (or equivalence class) level. The only scenario where simple similarity metrics may be meaningful is when the original dataset is of high disclosure vulnerability and equivalence class sizes do not differ in a marked manner, which is not likely to be true in uncurated real datasets.

## 2.2.6   Identity Disclosure in Synthetic Data is More Precisely Captured Under the Concepts of Attribute or Membership Disclosure

The concept of identity disclosure often encompasses the idea of singling out or linking records [7,88]. If we take k-anonymity or k-map as a common model for identity disclosure, this idea is reflected in equivalence classes where the size of the class translates into identity disclosure vulnerability [82]. In anonymized data, identity disclosure generally aligns with learning something new: when singling out a unique record based on quasi-identifiers, the adversary correctly learns revealed sensitive information; when matching a target with a record of the equivalence class size 3, then the adversary can correctly infer sensitive information with a minimum probability of 1/3. Note that their success can be higher if the variability of the sensitive information in this equivalence class is low, but it cannot be less than 1/3 (see Figure 7).

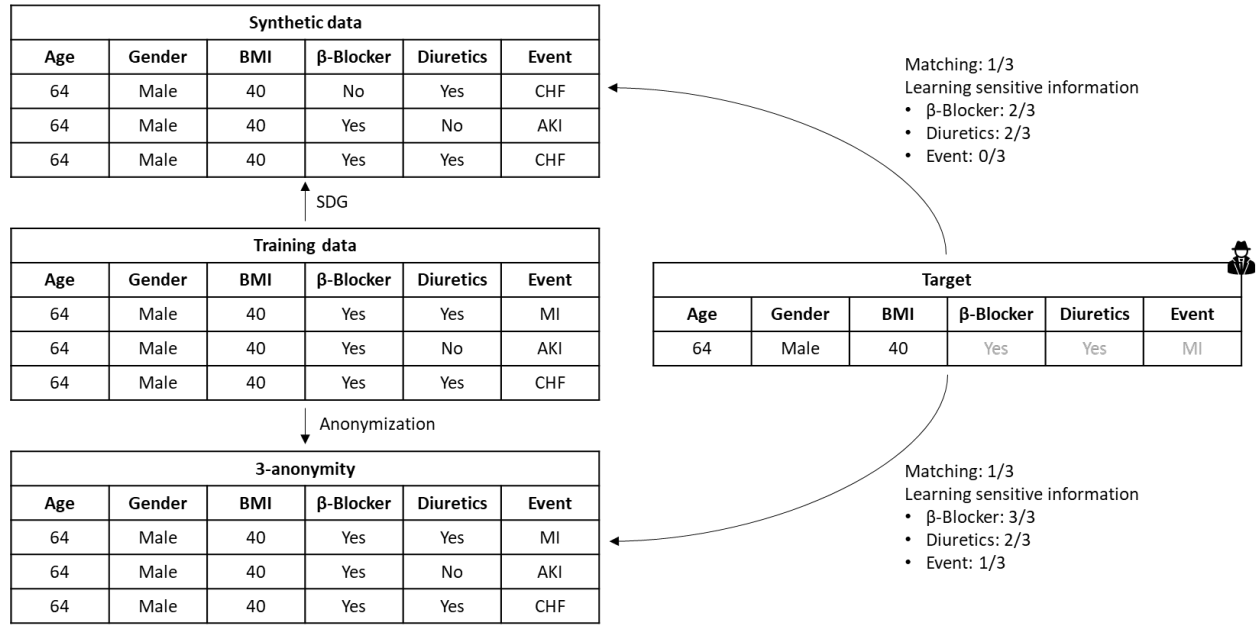

**Figure 7: Contrasting the Concept of Identity Disclosure in Anonymized and Synthetic Data**. An adversary is illustrated that matches a target to the anonymized and synthetic data. The probability of matching is negatively correlated to the equivalence group size and translates in k-anonymity to the identity disclosure vulnerability (i.e., 1/k). In this example, the equivalence group size is 3 (i.e., 3-anonymity). In anonymized data, the probability of correctly inferring sensitive information is limited downwards by the probability of matching. In synthetic data, the probability of correctly inferring sensitive information can be lower than the probability of matching. BMI: Body-Mass Index (in kg/m²), CHF: Congestive Heart Failure, AKI: Acute Kidney Injury, MI: Myocardial Infarction.

On the other hand, in synthetic data, attribute alignments are disrupted. This means that an identity disclosure may or may not come with learning sensitive information. If sensitive information of the synthetic record does not align with the training record, then the adversary does not learn correct sensitive information even if singling out a unique individual. In Figure 8, an adversary matches three targets of an equivalence group of size 1 to synthetic data. According to the common models for identity disclosure, this would translate to a vulnerability of 1. Intuitively, the vulnerability in the synthetic data qualitatively differs between the three targets. While all of them are singled out, sensitive attributes are learned in the first and to some extent in the third target but not in the second one. Membership is learned for all targets.

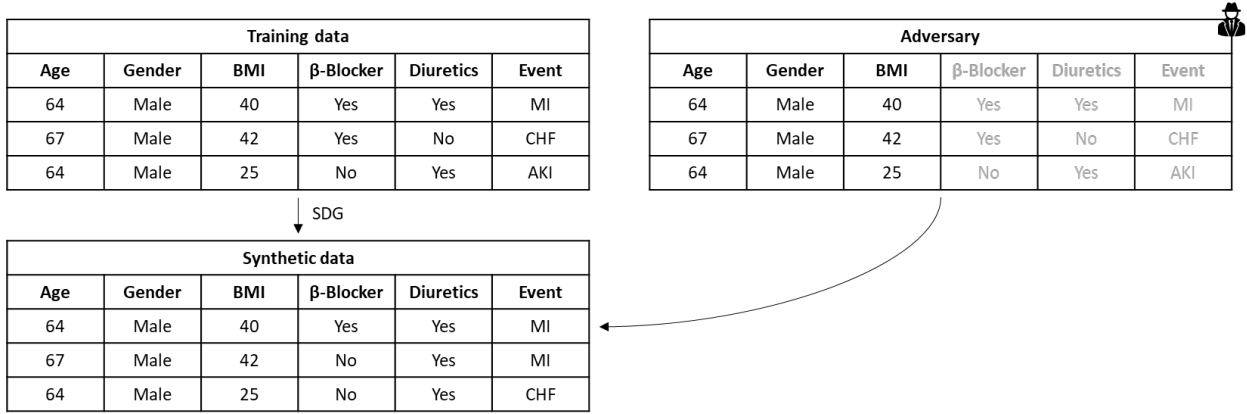

**Figure 8: Concept of Identity Disclosure in Synthetic Data**. An adversary is illustrated that matches three unique targets to synthetic data. The probability of matching is reversely correlated to the equivalence group size and translates in k-anonymity to the identity disclosure vulnerability (i.e., 1/k). In this example, the identity disclosure vulnerability would be 1. BMI: Body-Mass Index (in kg/m$^2$), CHF: Congestive Heart Failure, AKI: Acute Kidney Injury, MI: Myocardial Infarction.

These nuances make it difficult to define a concept of identity disclosure for synthetic data. An "improved" record-level similarity metric that successfully addresses the challenges discussed above may meet the definition of identity disclosure where learning new sensitive information is not an explicit assumption but the question then arises as to what this metric adds in measuring disclosure vulnerability in synthetic data: if we take the correctness of sensitive information into account, then this is the definition of an attribute disclosure metric (see 4.1); if we don't consider the correctness of sensitive information it is membership disclosure (see 3.1).

Consequently, a high record-level similarity contingent on identity disclosure of the original data is a privacy violation. This disclosure in synthetic data is, however, more precisely captured under the concepts of attribute or membership disclosure.

### 2.2.7   The Direction of Measurement and The Distance Metric Affect Results

What further complicates the interpretation of similarity metrics is the direction of calculation. When measuring TSD, there may be one synthetic record that is closest to multiple training records for example. When measuring STD, this effect would be reversed since the distance metrics are not symmetrical. Consequently, in both directions there may be records that are not used in the distance measurements and which records these are depends on the direction.

There are multiple distance metrics such as Euclidean distance, Hamming distance, cosine similarity or Manhattan distance. One would also get a different answer depending on the one that is implemented.

| **Summary Points from the Critical Appraisal of Similarity Metrics** |

- Absolute similarity metrics by themselves do not necessarily capture the concept of identity disclosure but are rather assessing reconstruction.

- Relative metrics can establish a link between a synthetic record and a real target only when certain conditions are met:

  1. Comparing TSD to TTD: The adversary knows that a target individual is in the training dataset and the record of that target individual is unique or an outlier in the training dataset

  2. Comparing STD to SHD: The metric needs to be calibrated to the size of the holdout dataset and the comparison is done at the individual record (not the dataset) level.

- The interpretation of such a link would be a disclosure that is captured under the concepts of membership disclosure or attribute disclosure.

- The interpretation of similarity metrics depends on the direction and metric of the measurement, which has been undefined in the literature.

## 2.3    Recommendations

The most important observation here is that synthetic data is not compatible with the concept of identity disclosure. Even if there were a similarity metric that was contingent on the identity disclosure vulnerability of the original data on a record-level, the question would remain as to what the adversary learns that is new from the similar record. If a synthetic record exactly matches an original record on quasi-identifiers and this original record has a high identity disclosure vulnerability, the correctness of the information learned can vary. Any "improved" metric is essentially more precisely classified as an attribute or membership disclosure metric.

Consequently, similarity metrics would not reveal a meaningful vulnerability beyond membership or attribute disclosure vulnerability in synthetic data, which calls into question their computation when used beyond these contexts to evaluate SDG privacy.

# 3.    Membership Disclosure

## 3.1    Definition

Membership disclosure is defined as the ability of an adversary to determine that a target individual was in the training dataset for the SDG model (a member of the training dataset). Predicting membership is a classification task with the labels members versus non-member.

There are two main ways to approach this problem. One is to use similarity-based methods (i.e., partitioning methods), the other one is the supervised training of a classifier. Supervised training of a classifiers requires labeled training data which is usually generated by the training of shadow models. The metrics for membership disclosure are calculated from the data controller's perspective and these metrics try to mimic the attack by an adversary.

Chen et al. categorize membership inference attacks in the context of GANs according to the knowledge available to an adversary [89]. They distinguish the following four scenarios starting with the one that makes the strongest assumptions about adversary knowledge on the SDG model:

1. **White-box GAN:** This scenario assumes access to the full SDG model including the discriminator and all hyperparameters.

2. **White-box generator:** This scenario assumes access to the generator component including its hyperparameters. Publishing the generator allows users to generate more samples.

3. **Partial black-box generator:** This scenario assumes access to the generator component but without its parameters. Yet, the learned representation of the input data is available to the adversary, so that they can observe input and output of the model but don't know the modeling process itself.

4. **Full black-box generator:** This scenario assumes access to the output (i.e., generated data) only.

These scenarios are also helpful for SDG models other than GAN. A full black-box scenario (i.e., a less knowledgeable adversary) with access to the synthetic data only could, for example, exploit that the synthetic data is very likely closer to training data (i.e., members) than to data not used in the training from the same population (i.e., non-members). Partitioning methods are one way to mimic such a scenario. Shadow model approaches are another way, but they can also make stronger assumptions (i.e., partial black-box or white box assumptions).

Beyond the access to the SDG model, both approaches make assumptions about adversary knowledge on the underlying distribution of the training data and the population where the target records are drawn from. These assumptions can be explicitly stated or implicitly derived from the calculation of the metric. We will illustrate this in the following using example metrics.

### 3.1.1 Partitioning methods

There are different approaches that have been used to operationalize the partitioning method. These differ in terms of the assumptions about the adversary, the distance measure that is used, and how they report results. Three examples are described below (the basic approach is illustrated in Figure 9).

In [90], the so called presence disclosure is measured using the partitioning method. The authors split their dataset into training (member) and holdout (non-member) sets by a 4:1 ratio. The attack dataset is then built by randomly drawing attack records equally from both datasets. The proportion of the attack records that are members can be denoted as $t = 0.5$. Synthetic data is generated using the training dataset. The distance between the attack records and synthetic records (ASD) is then calculated using the Hamming distance. The Hamming distance is generally the most used distance metric in partitioning methods [31,90–93], however, cosine similarity, Euclidean distance or Maximum Mean Discrepancy have also been described [91,94].

A match occurs when that distance falls below a certain threshold for the ASD and the matches are assumed to be proportional to the matches in the training dataset. An attack record is then evaluated for its membership status: a correct guess can be a matching member (true positive) or a non-matching non-member (true negative). A wrong guess would be a non-matching member (false negative) or a matching non-member (false positive) [31,92]. From these numbers, precision and recall are calculated and reported. Varying the attack dataset sizes (up to 10,000) in [90] did not alter the results. When the Hamming distance thresholds were increased (between 0 and 20) the recall increased.

The absolute membership disclosure metric can then be reported as precision and recall of the attack, or as their balanced *F1*-score [94]:

$$F1 = \frac{2 \times precision \times recall}{precision + recall} \qquad (9)$$

The authors in [91] follow the same steps but split the original dataset using a 70:30 ratio into a training and holdout dataset and then include all training (member) and holdout (non-member) records in the attack dataset. Consequently, the splitting ratio between training and holdout dataset translates into the proportion of attack records that are members and non-members ($t = 0.7$). Distance is measured by the Euclidean distance.

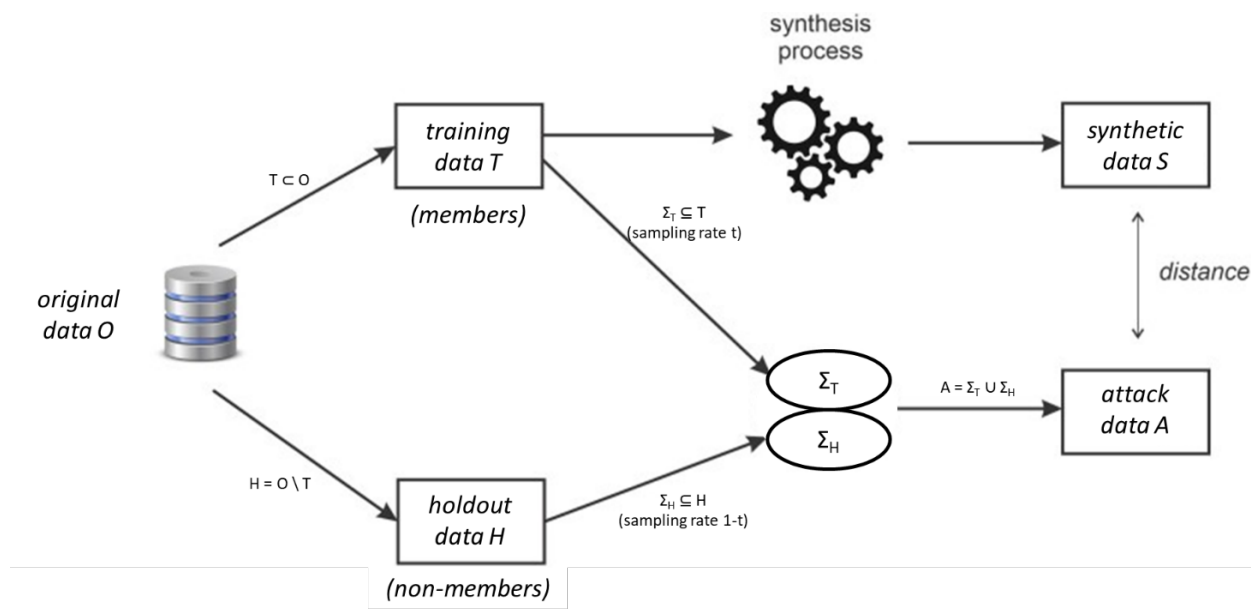

**Figure 9: The partitioning method**. The original dataset is randomly split into the training and holdout dataset. Synthetic data is generated by an SDG model trained on the training dataset. The attack dataset is drawn from both, the training and the holdout dataset, and thereby includes both – members and non-members. A match between the attack and the synthetic dataset is defined as a distance smaller than a certain threshold. The membership disclosure metrics are then calculated based on the number of matches that involved a true member. ⊆: subset (could be an equal dataset or a proper subset which of randomly drawn or selected records), ⊂: proper subset, U: union of two sets, \: complement (e.g. H = O \ T means that H consists of all elements of O except for T), A: Attack dataset, H: Holdout dataset, O: Original dataset, S: Synthetic dataset, T: Training dataset, Σ: drawn samples from training and holdout dataset. This figure us adapted from [31].

The partitioning methods presented in the literature generally assume a black box scenario [31,90,95–97]. Stronger assumptions about the access to the SDG model can improve distance measurements which is demonstrated in [89]. The authors present a black box, partial black box and white box scenario for partitioning methods. In the black box scenario, the distance is measured from a target record to the closest synthetic record, in the partial black box scenario it is measured to the closest latent representation of synthetic records and in a white box scenario to the closest reconstructed record. Distance measurements can also be improved via contrastive representation learning which has been proposed by Zhang et al. [96], or via principal component analysis [98].

Another general assumption is that the adversary's attack dataset is constituted of 50% members (i.e., t = 0.5) [72,90,91,96,97,99]. In [92], the authors extend the so far known partitioning methods by explicitly modeling the distribution of the attack dataset that is used to evaluate the performance. In scenarios where target records are drawn from the same population as the training data, they propose using $t = \frac{n}{N}$ where $n$ is the number of records in the original dataset and $N$ the size of the population. Accordingly, $t$ should be equal to the original data sampling fraction. We will go into this aspect in more detail in the critical appraisal.

A naïve baseline $F_{naïve}$ has been introduced in [92]. This is derived from a scenario where the adversary has no access to the synthetic data and all records in the attack dataset are simply classified as members. This can be denoted as follows:

$$F_{naïve} = \frac{2 \times p}{1 + p} \tag{10}$$

Where $p$ is the proportion of the attack records that are members, which would be equivalent to the sampling fraction of the original dataset from the population. A relative metric for membership disclosure vulnerability is proposed as the relative F1 score $F_{rel}$:

$$F_{rel} = \frac{F1 - F_{naïve}}{1 - F_{naïve}} \tag{11}$$

This metric is similar to Cohen's Kappa [100] in the way that it normalizes the difference to $F_{naïve}$ by the maximum possible vulnerability beyond the naïve baseline.

### 3.1.2 Shadow Model Approaches

Shadow model approaches involve the training of a shadow model that mimics the SDG model [21,63–65,101–105]. The idea is to generate labeled training data from the same or similar distribution as the training data of the SDG model. Data used in the shadow model training is then labeled as "member" and the one not used in the training as "non-member". A classifier trained on this labeled data can then determine if a target was part of the training dataset of the SDG model or not. A wide range of classifiers (e.g., SVM, Random Forest, Decision Tree, GBM) have been used to perform the task [101]. It can also be the discriminator component of the SDG model (assumed to be a GAN) itself [101,103].

A shadow model approach generally assumes a more knowledgeable adversary. While the assumptions about having access to the SDG model range from black- [63,63,65,101,105] to white-box scenarios [102,104,105], most approaches assume adversary knowledge of the distribution of the training data of the SDG model (i.e., auxiliary data) [63–65,101–103,105,106].

A black-box scenario using auxiliary data is described in [63]. This method starts with a target record that the adversary wishes to assess the membership disclosure for. The method requires that the adversary has access to the training algorithm $G$, the size of the input and output datasets and a large enough dataset from the same population as the training dataset that may be overlapping with the training dataset (i.e., auxiliary data). The adversary then trains two shadow-models. For the first one, the adversary samples multiple subsets from the auxiliary data ensuring that they do not have the target record, train a shadow-model using the known generative algorithm $G$, and for each constructs a synthetic dataset $S_j$ for $j = 1 \ldots J$ where $J$ is the size of the output dataset. The adversary then samples another set of records from that dataset with the target record in each of them, trains the second shadow-model using the known generative algorithm $G$ and creates a synthetic dataset $S_j{}'$, for $j = 1 \dots J$. The $S_j$ datasets are given a label of 0 (i.e., non-member) and the $S_j{}'$ datasets a label of 1 (i.e., member). Feature extraction techniques are used to characterize each dataset, such as the marginal frequency counts for each attribute. Using these features a classifier is constructed to distinguish between synthetic datasets that have the target record and those that do not. Once there is a trained classifier it can be applied on the actual synthetic dataset to determine whether the target record was used for its training. The data custodian would follow the same method to estimate the success rate of the adversary. This metric can also be called a targeted membership disclosure vulnerability metric [65] as it introduces the concept of a counterfactual world meaning that SDG is conducted with (i.e., member) and without (i.e., non-member) for a certain target. This concept is also implemented in [105].

In [101], a partial black-box generator approach is described where the adversary has access to the trained SDG model (GAN) and thereby to multiple synthetic datasets. They have also knowledge of its architecture and as in the first example an auxiliary dataset that comes from the same population as the training dataset but does not overlap with the training dataset. The adversary then trains multiple shadow-GANs using part of the auxiliary dataset as training input (i.e., members) and then uses the discriminator part with its predicted probabilities for the auxiliary data being a member to create a labeled training dataset for the classifier. This classifier exploits the weaknesses of the GAN and can be applied to the actual target records to reveal their membership status.

### 3.1.3   Application of Membership Disclosure Metrics

There is growing evidence that sometimes SDG models can memorize the training data and overfit, which makes synthetic data vulnerable to membership attacks [96,98,103,107–110]. Many authors report on the membership disclosure vulnerability when evaluating SDG models, with the partitioning method being the most implemented approach [31,94]. For example, Yan et al. include membership disclosure as the F1-score from a partitioning method using Euclidean distance in their benchmarking framework [91]. El Kababji et al. evaluate membership disclosure in synthetic clinical trial datasets and report a relative metric that includes the idea of a naïve guess (see equation (11)) [92]. Membership disclosure vulnerability is also usually reported when new SDG models are introduced (e.g., medGAN, Diffusion Models), and again most often by a partitioning method [80,90]. Fewer authors implement shadow models to determine membership disclosure [59,63,101,105]. A shadow model approach is, for example, part of the Toolbox for Adversarial Privacy Auditing of Synthetic Data (TAPAS) [105].

## 3.2   Critical Appraisal

For membership disclosure metrics, one of the key challenges lies in the attack dataset composition.

### 3.2.1   The Narrative Behind The Calculation Does Not Match What The Metrics Are Actually Measuring

The calculation of membership disclosure metrics requires making assumptions about the relationships between an adversary's attack dataset and the population where the original dataset is drawn from. This assumption is generally not explicitly communicated but can implicitly be drawn from the way metrics are calculated. For example, training a classifier based on auxiliary data from the same population as the training data suggests that targets are also drawn from this population. For other metrics, it is reasonable to infer this assumption as it is a straightforward approach to model the distribution of the attack dataset. We've reviewed the membership disclosure metrics of the four recently published systematic reviews focusing on their assumptions and narratives [19–22]. Among these, 1/21 makes the explicit assumption on the distribution of the attack dataset [31], the other metrics (20/21) implicitly assume that the target individual is sampled from the same population that the training dataset used for SDG is sampled from [63–65,89,90,95–97,99,101–106,106,108,111–113].

The problem is that this assumption is contradictory to the implicit assumption that can be drawn from the exemplar scenario often used in the literature (see below) [64,89,95,96]. We will describe some of these common narratives that are not aligned with the assumption of drawing targets from the same population as the training data in the following and present narratives that would be covered by what these metrics actually quantify.

Zhang et al. give an archetypical example scenario for membership disclosure attacks. The scenario is described as follows

> *"Imagine that a malicious attacker Mallory gained access to a patient Bob's health record history (e.g., via a data broker, self-disclosure by the patient themselves, or a breach of a data warehouse). At some later point in time, Bob received diagnosis x (e.g., HIV-positive) and was treated at a healthcare facility, which Bob intends to keep confidential. Then, a researcher at the facility makes public a synthetic cohort of individuals with diagnosis x based on its set patient records. Now, imagine that Mallory applies a membership inference strategy to learn that Bob's record was included in the records relied upon to generate the synthetic cohort. At this point, Mallory learns Bob was diagnosed with x, which further compromises Bob's privacy."* [96].

Here the adversary has background information about a target record. The adversary also has a synthetic dataset of HIV patients. The argument is that by predicting the membership in the training dataset correctly, the adversary can infer the information that the target is HIV positive, which may harm the individual [96]. Hayes et al. describe another harmful situation for persons from a criminal database who were part of the training set of face generation algorithms [103]. The adversary learns sensitive biographical information by membership disclosure.

The assumptions that are made for the attack datasets are crucial to understand what membership disclosure metrics quantify. Let's consider the HIV example and consider potential attack datasets.

The original dataset used for training the SDG model is presumably drawn from people with HIV, e.g., all people with HIV in Ottawa: the population of the original dataset would be "people with HIV in Ottawa". The calculation of membership disclosure metrics makes the assumption that the target record is sampled from the same population as the original data, in this case, from all people with HIV in Ottawa. Then the attack dataset consists of people with HIV only. In this case membership disclosure *does not* reveal the sensitive diagnosis because the adversary would *already know* Bob's HIV diagnosis, and in this case membership in the HIV database is the only information gain. Membership can come with further information, for example an intervention or the use of a certain drug. This can is a valid attack scenario for certain datasets but does not match the presented narrative.

If the adversary draws targets from a different population than the original data, for example, from all people who live in Ottawa, then the adversary would learn the diagnosis of HIV by membership disclosure. This would match the narrative of the example. However, in that case the membership disclosure vulnerability metrics would give the incorrect estimate of the vulnerability. This is because the attack dataset in membership disclosure vulnerability metrics is drawn from the same population as the original data.

Let's construct an alternative narrative where the assumptions for the metrics are met and where the measurement of membership disclosure vulnerability is meaningful accordingly. This could be a scenario where the adversary is trying to determine membership in a vaccination study and the adversary is a vaccination opponent. A narrative may be that harm results from the vaccination opponent inferring the pro-vaccine stance of a target individual based on study participation. Here the population of the original dataset are all people living in Ottawa since recruitment for the study is from the general population. In that case if the adversary draws a target from anyone living in Ottawa, then that would match the population of the original data and the way that membership disclosure vulnerability is measured.

In the examples below we illustrate how the wrong assumption for the attack dataset can result in incorrect membership disclosure vulnerability calculations.

### 3.2.2 Using Membership Disclosure Metrics in The Wrong Context Substantially Overestimates Vulnerability

An attack dataset drawn from the population of the original dataset can give higher vulnerabilities than an attack dataset drawn from a different population. Let's consider the HIV example again. A scenario that matches the narrative in the literature would be an attack dataset drawn from a different population than the original data, let's say from all people who live in Ottawa. However, not everyone in this population has HIV and having a correct determination of membership would be less likely than when the attack dataset is drawn from people with HIV. We conducted a simplified simulation with different attack scenarios to illustrate what "less likely" can actually mean.

The simulation is illustrated in Figure 10 and further detailed in the sidebar "Membership Disclosure Simulation Methodology", and contrasts two narratives: scenario A where an adversary draws targets from the same population as the training dataset and scenario B where they draw from a different population. Precision was 460/1,000 in scenario A but dropped to 3/1,000 in scenario B.

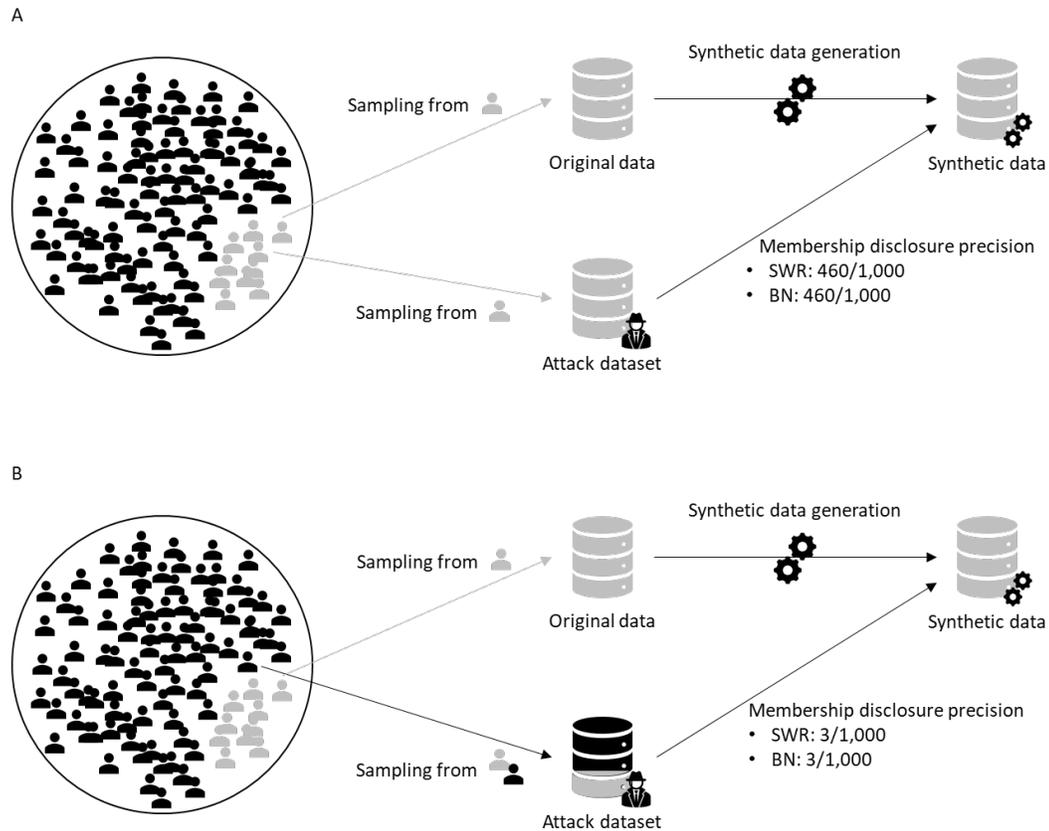

**Figure 10: Different adversary's attack datasets**. The original data consisted of 1,000 people with HIV. It was randomly drawn from the population of people with HIV in Ottawa (2,172) highlighted in grey. The attack dataset of 1,000 people was drawn from the same population as the original data, so from young people with HIV in Ottawa (A), or from all young people in Ottawa (B). The population of young people in Ottawa was set as 322,835. Membership disclosure is given as the precision of the attack. SWR: sampling with replacement; BN: Bayesian Network.

This simple simulation illustrates that vulnerabilities differ by a wide margin depending on the sampling assumptions. In scenario A the 0.46 precision reflects the accuracy of an adversary who draws targets from the same population as the training dataset and thereby already knows that the target has HIV. They learn that the targets are in the training dataset. In scenario B the adversary draws targets from a different population and thereby does not know that the target has HIV. The 0.003 reflects the adversary's ability to learn that highly sensitive information from membership disclosure. The current literature on membership disclosure operationalizes scenario A, but tells the narrative of scenario B.

## Membership Disclosure Simulation Methodology

In this simulation, we built two different populations where an adversary can potentially draw targets from. We thereby assess two different narratives: (A) an adversary who draws targets from the same population as the training data (i.e., HIV-positive individuals) and (B) an adversary who draws targets from a different (super) population.

The following background information supported us in generating the populations:

- The number of HIV-positive individuals in Ottawa is about 4,000 individuals ([https://ottawacitizen.com/news/local-news/fight-against-hiv-aids-stigma-remains-say-officials](https://ottawacitizen.com/news/local-news/fight-against-hiv-aids-stigma-remains-say-officials)).

- The incidence in the age group of 20-39 years accounts for the highest proportion (i.e., 54.3%) among all age groups ([https://www.ottawapublichealth.ca/en/reports-research-and-statistics/resources/Documents/hiv_stats_en.pdf](https://www.ottawapublichealth.ca/en/reports-research-and-statistics/resources/Documents/hiv_stats_en.pdf)).

- The incidence for male individuals in this age group is 3.2 times higher than for female individuals in the same age group ([https://www.ottawapublichealth.ca/en/reports-research-and-statistics/resources/Documents/hiv_stats_en.pdf](https://www.ottawapublichealth.ca/en/reports-research-and-statistics/resources/Documents/hiv_stats_en.pdf)).

- The number of individuals of 20-39 years in Ottawa is estimated to be 322,835 in 2025 ([https://www.ottawapublichealth.ca/en/reports-research-and-statistics/sociodemographics.aspx](https://www.ottawapublichealth.ca/en/reports-research-and-statistics/sociodemographics.aspx)).

As a first step, we mimicked the population of individuals between 20 and 39 years in Ottawa by generating 322,835 records with an age and gender distribution according to the sociodemographic data from Public Health Ottawa. As a next step, a population of HIV-positive individuals was gender-weighted randomly sampled from this super-population. The size of this population was estimated to be 2,172 (54.3% of HIV-positive individuals in Ottawa). We then drew a random sample of size = 1,000 from the population of HIV-positive individuals. This sample was used to generate synthetic data and is referred to as original dataset. Synthetic data was generated of the same size as the original dataset using (1) random sampling with replacement and (2) a Bayesian Network SDG model as implemented by Synthcity [56]. Sampling with replacement mimics a poorly generalizing SDG model.

We randomly sampled two different attack datasets each of 1,000 records from the populations as illustrated in Figure 10. We assumed an adversary that matched their target's age and gender with the synthetic records and guesses membership for the matching ones. This guess could be:

- true positive: matches the synthetic dataset and is a member of the SDG training dataset, or

- false positive: matches the synthetic dataset but is a non-member of the SDG training dataset.

From these numbers, we calculated membership disclosure as precision. The random sampling of the attack dataset was repeated 1,000 times per scenario, and the precision averaged across these iterations. This was done for both synthetic datasets.

This simulation makes several simplifying assumptions. We did not account for mortality when calculating prevalence using the proportion of males HIV incidence and used historic HIV information by Public Health Ottawa. In our attack, matching did not improve the adversary's success over a naïve guess. Further quasi-identifying information (e.g. ethnicity) may help with matching. However, these simplifications are not likely to alter the overall conclusions.

### 3.2.3 Membership Disclosure Metrics Are Only Stable When Considering the Sampling Rate

It is known that metrics built upon sensitivity and specificity vary with prevalence [114,115]. Membership disclosure metrics generally report vulnerability via precision, recall or the *F1* score [31,90,96,97,99,101,103,104,112]. So, the distribution of the attack dataset (i.e., member prevalence) that is used to determine membership disclosure affects the reported vulnerability [31,116]. The same metric can vary substantially when measured on two differently distributed attack datasets as shown in the simulation (scenario A versus scenario B).

None of the reviewed membership metrics (except for [31]), however, comes with an explicit model of the distribution of the attack dataset. Most metrics presented in the literature build an attack dataset that includes 50% members (i.e., t = 0.5) [90,96,97,99,101,104,108]. Some use another arbitrary fixed value [91,103,106]. At the same time, authors make the assumption that target records are drawn from the same population as training records [31,63–65,89,90,95–97,99,101–106,106,108,111–113].

In [31], this assumption is formalized in a hypergeometric distribution. The authors show that the calculated *F1* score varies with the member prevalence in the attack dataset (see Figure 11).

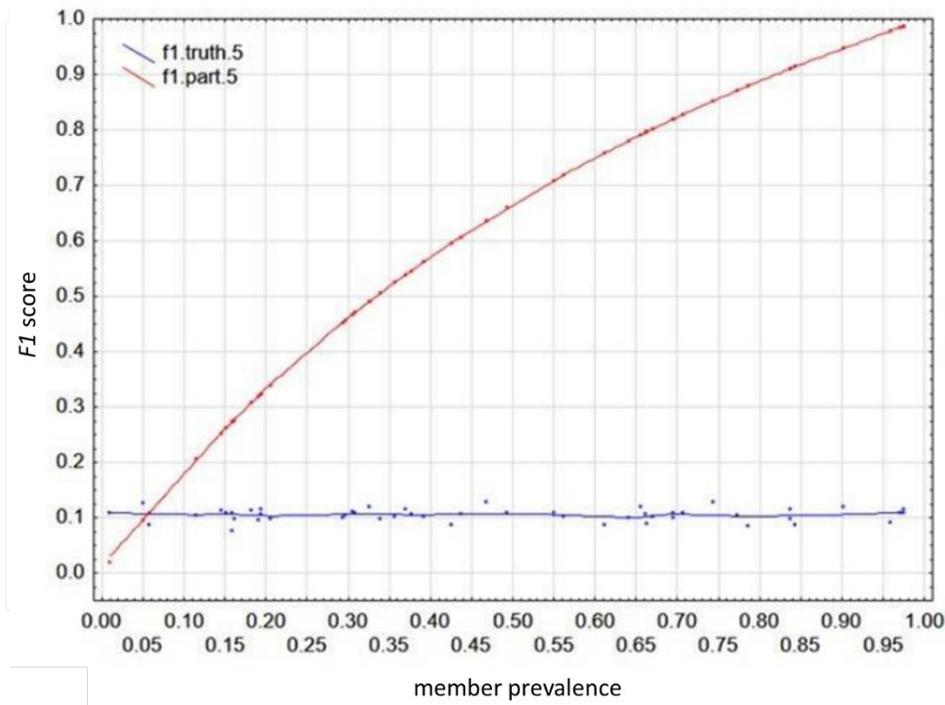

**Figure 11: *F1* score dependency on member prevalence**. *F1* scores for a dataset are calculated using the partitioning method when varying the member prevalence (i.e., t) in the attack dataset (red line). The ground truth is calculated from a simulation where targets are drawn from the same population as the training dataset and the sampling fraction of the training dataset was 0.055. This plot is taken from [31].

Consequently, the attack dataset needs to account for the member prevalence which in this scenario would be the sampling fraction. All other distributions of the attack dataset do not reflect the ground truth vulnerability when targets are drawn from the same population. When partitioning approaches take a balanced number of members and non-members (*t* = 0.5), the calculated *F1* score only applies to scenarios where there is an actual sampling fraction of 50%. Real-world scenarios have lower or higher sampling rates with respective different chances of being a member. An *F1* score that is based on a balanced partition would then not appropriately account for the effect of false positives.

### 3.2.4   The Relative F1 Score Metrics Are More Informative Than the Absolute F1 Score

Since the absolute *F1* score depends on the sampling fraction when the targets are drawn from the same population as the training data, it is not uniformly interpretable across different datasets and does not provide meaningful guidance for decision-making. Let's consider, for example, a data controller and their decision whether to release a synthetic data. In these examples, we assume that targets are drawn from the same population as the training data.

A. Consider a synthetic dataset with *F1* = 0.9505. The training dataset comprises almost all records of its population (e.g., census data). Here the membership disclosure vulnerability is quite high. This suggests a high vulnerability and it is true that an adversary could quite successfully infer membership. It is, however, not a vulnerability introduced by the release of the synthetic data but attributable to the high sampling fraction. Consequently, membership disclosure is independent of whether or not to generate and release synthetic data.

B. Now, consider a synthetic dataset with *F1* = 0.5. Here membership disclosure according to *F1* is comparably low. The training dataset, however, still compromises almost all records of its population (e.g., census data). While SDG may have diminished the adversary's success when conducting a membership computation, the worst-case assumption is that the adversary relies on the naïve guess rather than conducting a computation. Then, the *F1* conveys a wrong sense of security. Yet, the membership disclosure by the naïve guess is independent of whether or not the data controller should generate and release synthetic data.

C. Next, let's consider a membership disclosure evaluation where *F1* is quite low, for example, *F1* = 0.3. The sampling fraction in this case, however, is also very low (e.g., 0.005). Then the incremental vulnerability of releasing synthetic data may be considered as high.

In all examples, the interpretation of the *F1* score depends on the prevalence and the sampling fraction adds valuable information. This is captured by the idea of the naïve baseline in in [31]. The naïve baseline would be 1.0 when the training dataset is identical with the population (p = 1) and gets close to zero the smaller the sampling fraction. An adversary would be able to determine the value of $F_{naive}$ without using the synthetic data. Consider, for example, a simple press release which states that 50,000 individuals living in Ottawa participated in a vaccination study. This would give an $F_{naive}$ = 0.095 for any target from Ottawa. Therefore, it is independent of the PET applied but is dependent on the sampling fraction of the original dataset from its population. Considering $F_{naive}$, the scenarios above showcase various points:

- **Example A** ($F_{naive}$ = 0.95)**.** By itself, a high *F1* is not sufficient to make a decision on the release of synthetic data.  If $F_{naive}$ is also high, then it is not the access to the synthetic data that makes the adversary successful but a high sampling fraction. The success is then independent of the data release.

- **Example B** ($F_{naive}$ = 0.95)**.** A low *F1* gives an incomplete picture of membership disclosure vulnerability.  If $F_{naive}$ is high, then the adversary may not make the effort to conduct a membership disclosure attack but would just take a simple guess with a success rate reflected in $F_{naive}$. The most conservative approach is to adopt the higher of the two values to describe *F1*. The success, however, is again independent of the data release.

- **Example C** ($F_{naive}$ = 0.01). A low *F1* is not sufficient to take a decision on the release of synthetic data. If $F_{naive}$ is still lower, then the increase in vulnerability may be meaningful.

- **Example A and B**: If $F_{naive}$ is high to start off with, a higher or lower *F1* does not change the high membership vulnerability that is present independent of the synthetic data release.

Accordingly, the absolute metric *F1* comes with the challenge that it does not always provide a good sense of vulnerability but depends on the member prevalence and needs the naïve baseline to be more meaningful. There are two ways to deal with this:

1. Calculation of the relative metric $F_{rel}$ (see equation (11)): this metric provides a uniform interpretation across different scenarios. A common threshold can then be applied.

2. Calculation of the absolute metric *F1*: this metric is not uniformly interpretable, but thresholds can be adjusted according to the naïve guess.

Both cases require the estimation of $F_{naive}$, and in both cases it should be encouraged to report $F_{naive}$ alongside with the vulnerability of the synthetic data. If the objective of the SDG exercise is to make a data sharing decision (i.e., share or not) and $F_{naive}$ is high, then the calculation of the membership disclosure vulnerability of the synthetic data may not be a driver in the release decision. For example, in one study the population was defined as other clinical trials in the same geographies and period as a particular trial, and the trial represented 57% of that population [92], which gave $F_{naive}$ = 0.726. In that case, there would be no need to compute another membership disclosure metric to conclude that it exhibits high membership disclosure. Even if the synthetic data does not increase membership disclosure vulnerability, the vulnerability is high from the data subject perspective. Such considerations should be part of the overall risk management process (see 1.1.5) and require knowledge of $F_{naive}$.

### 3.2.5 The $F_{0.5}$ or $F_2$ Score May be More Appropriate in Certain Scenarios

Predicting membership disclosure is a typical classification task, with membership being the "positive" class. For a classification task, the confusion matrix (see Figure 12) has the following cells: true positives (i.e., correctly classified members), false positives (i.e., non-members wrongly classified as members), true negatives (i.e., correctly classified non-members) and false negatives (i.e., members wrongly classified as non-members).

|  |  | **Prediction by adversary** |  |  |
|---|---|---|---|---|
|  |  | Member | Non-member |  |
| **Ground truth** | Member | TP | FN | $recall = \dfrac{TP}{TP + FN}$ |
|  | Non-member | FP | TN | $specificity = \dfrac{TN}{TN + FP}$ |
|  |  | $precision = \dfrac{TP}{TP + FP}$ | $NPV = \dfrac{TN}{TN + FN}$ | $accuracy = \dfrac{TP + TN}{TP + TN + FP + FN}$ |

**Figure 12: Confusion Matrix in Membership Disclosure Vulnerability**. FN: false negative; FP: false positive; NPV: negative predictive value; TN: true negative; TP: true positive.

The *F1* score calculated to measure membership disclosure vulnerability uses recall and precision to describe classification performance and is thereby an incomplete metric, meaning that it ignores the portion of true negatives [115]. This choice of performance metric makes sense in the context of membership disclosure, where a consideration of true negatives would inflate the vulnerability. The *F1* score, however, also gives equal weight to precision and recall. While we did not encounter literature where this weight was discussed or justified for membership disclosure measurement, there may be situations where a data controller would prioritize precision over recall, or vice versa.

Let's assume an example where $F_{naive}$ is 0.2 (i.e., sampling fraction of 0.1). Now, a data custodian calculates an *F1* score of 0.2 for a synthetic dataset. Because precision and recall are equally weighted, the underlying precision of this *F1* score could be 1.0 and the recall 0.1, or vice versa. The *$F_{rel}$* would be zero in both cases. If a data custodian assumes that false negatives (i.e., members wrongly classified as non-members) are not relevant to an adversary, then they would need to put more weight on precision to adequately reflect their scenario. Such a more general formula for the *F* score can be described as follows:

$$F_\beta = \left(1 + \beta^2\right) \frac{precision \times recall}{\beta^2 \times precision + recall} \tag{12}$$

where $\beta$ is the factor by which recall is prioritized over precision. The more general notation of the $F_{naive}$ is:

$$F_{naive} = \left(1 + \beta^2\right) \frac{p}{\beta^2 \times p + 1} \tag{13}$$

and for the *$F_{rel}$*:

$$F_{rel} = \frac{F_\beta - F_{naive}}{1 - F_{naive}} \tag{14}$$

whereby the factor $\beta$ (i.e., weight) is the same for $F_\beta$ and $F_{naive}$. In the balanced *F1* score, the weight is 1; a weight of 0.5 favors precision and a weight of 2 favors recall.

The question is when would a data controller want to change the weights in practice ? Multiple different adversaries have been identified in the literature [86,117], and depending on their objectives the weight can be adjusted. In [117], 6 categories of adversaries are defined based on their motivation: personal motives, public recognition, selection or discrimination, profit motives, enforcement or control and self-interest. These categories can be roughly divided into those that are targeting a specific individual (i.e., personal motives, selection or discrimination, enforcement or control, self-interest), those that are targeting an arbitrary individual (i.e., public recognition), and those that are targeting a wide array of individuals (i.e., profit motives). In the example above, the adversary may be one that targets an arbitrary individual out of public recognition. Then, the data custodian may rather choose the $F_{0.5}$ score that prioritizes precision over recall than the $F_1$ score. In the example above, the $F_{naive}$ would then be 0.12, the $F_{0.5}$ score 0.36 and the $F_{rel}$ 0.27. This better matches the assumed adversary.

More generally, if the category of adversaries can be assumed for a data release scenario, then the weight should be chosen accordingly. If such an assumption cannot easily be made, the maximum out of $F_1$, $F_{0.5}$ and $F_2$ score may be a good strategy.

### 3.2.6 Comparing Membership Disclosure Vulnerability of the Synthetic Data to the One of the Original Data Does Not Provide Guidance for Decision-Making

Another idea is the introduction of an upper baseline based on the membership disclosure derived from the original data [59,63]. In such a case, the membership disclosure metric is also calculated between the attack and the original dataset, thereby giving a membership disclosure vulnerability that would arise when releasing the original data. This relative metric is not very likely to support a decision-making process on whether to release synthetic data as a membership disclosure vulnerability that is lower than this baseline may still be high in absolute terms.

### 3.2.7 The Choice of Classifier Determines Accuracy in Shadow-Model Approaches

In shadow-model approaches the choice of the classifier has an important influence on its accuracy which, ultimately, is where membership disclosure vulnerability is based on [59,63,64]. To eliminate this source of variability all implementations should standardize on one modeling approach and implementation.

### 3.2.8 White-Box Attacks or Large Auxiliary Datasets Are Not Reasonable From the Adversary Perspective

Shadow models are more complex overall and make stronger assumptions on the adversary's knowledge, which might account for their less frequent implementation. Some assumptions of such membership inference attacks are not likely to be accurate in most settings. For example, a model's specifications are usually not shared when sensitive data is used as training data, which is a requirement for white-box attacks [99,108]. While black-box access would be a more reasonable scenario, these metrics generally still rely on a large amount of reference data to train the shadow model [63,65,101–103,105]. This is hard to achieve in most settings in practice, for example in the healthcare setting [63] – one of the main problems that SDG is intended to solve is data scarcity.

### 3.2.9 Membership Disclosure is Not Linkability

Giomi et al. criticize that membership disclosure is often used as a proxy for linkability [62]. The authors point out that membership disclosure in synthetic data does not necessarily come together with the information of the synthetic record that can be matched on. So, linkability requires more than membership disclosure to occur.

### 3.2.10 The Adversary's Strategy for Using Their Available Information Determines Membership Disclosure Vulnerability

As illustrated in Section 1.4.1 through examples and simulations, the membership disclosure vulnerability may decrease the more attributes are used in calculating membership disclosure vulnerability. This is because with more attributes it is more likely to have a non-matching synthetic value. Therefore, an adversary would benefit from using subsets of the attributes that are known to them. Even within these subsets, different combinations of attributes, and their generalizations, will have different vulnerability values.

The data controller cannot anticipate a priori which combination of attributes and generalizations an adversary will try. It is therefore prudent to perform a search for the maximum vulnerability considering all the subset and generalization combinations of the attributes that are available to the adversary. For example, if an adversary knows all attributes for a target, then the highest (i.e., worst case) membership disclosure vulnerability for this scenario can be determined by measuring vulnerability for all combinations of subsets and generalizations the adversary could use in their attack.

**Summary of Key Points from the Critical Appraisal of Membership Disclosure Metrics**

- Current membership disclosure metrics explicitly or implicitly make the sampling assumption that targets are drawn from the same population as the training dataset. This does not match the very common membership scenario, where an adversary aims at learning the attribute that defines the population where the training dataset is drawn from by membership disclosure (e.g., HIV in a training dataset drawn from an HIV population). In such cases, current metrics give an incorrect vulnerability result.

- Notwithstanding the point above, the calculation based on current membership disclosure metrics is only correct if the attack dataset has the same base rate as the training data's sampling fraction, given the assumptions inherent in these metrics.

- The commonly used *F1* score is dependent on prevalence, and therefore is difficult to interpret in a general manner. A naïve guess reflects the prevalence and thereby helps to interpret the *F1* score. Also, the equal weights on recall and precision in the commonly applied *F1* score does not adequately reflect all attack scenarios.

- Shadow model approaches often rely on a large amount of reference (i.e., auxiliary) data to train the shadow model and/or the classifier. This is a strong and generally unrealistic assumption regarding the adversary in practice.

- Calculating membership disclosure metrics with all attributes that are available to an adversary is unlikely to result in the highest possible vulnerability. Higher vulnerabilities can be obtained when using combinations of attributes at different subset sizes, and generalization of attributes. Current metrics do not perform such searches.

## 3.3 Recommendations

### 3.3.1 Making Membership Disclosure Vulnerability Metrics More Meaningful by Explicitly Modeling the Underlying Distribution of the Attack Dataset

When calculating membership disclosure, it is relevant whether the assumptions behind membership disclosure have been met for the attack dataset. Current membership disclosure metrics are only capturing one specific attack scenario (scenario A in Figure 10) where targets are drawn from the same population as the training data. If this is consistent with the attack assumptions, then the partitioning method should explicitly model the attack dataset accordingly (i.e., with $t = \dfrac{n}{N}$).

In situations, where an adversary is more likely to draw its targets from another population than the training dataset (scenario B in the simulation) to learn, for example, the HIV diagnosis from membership, models that correctly capture the underlying distribution of the attack dataset need to be developed. It may be an option to conduct an attack simulation similar to the one we have run in this section. Note that this would still be an inflated vulnerability estimate as sampling with replacement is a worst-case scenario for an SDG model and a population constituted of unique records a worst-case scenario for a training dataset. It is, however, a more realistic estimate than the one computed from the vanilla membership disclosure metric in such a situation.

### 3.3.2 Defining Acceptable Membership Disclosure Vulnerability Using Relative (i.e., $F_{rel}$) Metrics

Given that the metric correctly models the distribution of the attack dataset, the determination of an acceptable membership disclosure vulnerability would then follow the considerations of $F_{rel}$:

- $F_{rel} \leq 0.2$: acceptable vulnerability
- $F_{rel} > 0.2$: high vulnerability

Note that $F_{rel}$ is sufficient to assess the vulnerability of the synthetic dataset. The absolute $F1$ score does not have a uniform interpretation on its own due to its dependency on prevalence. $F_{naive}$ by itself, however, still provides meaningful information for the overall risk management process and should be reported alongside with $F_{rel}$. The thresholds that have been used in the literature are 0.2 for $F_{rel}$ [30–32]. These would continue to be used and may be adjusted up or down based on the context (see 1.3.4).

The $F1$ score is the balanced combination of precision and recall. In some scenarios, unequal weights may be more appropriate, and the data custodian can tailor the $F1$ score accordingly when justified.

# 4.    Attribute Disclosure

## 4.1    Definition

As a disclosure vulnerability, attribute disclosure has also been a concern for machine learning models in general [118–120]. This concern is particularly acute for deep learning models because they can more readily overfit the training data, which could allow extracting those data from the trained model.

In the context of anonymization, or statistical disclosure control more broadly, attribute disclosure has been defined as when an adversary can infer sensitive information about a target individual [38,121,122]. Attribute disclosure may occur without re-identifying the target individual's record. Multiple attribute disclosure attacks have been proposed, including a homogeneity attack, background knowledge attack, skewness attack, similarity attack, proximity breach, and minimality attack [121].

Attribute disclosure is also an important type of vulnerability for synthetic data and entails the adversary constructing models from the available synthetic data to predict a sensitive attribute for a target individual, sometimes complemented with prior information. Assume that the adversary uses the synthetic data to train a model using $(T,K)$ where $T$ is the sensitive (target) attribute and $K$ are non-sensitive (key) attributes, with $T$ as the outcome. In the literature, the $K$ attributes may be the complete record (all attributes except for $T$), quasi-identifiers only or may constitute other types of attributes that are not quasi-identifiers. The adversary uses the trained model to predict the value $\hat{t}$ of an unknown sensitive attribute $t$ out of $T$ from a set of known attributes $k$ for a target individual using that model. If a reasonably accurate predictive model can be constructed, then that may be deemed to be an undesirable attribute disclosure vulnerability from the dataset.

Attribute disclosure on synthetic data does not assume re-identification. In fact, as discussed in the section on record-level similarity, identity disclosure is not a suitable way to evaluate privacy for synthetic data because, given the generative character of the method, a match between the synthetic records and the data subjects cannot be translated into a one-to-one mapping.

### 4.1.1    Attribute Disclosure Techniques

The adversary would know the true values on $k$ for a target individual. One approach for predicting the sensitive attribute $t$ for the target is for the adversary to (exactly or approximately) match with the synthetic data on the true $k$ values, and then the sensitive attribute for the closest synthetic record is used as the predicted value $\hat{t}$ for the target, or the distribution of the closest set of synthetic records is used to predict the target sensitive value $\hat{t}$ [25–28,62]. There are various options how an adversary may handle non-matches in such a scenario. One option is to use a nearest-neighbor model that does probabilistic matching on $K$ [27,28,62], another one is to reduce the number of predictors to increase the likelihood of a match or generalizing the predictors [25,26,57]. Some calculations combine various possibilities [62]. All of these strategies aim at increasing the pool of matched synthetic records. A very different option to handle non-matches is to directly assign an incorrect guess to non-matches or to count them as undefined [25,26].

Beyond such nearest neighbor methods, other statistical and machine learning models can be trained on the synthetic data (with $K$ as the predictors and $T$ as the outcome) [59,63]. The adversary would then use the true values on $k$ to predict the sensitive value $\hat{t}$ using the trained model.

Domingo-Ferrer et al. introduce an alternative approach where the synthetic data is checked for correlations that are also present in the training data by canonical correlation analysis [123]. Attribute disclosure is then assumed when strong correlations of the training data are maintained in the synthetic data.

### 4.1.2  Absolute and Relative Metrics

Attribute disclosure has been reported as a simple (absolute) accuracy metric for the predicted sensitive attribute [25–28]. The underlying correctness of a prediction is then defined differently when drawn from categorical versus numeric values. In the latter, authors propose a certain closeness to the correct value [26,59,62,63].

Relative metrics have been proposed with baselines similar to those encountered in the other types of privacy metrics have been proposed. Encountered baselines for attribute disclosure vulnerability are listed in the following:

- An upper baseline was  calculated from the accuracy when using a model trained on original data to predict a target's sensitive attribute [25,26,59,63].

- A lower baseline was defined by a naïve guess derived from the univariate distribution of the sensitive attribute [59,62].

- Another lower baseline has been introduced as a non-member baseline [12,62,63]. This means that the accuracy using only data from individuals that were not part of the training dataset for SDG (i.e., non-members) is calculated and compared to the accuracy using only data from individuals that were part of the training set for SDG (i.e., members).

### 4.1.3  Privacy Violation versus Knowledge Generation

Some authors have argued that an individual's privacy can only be violated when that individual is in the dataset (i.e., member) [12,124,125]. Therefore, accurately predicting the sensitive value $\hat{t}$ of a target non-member would not be considered a privacy violation and hence not an attribute disclosure, whereas accurately predicting the sensitive attribute for a member target would be considered an attribute disclosure, especially if the accuracy benefited from being a member. This is based on the concept that the presence or absence of a single individual in a dataset should not affect the outcome which has a certain similarity to differential privacy [126]. For a data subject, an attribute disclosure occurs from being part of a dataset rather than from being part of the population where the dataset is drawn from. This concept is not meant to address harm. Harm may also occur in individuals that are not members of the dataset (see 4.3.4).

It should be noted that while it is desirable to get similar accuracies for non-members as for members (i.e., external validity of results), it is not always likely to be the case. In machine learning, it is commonly understood that models are generally not capable of performing as well on unseen data (non-members) as they do on the training data (members) [127–129]. While it is not impossible to achieve similar accuracy on non-members, it is more likely that a difference will always remain [62,63]. In machine learning models, such a difference has also been used when evaluating privacy and is thought to reflect the model's ability to generalize and the diversity of the training data [37,112].

Developing models from data with high predictive accuracy is also the aim of scientific research in the sense of *knowledge generation*. Knowledge generation typically involves the identification of general patterns or trends and is not intended to reveal attributes of a specific individual.

The line between attribute disclosure and knowledge generation can be subtle. In the literature, different ways have been proposed to separate attribute disclosure from knowledge generation closely related to the above described baselines [57,62,63]. These ways can serve as subcategories to group attribute disclosure metrics that share common considerations. Note that the subcategories are not presented in the literature but are introduced in this report for the sole purpose of enabling a more focused critical appraisal that allows us to better understand the complexity of attribute disclosure metrics. The following subcategories have been defined:

1. **Undifferentiated absolute attribute disclosure metrics:** The first subcategory are absolute attribute disclosure metrics without separating knowledge generation from disclosure vulnerability [28,57,90,99].

2. **Undifferentiated relative attribute disclosure metrics:** The second subcategory are relative metrics with baselines that, however, do not differentiate between knowledge generation and disclosure vulnerability [25–27,59]. .

3. **Attribute disclosure metrics with non-member baseline:** The third subcategory are relative metrics that establish a non-member baseline as a proxy for knowledge generation which gives a reference point for no privacy violation [62,63].

The key factors that divide the metrics into these subcategories is their choice of baseline and the composition of the attack datasets. The latter is illustrated for the example metrics in Figure 13.

### 4.1.4 Application of Attribute Disclosure Metrics

Attribute disclosure metrics have been used in research evaluating the disclosure vulnerability of synthetic data. Multiple authors report attribute disclosure when presenting new SDG methods [90,99,130]; in several evaluation frameworks it is part of the privacy evaluation set-up. For example, Yan et al. present a comprehensive benchmarking framework for synthetic data with multiple (weighted) privacy and utility metrics [23]. For privacy evaluation, the authors implemented an undifferentiated absolute attribute disclosure metric that finds the closest (by Euclidean distance) records to the target and calculate the attribute disclosure vulnerability as the entropy-weighted sum of multiple sensitive attributes. The method itself was introduced in [99]. Yan et al. also include attribution (or meaningful identity) disclosure in their benchmarking framework. This approach also counts to undifferentiated absolute attribute disclosure metrics and has been introduced by El Emam et al. [57]. It is unique in the sense that it takes the equivalence classes in the original data and the population into account.

Utilizing attribute disclosure is common when ranking different SDG models. For example, Goncalves et al. conduct a systematic evaluation of multiple SDG techniques and use the majority vote among k nearest neighbor in the synthetic data, so an undifferentiated attribute disclosure metric, as a measure for attribute disclosure vulnerability [93]. The recently published evaluation tool Syndat by the German

National Research Data Infrastructure for Personal Health Data Initiative implemented three privacy metrics to match the legal terminology of singling out, linkability, and inference by the Article 29 Working Party [62,131]. The inference metric they implemented to learn sensitive information is best categorized as an attribute disclosure metric with a non-member baseline. TAPAS, another evaluation tool for synthetic data, implemented an undifferentiated attribute disclosure metric [105] that was initially introduced in [26].

## 4.2    Example Metrics

While the definition of attribute disclosure was primarily from the adversary perspective, the calculation of the example metrics is described from the data controller's perspective because it is their responsibility to evaluate the synthetic data. While doing so, assumptions need to be made about the adversary's strategy based on the data available to the data controller.

In this sense, the terms attack record and attack dataset are used to refer to the data controller's effort to mimic the adversary's target record or target dataset. To be consistent with the critical appraisal, example metrics are presented in the subcategories mentioned earlier.

A

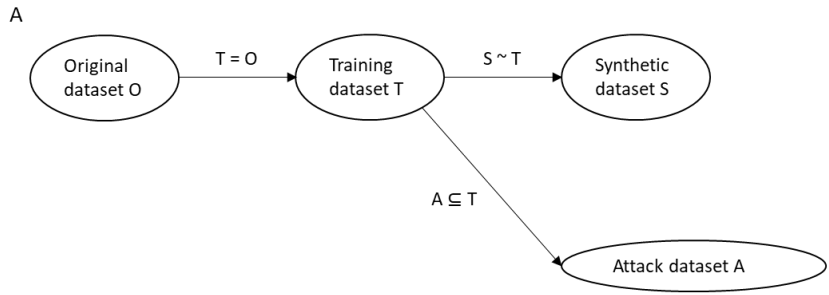

B

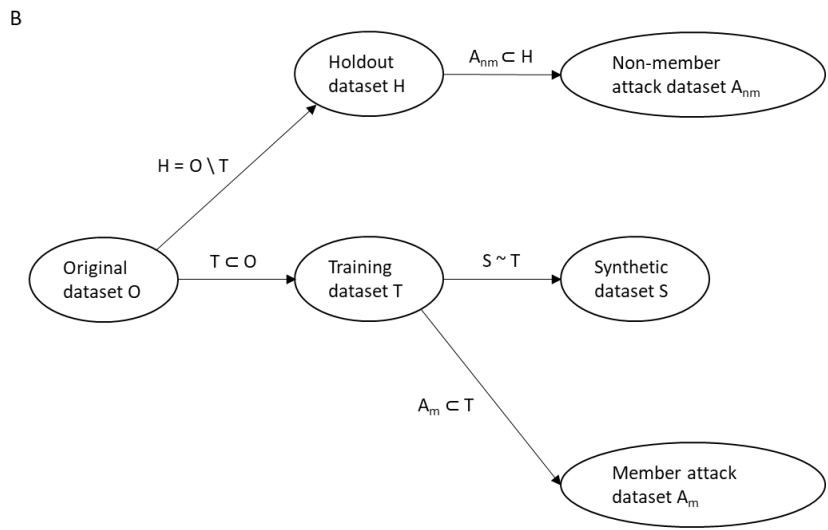

C

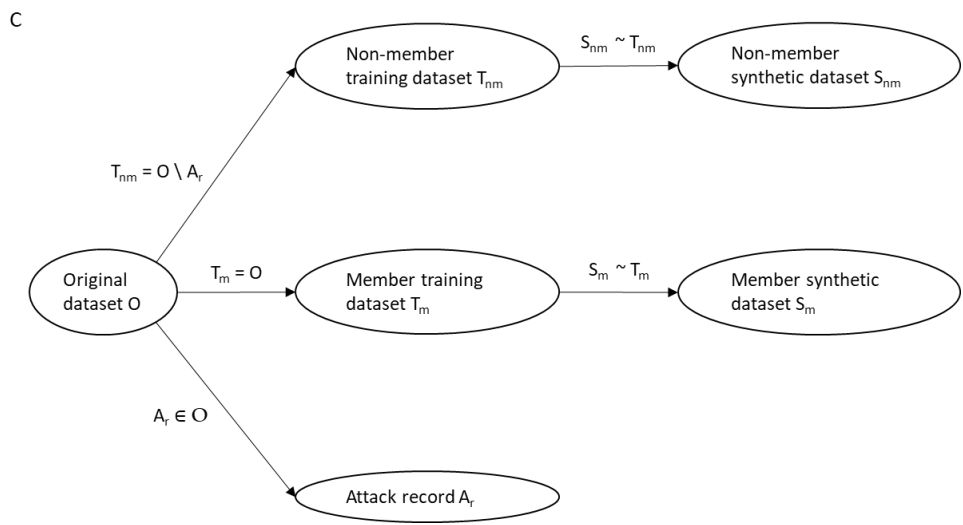

**Figure 13: The constitution of training and attack datasets**. Attack datasets may be a subset of a training dataset that is the same as the original data denoted as attack dataset A ⊆ T (panel A) or a proper subset of a training dataset that by itself is a subset of the original dataset denoted as member attack dataset $A_m$ ⊂ T (panel B). Attack datasets may also be drawn from a holdout dataset denoted as non-member attack dataset $A_{nm}$ ⊂ H (panel B) or include one single record from the original dataset denoted as attack record $A_r$ ∈ O (panel C). Attribute disclosure metrics can be undifferentiated (panel A) by being simple absolute metrics using the synthetic and the attack dataset only, or by including a baseline derived from the training dataset. Attribute disclosure metrics with membership baseline can use a holdout (panel B) or a non-member training dataset (panel C). When attack datasets are built, different subset sizes are described ranging from all records to a subset of randomly drawn records or few selected individual records. ⊆: subset (could be an equal dataset or a proper subset of randomly drawn or selected records), \: complement (e.g. H = O \ T  means that H consists of all elements of O except for T), ∈: element of,  ~: SDG, A: Attack dataset, $A_m$: Member attack dataset, $A_{nm}$: Non-member attack dataset, $A_r$: Attack record, H: Holdout dataset, O: Original dataset, S: Synthetic dataset, $S_m$: Member synthetic dataset, $S_{nm}$: Non-member synthetic dataset, T: Training dataset, $T_m$: Member training dataset, $T_{nm}$: Non-member training dataset.

### 4.2.1  Undifferentiated Absolute Attribute Disclosure Metrics

Undifferentiated absolute attribute disclosure metrics generally have one attack dataset drawn from the training dataset (see Figure 13 panel A). The training data is in this case equal to the original data.

One example for an absolute undifferentiated attribute disclosure metric has just recently been published by Kwatra et al. [28]. In the attack, the sensitive target value $\hat{t}$ is retrieved from the nearest synthetic neighbor of the attack record. The value is then evaluated for correctness (or in case of numeric values for a certain closeness to the real value). The Inference Accuracy $I.A.$ is then calculated as follows:

$$I.A. = \frac{\#\left\{\hat{S}_j : O_j = S_j, j = 1...n\right\}}{n} \tag{15}$$

where # means count, $\hat{S}$ a successful guess, $j$ one attack record out of an attack dataset of size $n$, $O_j$ the correct target value $t$ and $S_j$ the inferred value $\hat{t}$ from the synthetic data. Note that the notation in the publication used $O_j == S_j$ for the equality test but we changed it to $O_j = S_j$ to be consistent with the other example metrics. In this method, the attack dataset is the same as the training dataset except for target attribute $T$. Accordingly, $I.A.$ measures the correct inference across all records of the training dataset.

Attribution disclosure, or meaningful identity disclosure, as proposed by El Emam et al. presents another absolute metric in this subcategory [57]. The metric has three conditions:

1. A synthetic record is similar to an original record on the quasi-identifiers,

2. The original record has a high identity disclosure probability, and

3. The adversary would learn sensitive information about the specific target individual

This means that attribution disclosure occurs if there is a synthetic record that is similar to a real person, not only to an original record [57]. In this metric, this distinction is drawn because the original records can represent a sample from the population, and therefore, matching to a unique original record does not mean matching to a unique real person. The implicit assumption is an adversary that does not know whether their target is in the original data.

In the first condition, similarity is assessed by exact matching on quasi-identifiers. By only using a subset of the quasi-identifiers or generalizing them, the pool of matched synthetic records can increase. The attack dataset is the same as the training dataset except for the target variable $T$.

In the third condition, the sensitive attribute of the matching record is not only assessed for correctness but also for being atypical. This is intended to quantify information gain. For example, if a sensitive attribute in the real data is unimodal and has a small standard deviation, and the sensitive attribute of the target individual is equal to the mean, then according to the authors little is learned: an adversary that just guesses the mean would be almost correct. However, if the matched sensitive attribute is in the extreme tail of the distribution, then what is learned by the adversary is atypical, which is then considered as meaningful information gain. These distinctions are illustrated in Figure 14.

| | | Similarity Within Original Data | |
| --- | --- | --- | --- |
| | | Individual is Similar to Others (typical) | Individual is an Outlier (atypical) |
| **Similarity Between Original & Synthetic Records** | **Individual's Synthetic Information Similar to Real Information** | Low | High |
| | **Individual's Synthetic Information Different from Real Information** | Low | Low |

This table only applies to records that match between the synthetic and real data, and hence have passed the first test for what is defined as meaningful identity disclosure.

**Figure 14: Quantification of information gain.** This shows the relationship between an original record to the rest of the original data and to the matched synthetic record, and how that can be used to determine the likelihood of learning something new. Figure from [57].

The overall attribute disclosure vulnerability is calculated as:

$$\max\left( \frac{1}{N} \sum_{s=1}^{n} \left( \frac{1}{f_s} \times \lambda'_s \times I_s \times R_s \right), \frac{1}{n} \sum_{s=1}^{n} \left( \frac{1}{F_s} \times \lambda'_s \times I_s \times R_s \right) \right) \qquad (16)$$

where $N$ is the size of the population, $n$ the size of the original dataset, $s$ the attack record, $f_s$ the equivalence class size of the attack record in the original data, $F_s$ the one in the population, $\lambda'_s$ an adjustment factor for data errors and verification, $I_s$ a binary indicator whether there was a match and $R_s$ a binary indicator whether information gain was considered as high. The identity disclosure vulnerability of the original dataset is measured from two directions in this equation. The population-to-sample attack is reflected in the following notation:

$$\frac{1}{N} \sum_{s=1}^{n} \frac{1}{f_s} \qquad\qquad (17)$$

The sample-to-population attack denoted as:

$$\frac{1}{n} \sum_{s=1}^{n} \frac{1}{F_s} \qquad\qquad (18)$$

### 4.2.2 Undifferentiated Relative Attribute Disclosure Metrics

As undifferentiated absolute metrics, these metrics generally have one attack dataset drawn from the training dataset (see Figure 13 panel A) but use additional baselines drawn from a naïve guess or the training data in these metrics.

Elliot et al. and Taub et al. describe the concept of Correct Attribution Probability (CAP) [25,26]. The authors conduct several attacks to account for a baseline to compare against, making this a relative metric. The main attack finds exact matches between the attack dataset and the synthetic dataset. There may be one, multiple or no synthetic match. This can vary depending on how $K$ is chosen. Per record, the accuracy is given as the correctness of the retrieved sensitive synthetic value $_{\hat{r}}$ in case of one matching record, or as the accuracy across all retrieved values in case of multiple matching records. It is calculated as follows:

$$CAP_{s,j} = \Pr(T_{o,j} \mid K_{o,j})_s = \frac{\sum_{i=1}^{n} \left[ T_{s,i} = T_{o,j}, K_{s,i} = K_{o,j} \right]}{\sum_{i=1}^{n} \left[ K_{s,j} = K_{o,j} \right]} \qquad (19)$$

where $CAP_{s,j}$ is the CAP for synthetic data, $Pr(T_{o,j}|K_{o,j})_s$ the conditional probability of the sensitive target values $T_{o,j}$ of an attack record $j$ given its key values $K_{o,j}$, $n$ the number of attack records, $T_{s,i}$ the sensitive target values of the matched synthetic record $i$ and $K_{s,i}$ its synthetic key values. The denominator counts the matches for that attack record.

Non-matches (i.e., denominator of 0) can count as an accuracy of 0 or as non-defined. This makes a difference when calculating the accuracy across the whole attack dataset. The attack dataset in this method can be the same as the training dataset or a subset of selected records, both options take the complete record except for the target attribute $T$. The overall accuracy is measured via averaging and the total amount of records that are used in averaging is reduced by the number of non-matches when these are counted as non-defined.

The authors use a baseline that is calculated as CAP for the original dataset and one that comes from the majority class of the univariate distribution of the sensitive target attribute $T$ (naïve guess). They report all three estimates, the CAP of the synthetic, the CAP of the original dataset and the naïve guess.

Hittmeir et al. extend CAP to a generalized CAP (GCAP) by using a nearest neighbor approach that handles non-matches [27]. As for CAP, the authors introduce a baseline derived from the original data to support interpretability. The actual attack finds exact matches between the target record and the synthetic dataset. Again, there may be one, multiple or no match. In case of a non-match, the number of attributes in $K$ is reduced until one or multiple matches are identified. The subsequent calculation of the absolute

GCAP is identical to the CAP by Elliot et al. and Taub et al. [25,26]. Again, there is no composite metric, but the baseline built upon the original dataset is given as additional information.

Stadler et al. proposed two frameworks to measure privacy vulnerabilities in synthetic data, both of which include an attribute disclosure module and both of which follow a concept that is best described as nested baselines [59,63]. In that way, they differ from the metrics presented thus far by reporting a relative metric only.

The first published framework fits into the subcategory of undifferentiated attribute disclosure metrics. It calculates the composite metric Privacy Gain (PG) out of the Privacy Loss (PL) in the original data and the one in the synthetic data [59] as follows:

$$PG_t(S,R) = \frac{PL_t(R) - PL_t(S)}{2} \tag{20}$$

where $PG_t$ is the PG for a specific attack record, $S$ is the synthetic dataset and $R$ the original dataset, and $PL_t$ their respective PL for that record. In this metric, $PL_t(R)$ serves as an upper baseline and a $PL_t(S)$ equal to or larger than $PL_t(R)$ is deemed to be too high. A naïve guess is further incorporated as a nested baseline when calculating PL:

$$PL_t(X) = P\left[\hat{t}_s = t_s \mid X\right] - P\left[t_s\right] \tag{21}$$

where $X$ can be the original or the synthetic dataset, $P\left[\hat{t}_s = t_s \mid X\right]$ is the probability that the presumed sensitive target value $\hat{t}$ is correct given the respective dataset and $P[t_s]$ is a naïve guess according to the adversary's prior beliefs. $P[t_s]$ as nested baseline, however, gets cancelled out when calculating the PG between the original and the synthetic data as it is assumed to be the same for both datasets. Therefore, no details are given on how it may be computed. $P\left[\hat{t}_s = t_s \mid X\right]$ is calculated by training a prediction model on the respective dataset $X$ as a first step. The model is then used to predict the sensitive target value $\hat{t}$ given the key attributes $k$ of an attack record. The attack dataset in this method is a small random or selected subset of the training dataset.

### 4.2.3 Attribute Disclosure Metrics with Non-Member Baseline

Two exemplar metrics of the third subcategory are presented in the following both of which are part of more comprehensive frameworks for measuring disclosure vulnerability [62,63].

Giomi et al. introduce a non-member baseline that is derived from a holdout dataset (see Figure 13 panel B) [62]. It serves as a lower baseline and is used to calculate attribute disclosure vulnerability. The actual attack is based on nearest neighbor modeling for attack records derived from the training (i.e., member) dataset. The non-member baseline is realized by repeating this process on the holdout (i.e., non-member) dataset. The disclosure vulnerability is then composed as follows:

$$R = \frac{r_{train} - r_{control}}{1 - r_{control}} \tag{22}$$

where $r_{train}$ is the metric obtained from the attack using members as attack records and $r_{control}$ the one using non-members. The components $r_{train}$ and $r_{control}$ are calculated as the accuracy for the retrieved sensitive target value $\hat{t}$ of being correct. The predictions are drawn from the nearest synthetic neighbor for each attack record based on their key attributes $k$. In case of multiple nearest synthetic neighbors, the proportion of correct synthetic target values among all neighbors is used. The accuracy for all members is averaged into $r_{train}$, the one of non-members (from the holdout set) into $r_{control}$. There is the same number of member and non-member attack records, and both attack datasets are random subsets of the training and holdout dataset respectively.

While this metric does only report a relative metric, its structure is similar to Cohen's Kappa [100] in the way that it normalizes the difference to $r_{control}$ by the maximum possible vulnerability beyond $r_{control}$ and thereby takes the scale into account. $R$ can be calculated in multiple ways: using each attribute of the dataset as $T$, reducing the number of attributes in $K$ or increasing the number of neighbors. $R$ may reflect one of these set-ups but can also be provided as the average across them. The authors further introduce a weak attack by calculating the Attack Strength with the following formula:

$$s = r_{train} - r_{naive} \qquad (23)$$

where $r_{naive}$ is a naïve guess where the sensitive target value $\hat{r}$ is randomly drawn from all possible values in its domain. A weak attack occurs when the naïve guess $r_{naive}$ exceeds the member attack $r_{train}$. Weak attacks are invalid and may indicate a poorly chosen prediction model. In this context, the naïve guess is not a baseline but used to ensure a certain stability of results. To understand the interplay of the different components, we can look at exemplar scenarios:

1. Scenario 1: Members are better predicted than non-members. $r_{train}$ = 0.9 (members), $r_{control}$ = 0.5 (non-members), $r_{naive}$ = 0.5 (naïve guess). The attribute disclosure metrics are $R = 0.8$ and $s = 0.4$.

2. Scenario 2: Members are better predicted than non-members. $r_{train}$ = 0.5 (members), $r_{control}$ = 0.1 (non-members), $r_{naive}$ = 0.5 (naïve guess). The attribute disclosure metrics are $R = 0.84$ and $s = 0$.

3. Scenario 3: Members are better predicted than non-members. $r_{train}$ = 0.95 (members), $r_{control}$ = 0.9 (non-members), $r_{naive}$ = 0.5 (naïve guess). The attribute disclosure metrics are $R = 0.5$ and $s = 0.45$.

4. Scenario 4: Members and non-members are equally poorly predicted. $r_{train}$ = 0.5 (members), $r_{control}$ = 0.5 (non-members), $r_{naive}$ = 0.5 (naïve guess). The attribute disclosure metrics are $R = 0$ and $s = 0$.

5. Scenario 5: Members and non-members are equally well predicted. $r_{train}$ = 0.9 (members), $r_{control}$ = 0.9 (non-members), $r_{naive}$ = 0.5 (naïve guess). The attribute disclosure metrics are $R = 0$ and $s = 0.4$.

Among these scenarios, the second and fourth scenarios would provide invalid results according to the attack strength. In all other scenarios, the relative metric $R$ would be used to report attribute disclosure vulnerability.

Stadler et al. propose another framework that includes a nested non-member baseline (see Figure 13 panel C) [63]. Again, the relative metric Privacy Gain (PG) is calculated out of estimates from the original and the synthetic data:

$$PG = Adv^i(R, r_t) - Adv^i(S, r_t) \qquad (24)$$

where $Adv^i$ is the Adversary's Advantage for attribute disclosure, $R$ the original dataset, $r_t$ the attack record and $S$ the synthetic dataset. The Adversary's Advantage of the original dataset ($Adv^i(R, r_t)$) serves as an upper baseline and an $Adv^i(S, r_t)$ equal to or larger than $Adv^i(R, r_t)$ is deemed to be too high. A non-member baseline is then incorporated as nested baseline when calculating $Adv^i$:

$$Adv^I \left( X, r_t \right) = P\left[ \hat{r}_s = r_s \mid s_t = 1 \right] - P\left[ \hat{r}_s = r_s \mid s_t = 0 \right]$$ (25)

where $X$ can be the original or the synthetic dataset, $P\left[ \hat{r}_s = r_s \mid s_t = 1 \right]$ is the probability that the presumed sensitive target value $\hat{r}$ (i.e., $\hat{r}$) is correct given that the attack record is a member ($s_t = 1$) and $P\left[ \hat{r}_s = r_s \mid s_t = 0 \right]$ the probability given it is not a member ($s_t = 0$). The accuracy for the record is the classification accuracy of a random forest model trained on the synthetic dataset. A linear regression model is used in case of numeric target values.

This formulation of the non-member baseline requires that for each attack record two synthetic datasets are generated, one with a training dataset containing the record (member training dataset) and one without it (non-member training dataset). The non-member training dataset contains a duplicate of a random record to maintain the same training dataset size. The authors choose 5 records that serve as attack records. These records were considered as vulnerable because of having rare (outside 95% quantile) attributes. Apart from this deliberate preselection, it also seems difficult to apply the method on a large scale considering its computationally intensive workload.

To understand the metric's behavior, two distinctive influencing factors may be relevant: the replication of members in the synthetic dataset (i.e., SDG process overfitting) and the overfitting of the prediction model trained on the synthetic dataset. The following example scenarios can help to understand the metric, and they are defined to assume that the extent of overfitting or generalizing of the prediction model is exactly the same for both, the one trained on member synthetic data and the one trained on non-member synthetic data. However, random forest classifiers may not behave the exact same way, which would add an additional layer of complexity in interpretation:

1. Scenario 1: The SDG process and the predicting models overfit. Members are better predicted than non-members in both predicting models. $P\left[ \hat{r}_s = r_s \mid s_t = 1 \right] = 0.9$ (members), $P\left[ \hat{r}_s = r_s \mid s_t = 0 \right] = 0.5$ (non-members), $Adv^I \left( S, r_t \right) = Adv^I \left( R, r_t \right) = 0.3$, $PG = 0$.

2. Scenario 2: The SDG process overfits and the predicting models do not overfit. Members are not better predicted in both models. $P\left[ \hat{r}_s = r_s \mid s_t = 1 \right] = 0.8$ (members), $P\left[ \hat{r}_s = r_s \mid s_t = 0 \right] = 0.8$ (non-members), $Adv^I \left( S, r_t \right) = Adv^I \left( R, r_t \right) = 0$, $PG = 0$.

3. Scenario 3: The SDG process and the predicting models do not overfit. Members are not better predicted in both models. $P\left[ \hat{r}_s = r_s \mid s_t = 1 \right] = 0.8$ (members), $P\left[ \hat{r}_s = r_s \mid s_t = 0 \right] = 0.8$ (non-members), $Adv^I \left( S, r_t \right) = Adv^I \left( R, r_t \right) = 0$, $PG = 0$.

4. Scenario 4: The SDG process and the predicting models overfit. Members are better predicted than non-members in both predicting models. $P\left[ \hat{r}_s = r_s \mid s_t = 1 \right] = 0.5$ (members), $P\left[ \hat{r}_s = r_s \mid s_t = 0 \right] = 0.1$ (non-members), $Adv^I \left( S, r_t \right) = Adv^I \left( R, r_t \right) = 0.4$, $PG = 0$.

All of these scenarios would result in a PG of zero. It is unclear if a PG of zero means that the attribute disclosure vulnerability is deemed to be too high or merely that there was no improvement on the original data. The relative metric PG does not explicitly give the difference between non-members and members of the synthetic data but compares it to the original data's difference between non-members and members. In this sense, PG ultimately takes the original data as a reference point and can formally be classified as an undifferentiated metric. Its component $Adv^{\sqrt{}}(S,r_t)$, however, is an attribute disclosure metric with non-member baseline. We decided to categorize the whole metric as such because this is probably where the strength of the metric lies, and it gives another option of how a non-member baseline (i.e. $Adv^{\sqrt{}}(S,r_t)$) can be implemented. Looking at the $Adv^{\sqrt{}}(S,r_t)$ only in the scenarios draws a different picture than using PG. It may fit better our intuitive vulnerability assessment.

The proposed non-member baseline in [12] takes components of both metrics presented so far. The metric itself take the form of equation (22) but only when using members as attack records is taken from the synthetic data while the attack records for non-members are taken from the original data.

## 4.3    Critical Appraisal

There are several challenges with the way attribute disclosure is measured in the literature. We will describe common challenges first and then go into specific considerations for each subcategory.

### 4.3.1    Attribute Disclosure Vulnerability is Tied to Model Selection

The success of predicting a sensitive attribute (i.e., attribute disclosure) depends on the prediction model and attribute choice. Not every prediction modeling technique delivers the same accuracy. For example, a CART model is not likely to perform as well as a gradient boosted decision tree for prediction of the sensitive attribute. If the model is poorly chosen, the results of attribute disclosure evaluation will be lower vulnerability, even though the results would look completely different if a better prediction model was used [27]. Furthermore, some attributes may be predicted better than others. So, if the sensitive attribute $T$ has a strong relationship to the key attributes $K$, the prediction model performance would indicate high vulnerability while it would be very different for an attribute $T$ that is fairly uncorrelated to the predictors.

A low attribute disclosure vulnerability according to prediction model accuracy cannot be interpreted without taking the model, the key attributes $K$, and sensitive attribute $T$ choice into account. The decision on when a model performance becomes problematic needs to consider these factors. For example, some have argued that the prediction needs to have a "significant probability" of being correct [132]. This could mean that if accuracy is above 80% then disclosure would have occurred. But then it is unclear, if consequently an accuracy of 60% should be considered as low vulnerability.

Unstable results may also be present in relative metrics. For example, in [63], the attribute disclosure vulnerability of synthetic data is assessed by comparing the accuracies of a prediction model trained on two different synthetic datasets (see equation (25)): one where the training dataset for SDG contained a specific record (i.e., member) and another without that record (i.e., non-member)). Assuming an extreme scenario where SDG has resulted in a replicate of the training data, then attribute disclosure vulnerability of the synthetic data could be low in case the prediction model does not overfit.

However, if prediction model overfits or memorizes the training data, the vulnerability would be high. Therefore, the quality of the prediction model training and hyperparameter selection will also be relevant factors in deciding on the extent of attribute disclosure vulnerability.

One option is to factor out the influence of the prediction model by defining a relative metric where the baseline is derived with the same model choice [63]. This would work under the assumption that the extent of overfitting or generalizing in the prediction models can be controlled. Another way to get more stable results may be to standardize on a nearest neighbor model [62]. This has the advantage that it maximally overfits in the sense, that when there is one exact match and this match is a replicate, it will have an accuracy of 100%. In [62], the authors come up with the idea of an indicator derived from a naïve guess which has been defined as the Attack Strength where the absolute attribute disclosure for a member is measured against the naïve guess randomly drawn from all possible values (see equation (23)). A weak attack is then defined as a prediction that falls below the naïve guess which is deemed to be invalid and may indicate that the key attributes $K$ and sensitive attribute $T$ for this model are poorly chosen.

### 4.3.2 Privacy Should Focus on The Ability to Predict Not on Potential Harm of The Predicted Value

Various authors suggest only targeting a subset of selected (i.e., "vulnerable") records [25,59,63]. It is essential to clarify the meaning of "vulnerable" within this context. Vulnerable records may be those containing potentially harmful information (e.g., HIV diagnosis), or those being simply rare in their combination of the attributes $K$ or their sensitive value $t$.

Under the first definition, the main concern with attribute disclosure is whether the predicted sensitive value can be harmful to data subjects, rather than the ability to correctly predict a sensitive value [133]. A common narrative is that predicting a sensitive value can be used in a discriminatory manner. However, that pertains to the *use* of the predicted value rather than training a predictive model. Predicted sensitive attribute values can also be used for good. For example, predictions from a model that predicts the likelihood of men getting prostate cancer in five years can be used to deny bank loans or to initiate wellness and screening programs.

Related to this definition is the idea that an adversary would experience higher ("meaningful") information gain when a sensitive target value is atypical [57]. Such an interpretation, however, is highly contextual. Discovering a typical sensitive target value can also provide meaningful information gain. For example, consider a bank manager who approved a high-risk loan based on the belief that an individual had an atypical high income. Upon gaining access to a dataset that includes financial information about the individual, the manager discovers that the individual actually has a typical income level. This information gain is very likely to have consequences that are meaningful for the manager and the individual. Therefore, whether a typical sensitive target value represents significant information gain depends on prior beliefs and context.

More broadly, attribute disclosure should not exclude typical values from the assessment for privacy violation and make assumptions on the value of this information. The predicted value itself may or may not be harmful. Privacy laws, however, emphasize the personal nature of the information being processed rather than the harms that flow from this processing [132]. Disclosure vulnerability metrics should quantify the ability to predict rather than the potential harm resulting from those prediction. Determining potential harm is still part of the overall privacy risk assessment and management process, which includes the Invasion of Privacy construct to adjust the thresholds that are used as described in Section 1.3.4, and by implementing an ethical review on synthetic data uses.

### 4.3.3 There is no Clear Relationship Between Being Rare or Atypical and High Attribute Disclosure Vulnerability

The other definition of vulnerable considers rarity as a relevant characteristic for a record to be vulnerable to attribute disclosure. Rarity indeed may be relevant for the ability to predict but depends on which parts of a record are rare. The ability to predict would be affected differently when there is a rare or atypical combination of the key attributes $k$ versus a rare or atypical target sensitive value $t$ within all records versus a rare or atypical target sensitive value $t$ within the pool of matched synthetic records. The latter may be equivalent to rarity of the sensitive value $t$ within an equivalence class.

Rare key attributes $k$ can decrease the pool of matches but a clear consequence on the ability to predict cannot be drawn. A rare value $t$ could be hard to predict if there is no clear relationship to the key attributes *k* but may be easier to predict if there is. A rare value *t* within an equivalence class, however, is very likely protective to the target. The adversary would predict the majority class within the equivalence class with a resulting low disclosure vulnerability for that very target.

These examples show that assumptions on which records may have a high attribute disclosure vulnerability cannot be easily generalized and depend on the respective dataset. This underscores that a pre-selection of attack records may not be of much use. Relying on the assumption that rare records come with highest attribute disclosure vulnerability, can result in overlooking records that actually have a higher vulnerability. Rarity is not always a worst-case scenario because it can also protect from prediction. Accounting for a worst-case scenario should always involve evaluating all records in the first instance, and then filtering for those records that lead to the worst case scenario.

### 4.3.4  The Way Prediction Performance is Measured Requires Context-Specific Considerations

Attribute disclosure vulnerability is a prediction task where the sensitive target variable can be categorical (i.e., classification) or continuous (i.e., regression). There are many ways to quantify prediction performance for both of these. For example, classifiers can be ranked very differently for the same task depending on the chosen performance measurement [115], and this ranking may be totally different for another task. The choice of performance measure will depend on the nature of the sensitive attribute and will be problem specific.

Many prediction problems are framed as a binary classification task. A common measure of prediction performance for binary classification is the Area Under the Receiver Operating Characteristics Curve (AUROC). The AUROC evaluates whether positive classes are generally ranked higher than negative ones regardless of the actual scores which has the advantage of being threshold-independent. In attribute disclosure vulnerability, threshold-independence can be beneficial in situations where the prevalence of the target values in the attack dataset of an adversary is unknown and thereby the choice of a reasonable threshold not possible. The AUROC is also agnostic to the nature of prediction (e.g., which value of the target attribute is predicted) meaning that it treats positive and negative classes with equal relevance. We will therefore use the AUROC as an exemplar attribute disclosure vulnerability metric, analyze its interpretation in attribute disclosure vulnerability and present a way of using its precedents to inform possible thresholds. Such thresholds can be based on the interpretation of AUROC in the prognostic literature since in the context of attribute disclosure vulnerability, we are assessing prediction performance [36,134]. Different labels have been proposed when interpreting the AUROC ranging from failed or random to very good, high or excellent accuracy (see **Figure 15**) [36].

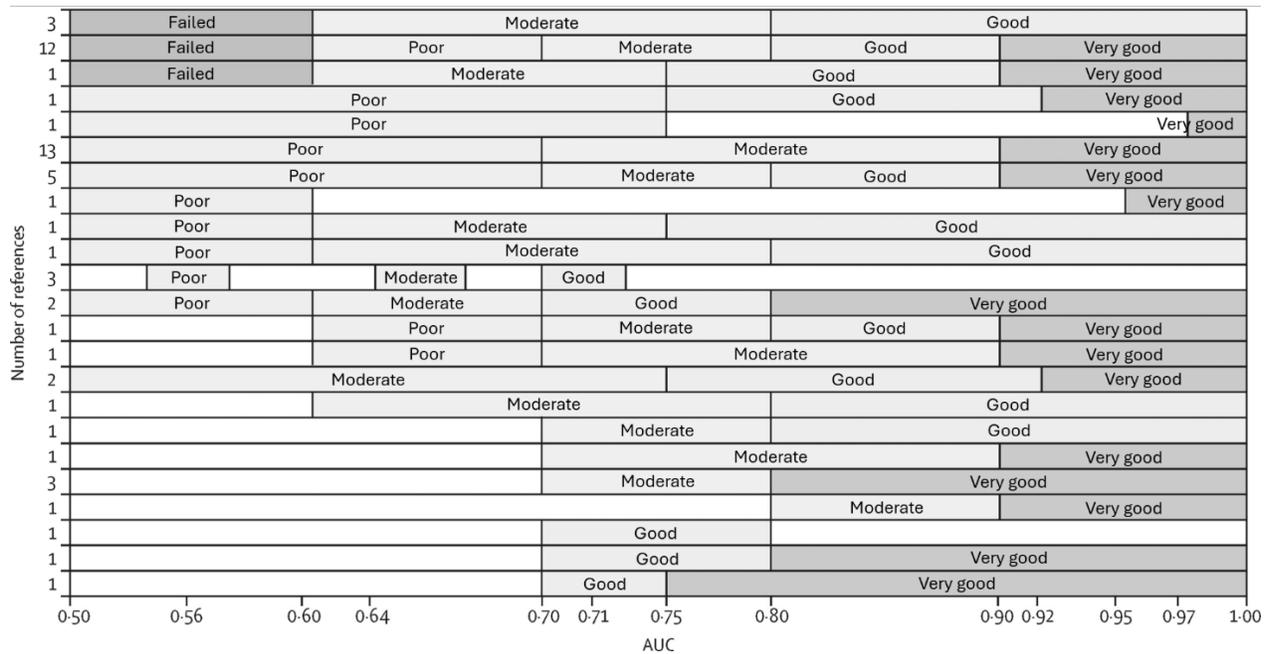

**Figure 15. Different labelling systems of AUROC from the literature**. Adapted from [36].

Most labeling systems in the literature use three or five categories for interpreting the AUROC result [36]. When three categories are used, poor accuracy is assigned to an AUROC lower than or equal to 0.7, moderate accuracy to an AUROC between 0.7 and 0.9 and very good accuracy to an AUROC higher than or equal to 0.9. Alternatively, in labeling systems with five categories, a model is deemed to have failed if its accuracy is 0.6 or lower. Following this, the remaining four labels distribute evenly across the range of 0.6 to 1.0, with each successive category differing by 0.1 AUROC points. Following this interpretation, a conservative threshold of 0.6 could be established for the absolute attribute disclosure vulnerability (i.e., $A_{members}$) meaning that only accuracies that exceed random or poor prediction (i.e., 0.6) can be meaningful personal information disclosure. For the relative metric (i.e., $A_{rel}$), the change in labels can be used to inform a threshold. These are typically after 0.1 to 0.2 AUROC points and setting a threshold of 0.15 appears reasonable. These labeling systems have not been specifically proposed for AUROC in attribute disclosure vulnerability, but they give meaningful guidance on how to interpret AUROC and which threshold to use as a starting point (see Figure 16). Note that certain scenarios may be considered as more privacy invasive, so that an upward adjustment of the thresholds is required (see 1.3.4).

|  | $A_{members}$ | |
| --- | --- | --- |
|  | ≤ 0.6 | > 0.6 |
| $A_{rel}$ ≤ 0.15 | Acceptable vulnerability | Acceptable vulnerability |
| $A_{rel}$ > 0.15 | Acceptable vulnerability | High vulnerability |

**Figure 17:** Defining acceptable attribute disclosure vulnerability based on thresholds for relative and absolute attribute disclosure accuracy.

The AUROC is just one prediction performance measurement and has been criticized for being insufficiently sensitive in imbalanced datasets [135]. The Area under the Precision-Recall Curve (AUPRC) has been proposed as an alternative in imbalanced datasets [135], this, however, puts a higher emphasis on false positives than false negatives [136] which needs to be taken into account.

### 4.3.5    Undifferentiated Absolute Attribute Disclosure Metrics

The major issue with these metrics is that model training and accuracy prediction are common tasks in scientific investigations, whether they are academic or commercial, and contribute to knowledge generation.

When the adversary provides the values of *K* about the target as input to the model, predict the sensitive value and the prediction accuracy is used to report attribute disclosure, then a high risk would simply be a high prediction accuracy of a model. High prediction model accuracy is identical to the very aim of research. If the synthetic data is sufficiently useful, then relationships are maintained, and the prediction is accurate. Considering high accuracy as a privacy issue implies that we should not retain important relationships in synthetic data to prevent attribute disclosure as defined by these metrics.

Provided that the prediction models that inform such metrics are generalizable, then undifferentiated attribute disclosure metrics would presumably be high even in targets who were not in the training data (i.e., non-members). They could also be high even if the synthetic data is not generated from original data. For example, the Synthea approach generates synthetic data based on guidelines and expertise, but no underlying data was input to train a generative model [1]. Synthetic data can also be derived from publicly available risk calculators [2]. Fitting an accurate predictive model on the generated synthetic data in this case would still be deemed to entail a high attribute disclosure vulnerability.

A legal analysis showed that such a scenario could indeed be interpreted as a privacy violation [132]. It states that sensitive information about an individual that was not part of a training dataset (i.e. non-member) might still be considered personal information under the definitions in, for example, the California Consumer Privacy Act [132]. The authors acknowledge at the same time that in the US this would probably raise First Amendment challenges. The First Amendment in the US protects freedom of speech and inference constitutes speech. Similarly, it has been proposed that the concept of privacy can be extended to groups or to collective privacy in the context of big data [137,138]. This means that a predictive model that can make predictions about groups could result in a privacy violation. However, unless there is high rate of memorization, predictive models will make predictions about groups. Under such concepts, privacy can be compromised even without the target individual having disclosed any information (non-member). Ethical policies such as the Tri-Council Policy Statement in Canada recognize group privacy but it remains unclear to what extent non-member individuals of the group are also included in this definition [139].

In technical privacy research, however, an increasingly accepted principle is to make sure that being a member in a dataset does not increase the likelihood of an adversary gaining sensitive information about an individual so that everything that can be learned about the individual can also be learned without them being a member of the dataset [12,13,124,125]. In [13], the first Cancer Prevention Study (CPS) is used as an illustrative example to explain these considerations. This study with data collected from 1959 to 1972 identified smoking tobacco cigarettes as a risk factor to die from lung cancer and coronary heart disease. These findings inform preventive medicine to reduce mortality but can also result in higher health and life insurance costs for smokers. Let's assume a smoker born after 1972. While they would be impacted by the findings of this study, it would not be considered as disclosure of their data but rather as knowledge generation. Another individual would be a smoker that participated in the study. They could suffer from the consequences of the study as their insurance costs might have increased for smokers as a reaction to the study. This, however, would have been the consequence whether they participated in the study or not, and should thereby not be considered as a privacy violation.

From this perspective, it follows that inferences about non-members cannot be attribute disclosure in the sense of a privacy compromise. This would be more consistent with the goals of research where inference about non-members results from publishing, risk calculators, aggregate statistics and prognostic results. It is also in line with authors questioning whether privacy laws should be extended to cover inferences of non-members [132,140]. This is not to say that the release of data, its analysis, and knowledge generation cannot be harmful to non-members, and the release may be discarded due to these reasons. The concept of harm, as noted earlier, should be considered separately from disclosure vulnerability.

### 4.3.6 Undifferentiated Relative Attribute Disclosure Metrics

The choice of baselines when defining relative attribute disclosure vulnerability metrics is important [59,62,63], which we consider below.

Most metrics include a baseline derived from the original data [25–27,59,63]. This is often framed as an upper baseline. Simply put, the idea is that a prediction model trained on synthetic data must not be as good as the one trained on original data [59]. For example, the first published definition of Privacy Gain by Stadler et al. defined it as the difference in model accuracy between a model trained on the original data and one on the synthetic data (see equation (20)) [59].

In general, replicability of analysis results is one of the characteristics of good synthetic data, and it means that the results from analyses performed on original data can be replicated on the synthetic data [32,61,141]. While Privacy Gain is a broader concept that explores whether or not synthetic data results in less disclosure vulnerability than original data, the intuitive interpretation would be that the disclosure vulnerability of original data is an upper threshold that must be reduced. With such an interpretation of attribute disclosure, replicability would also be considered as a "risky" situation. A negative Privacy Gain would even be considered as severe disclosure but may occur when the synthetic data model is simply better at capturing global properties with a subsequent better prediction model performance than the original data prediction model. It remains unclear why a prediction accuracy as good as the one from a model trained on the original data should be deemed problematic. This comes back to the matter of knowledge generation that has already been discussed and is clearly distinguished from privacy by other authors [62,124].

Another frequently used baseline is the one of a naïve guess derived from the univariate distribution of the sensitive attribute [25,26,59]. The idea behind this baseline is to compare the absolute vulnerability of the synthetic data to the one that would result from the common practice publishing univariate distributions which is in general deemed not to be a privacy violation [12,25,59]. While this choice is pragmatic, it also poses the risk of introducing arbitrary baselines that may give additional information but do not add value when it comes to making a decision (see 1.3.2).

## 4.3.7 Attribute Disclosure Metrics With Non-Member Baseline

Non-member baselines seem to be a reasonable way to solve the dilemma between attribute disclosure and knowledge generation. It is to be determined, however, how such a baseline can best be defined, and the relative metric best be formulated.

### 4.3.7.1 A Holdout Set May be a Pragmatic Proxy for a Non-membership Baseline

A precise implementation of a non-membership baseline is performed in [63] (see Figure 13 Panel C). Synthetic data is once generated from the complete training dataset (member training dataset) and once from a training dataset where the attack record is removed (non-member training dataset). The prediction accuracy for this very record is then compared between the two prediction models trained on different synthetic datasets. The authors only calculated the risk for selected records. To obtain a vulnerability at the dataset level, however, each record in the dataset needs to be iteratively held out. This comes with considerable computational cost which makes it an impractical solution to assess attribute disclosure at large scale [62,65].

A computationally more efficient approach is to approximate the non-member baseline by randomly drawing non-members into a larger holdout (non-member) dataset and evaluating them with the same type of prediction model as the members are evaluated (see Figure 13 Panel B) [62]. The idea is to detect how well the SDG model generalizes. If the synthetic data is a better reflection of the population than the training dataset, then attribute disclosure drawn from the synthetic data should be similarly accurate for non-members and members. Note that this is conceptually exceptional as, in such a case, there is no privacy utility trade-off but a privacy utility-alignment. Both, privacy and utility, strive towards better generalization. The holdout baseline is, however, only an approximation to the non-member concept and depends on the size of the holdout dataset. This can be drawn from the simulations conducted for record-level similarity, and from observations when performing singling out attacks [62]. In both examples, the vulnerability was dependent on the holdout dataset size. There is no compelling reason why this should not apply to any non-member baseline that is based on a holdout dataset. It therefore may require calibration on the sample size.

Another way to approximate the non-member baseline by a holdout dataset is proposed in [12] where accuracy for non-members is calculated from predictions using a model trained on the training dataset (without non-members) while accuracy for members is determined using a model trained on the synthetic dataset. This is meant to reduce the risk of introducing artificially low non-member baselines and avoiding overprotecting datasets. Artificially low non-member baselines may occur when a model is trained on anonymized data and used to predict non-members. In SDG that generalize well, however, we may encounter a higher prediction accuracy for non-members when training the model on the synthetic data than on the training data. Conducting an evaluation on the same model also comes with the benefit of factoring out the factors related to model choice (see 4.3.1).

### 4.3.7.2 Where on The Scale The Accuracy Lies Still Matters

Both presented attribute disclosure metrics have a relative metric only to report attribute disclosure vulnerability [62,63]. Consequently, the question arises whether the relative attribute disclosure vulnerability is sufficiently informative for decision-making processes. This is, in particular, relevant when thinking of attribute disclosure vulnerability as guidance to the decision whether to release synthetic data.

In such a decision, the data controller may want to consider the perspective of the adversary and also the one of the data subjects in such a decision-making process.

An adversary does not care whether or not a target record is a member but aims for a high prediction accuracy. A similar perspective would be taken by data subjects who presumably care more about a difference in high accuracy than about one in lower accuracy. So, it may be that not only the difference to the non-member baseline but where the difference is within the range of the scale matters. It may, for example, be acceptable if a member is predicted with an accuracy of 0.45 even if the non-member accuracy is substantial lower, let's say 0.2. In contrast, a situation would be presumably interpreted as unacceptable if the members were predicted with an accuracy of 0.85 and non-members with 0.6 even though the relative metric can be the same. Therefore, a more comprehensive approach would take the absolute as well as the relative metric into account.

Any form of absolute scale, however, would need to factor out influencing factors of the model choice as discussed earlier.

### 4.3.7.3 A Ratio Relating to The Maximum Vulnerability beyond the Non-Member Baseline Penalizes Relevant Knowledge Generation

A relative attribute disclosure vulnerability that is reported as a ratio does also reflect the absolute metric (i.e., the scale) to a certain degree. This is, for example, the case in equation (22) where the attribute disclosure vulnerability is given relative to the maximum possible attribute disclosure vulnerability beyond the non-member baseline [62]. The non-member is thereby calculated as shown in Panel B of Figure 13. The metric treats differences at the upper end of the scale as more severe than those at the lower end. At the same time, however, it penalizes high non-member baselines and thereby penalizes potentially relevant knowledge generation. Assuming, for example, a non-member baseline of 0.9 accuracy. This is commonly seen as a very good or excellent model [36]. The high accuracy on unseen data (i.e., non-members) suggests that relevant and externally valid insights have been derived from the data. A higher accuracy for members is very likely to occur (see 4.3.1). At this end of the scale, however, a slightly higher member accuracy of 0.95 would result in a high attribute disclosure metric of 0.5 as calculated from equation (22). With a commonly used threshold for ratios of 0.2 [30–32], this would be considered as an unacceptably high vulnerability. Figure 18 shows the behavior of the metric for selected non-member baselines ranging from very poor (A) to very good (D) model performance. As the non-member baseline increases, a slightly better accuracy for members already exceeds the threshold of 0.2. Such a relative metric thus comes with the challenge that potentially highly valuable insights would be considered as privacy violation even though the actual difference is negligible.

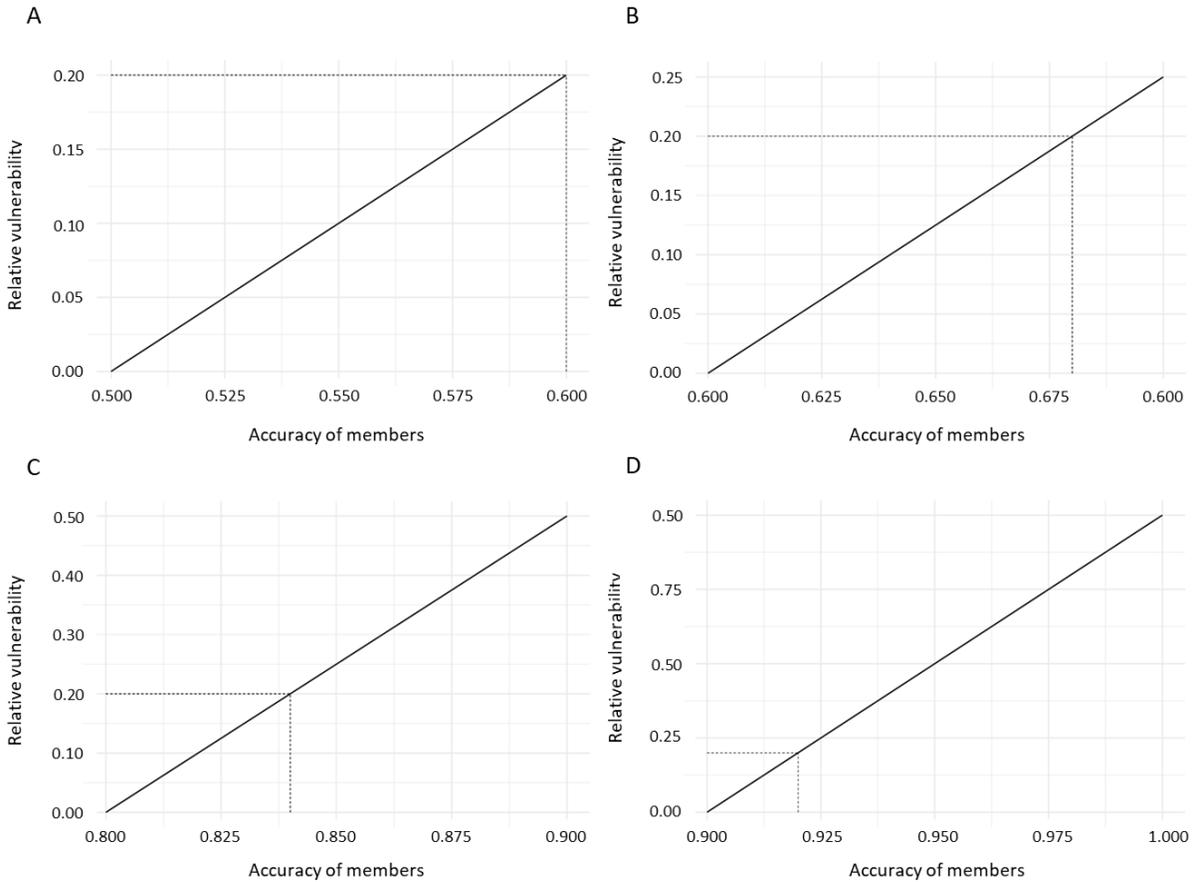

**Figure 18: Relative attribute disclosure vulnerability for selected non-member baselines**. The calculation of disclosure vulnerability follows equation (22). The non-member baseline was fixed at four points from very poor to very good accuracy: 0.5 (A), 0.6 (B), 0.8 (C) and 0.9 (D). Values of member accuracy cover a range of 1 accuracy point from the non-member baseline on. A relative attribute disclosure vulnerability of 0.2 with the respective tolerance for member accuracy is highlighted (dashed line).

---

**Summary of Key Points from the Critical Appraisal of Attribute Disclosure Metrics**

- Attribute disclosure metrics are calculated as a prediction task where the vulnerability estimates depend on the prediction model chosen.

- Pre-selecting records to calculate targeted attribute disclosure is failure prone as pre-selection must not necessarily capture the records with highest vulnerability.

- There are multiple ways to report prediction performance. For attribute disclosure vulnerability, an appropriate metric should be used that is consistent with the nature of the data and the target attribute being predicted.

- Undifferentiated attribute disclosure metrics may incorrectly classify knowledge generation as privacy violation.

- An attribute disclosure occurs from being part of a dataset rather than from being part of the population where the dataset is drawn from. A non-member baseline based on a holdout set is an effective way to simulate this idea.

- When using a non-member baseline, knowledge of the absolute AUC is still required to make a meaningful decision. Where the difference to the non-member baseline is within the range of the scale matters.

- Attribute disclosure vulnerability that is given as a ratio relative to the maximum possible attribute disclosure vulnerability beyond the non-member baseline penalizes high non-member baselines and thereby potentially relevant knowledge generation.

---

## 4.4    Recommendations

### 4.4.1    Distinguishing Attribute Disclosure Vulnerability from Knowledge Generation

It is clear from the discussion above that knowledge generation is something that needs to be accounted for when deciding on an attribute disclosure metric. A non-member baseline appears to be able to separate privacy violation from knowledge generation, and the implementation by Giomi et al. [62] using a holdout dataset serves as a pragmatic approximation.

### 4.4.2    Defining Acceptable Attribute Disclosure Vulnerability Using the Non-member Baseline and Member Accuracy

A metric that gives attribute disclosure vulnerability as a ratio relative to the maximum possible attribute disclosure vulnerability beyond knowledge generation (see equation (22)) penalizes high non-member baselines that provide potentially very relevant and externally valid insights derived from the data. A relative attribute disclosure vulnerability metric $A_{rel}$ can also be defined as pure difference to the non-member baseline meaning that the accuracy using only data from individuals that were not part of the training dataset for SDG (i.e., non-members) is subtracted from the accuracy using only data from individuals that were part of the training set for SDG (i.e., members):

$$A_{rel} = A_{members} - A_{non-members} \qquad (26)$$

where $A_{members}$ is the accuracy of members and $A_{non-members}$ the one of non-members. This is identical with the second component of the metric in [63] (see equation (25)).

In attribute disclosure vulnerability, both $A_{members}$ and $A_{rel}$ must be considered. An $A_{rel}$ above a certain threshold can only be considered as an attribute disclosure vulnerability if the $A_{members}$ is meaningful in absolute terms. The threshold itself is very likely to vary depending on the choice of performance measurement so that no global recommendation can be given. The choice must match the assumptions taken in the SDG scenario (see 4.3.4).

# 5. Differential Privacy

## 5.1 Definition

The main idea of differential privacy is that the findings from the data hardly change whether or not a data subject is in the data. Dwork et al. describe it as a protection "from any additional harm that they might face due to their data being in the private database x that they would not have faced had their data not been part of x" [142]. In this sense, differential privacy is an *a priori* feature of a process and different from privacy concepts discussed so far that are features of the dataset.

### 5.1.1 Reporting Disclosure Vulnerability as the Relative Parameter $\varepsilon$

Let *D* be a dataset and *M(D)* be a mechanism or algorithms applied to *D*. Then *M* is considered $\varepsilon$-differentially private if:

$$\Pr\big[M(D) \in S\big] \le e^{\varepsilon} \Pr\big[M(D') \in S\big] \qquad (27)$$

where *D* and *D'* differ with one record, and *S* is the range set of *M*. The parameter $e^{\varepsilon}$ is a relative parameter and gives a quantification of the maximum information gain. It is referred to as privacy budget. Note that even though information gain is often equated with $\varepsilon$, it is important to emphasize that it is in fact $e^{\varepsilon}$ so that small changes in $\varepsilon$ result in a much larger information gain than one might have expected. Noise (e.g., Laplace or Gaussian noise for numerical outputs; Exponential Mechanism as a generic mechanism) is added to satisfy the requirement. A weaker notation of differential privacy is *($\varepsilon$,$\delta$)*-differential privacy that is used to account for the uncertainty in satisfying $\varepsilon$-differential privacy in practice. The parameter $\delta$ adds another privacy option to reduce privacy in favor of better accuracy (less noise introduction). Numerous further differential privacy notations have been developed that mainly propose relaxations of the initial definition such as Rhenyi-Differential-Privacy [143], concentrated or rho-zero-concentrated [144,145].

### 5.1.2 Differentially Private Synthetic Data

Differential privacy gives theoretical guarantees that should be independent of any threat modeling and comes with the recognition of disclosure vulnerability being cumulative rather than static [146]. The definition of differential privacy is, however, meant for an interactive (query) scenario. This is disadvantageous for sophisticated analyses where researchers might spend the whole privacy budget for data pre-processing. Also, data users who are used to work with individual-level data might not show a high level of acceptance towards the interactive query scenario in differential privacy [147]. Differentially private SDG, in contrast, presents an opportunity to overcome this challenge by providing differentially private individual-level data.

There are several ways to achieve differentially private synthetic data. When looking at the NIST 2018 Differential Privacy Synthetic Data Challenge [148], various winners built upon noise introduction into conditional distributions (e.g. PrivBayes). Noise can also be introduced into the training process of a generative model, e.g., into the discriminator gradient updates of GANs before applying them to the weights. In the case of GANs, the generator will generally be differentially private according to the Post-Processing Theorem as it has no direct access to the original data but uses differentially private output of the discriminator (as described below).

### 5.1.3 Application of Differential Privacy

Since its introduction, differential privacy has gained a lot of attention. The product *OnTheMap* by the US Census Bureau [149] was one of the first practical implementations of differential privacy, followed by other deployments at companies such as Google, Apple and Microsoft [142,147,150]. The NIST 2018 Differential Privacy Synthetic Data Challenge demonstrated the interest of public authorities in a mathematical guarantee for privacy also in SDG [148], early in 2024 a differentially private version of the Israeli National Registry of Live Births was released to the public [151] and several authors have already named differential privacy as an SDG quality standard [63,142,150,152,153].

Various differentially private SDG models have been introduced and evaluated. For example, Jordon et al. claim to give privacy guarantees in SDG when introducing PATE (Private Aggregation of Teacher Ensembles)-GAN [154]. Rosenblatt et al. construct a comprehensive benchmarking framework for differentially private SDG to pave the way for their adoption [153]. Stadler et al. evaluate PATE-GAN and PrivBayes for utility and privacy [63]. Tao et al. found differentially private SDG based on marginal distributions to outperform the one based on GAN [155]. While this shows the interest in differential privacy in research, their implementation in practice seems to be low or limited to highly-funded companies or governments as shown by reviews on SDG [19,66].

## 5.2 Example Metrics

Examples of differentially private SDG build upon two theorems: Firstly, the Post-Processing Theorem that ensures that any output of an operation that takes differentially private data as input can be considered as differentially private as well and secondly, the Composition Theorem that states that the composition of two *(ε,0)*-differentially private mechanisms add up to *(2ε,0)*-differential privacy [142].

As is clear from the name, PATE-GAN is realized by means of a GAN [154]. The main components of a GAN are the generator and the discriminator. The former generates synthetic data samples from random noise and updates its generation mechanism to provide more realistic data throughout the training process. The latter is a classifier to distinguish between synthetic and real data. Throughout training, these competitive components are optimized. The key idea in PATE-GAN is to replace the discriminator by an ensemble of teacher discriminators and a student discriminator. The teacher discriminators are trained on different partitions of the dataset. By adding Laplace noise to the aggregated teacher-labelled generated samples during training, a differentially private mechanism is created. This mechanism, however, introduce non-differentiability of the teacher discriminators. Differentiability in the output is needed to serve as a loss function for the generator. A student model is therefore interposed. It approximates the noisy output of the teacher ensemble. This approximation is a differentiable function and allows for backpropagation of gradients throughout the training process. This means that the generator is no longer improved with respect to the teacher discriminators but to the student. The output of PATE-GAN can be considered as differentially private given the Post-Processing Theorem: the student model is trained on differentially private output of the teacher ensemble, which makes its output and subsequent training of the generator differentially private as well. The iterations between generator and discriminators can be seen as querying the differentially private teacher ensemble and is repeated until the privacy budget $\varepsilon$ is reached.

Zhang et al. present an approach built upon noise introduction into conditional distributions resulting in the differentially private Bayesian network PrivBayes [156]. Bayesian networks are used to retrieve the underlying structure and relationships between the attributes in a dataset. In the training process, the conditional probability distribution of each attribute are estimated from the training data. Synthetic data can then be generated by drawing samples from the learned distributions (i.e., the Bayesian network model). PrivBayes is based on a low-degree Bayesian network representing joined distributions that satisfy $\varepsilon_1$-differential privacy. With this network in place, $\varepsilon_2$-differentially private conditional distributions are modelled. Samples (i.e. synthetic data) are drawn from these noisy distributions. The sampling itself does not require any non-differentially private data, so that the overall SDG model can be considered as $(\varepsilon_1 + \varepsilon_2)$-differentially private given the principle on the composition of differentially private mechanisms. The authors recommend an equal partition between these two budgets or a slight greater proportion to $\varepsilon_2$ as derived from their empirical utility evaluation.

## 5.3    Critical Appraisal

Differential privacy is different from the other disclosure vulnerability metrics already covered in that it does not explicitly target a specific privacy concept. Despite being agnostic towards a specific disclosure concept, application areas provide indirect insight into the types of disclosure conceptss it addresses. Applying differential privacy to the US census, for example, is primarily driven by the objective of preventing identity disclosure [68,147]. It may also aim at mitigating attribute disclosure and membership disclosure [27,103].

To discuss differential privacy and the interpretation of its privacy budget in SDG, we built upon experiences gained with differential privacy on tabular data outside SDG. The experiences with differentially private SDG are very likely comparable due to the fundamental nature of the mechanisms and challenges. Applications of differential privacy outside SDG can further be generally considered as less demanding as they apply differential privacy on certain statistical queries where the privacy of the outputs of the queries is guaranteed. SDG is different in the way that an entire dataset needs to be private so that challenges outside SDG are unlikely to be resolved within SDG. For both reasons, an analysis of the strengths and weaknesses of differential privacy in general is relevant to our appraisal in the context of SDG. While our report does not cover text or image generation by LLMs, it should be noted that the literature on differentially private LLMs is rapidly growing reflecting the interest in differentially private generative models [157,158]. These experiences, however, are not easily transferrable to tabular data and are outside the scope of this report.

### 5.3.1    Large Privacy Budgets Are Not in Line With The Theoretical Concept of Differential Privacy

The resulting privacy preservation in differential privacy depends largely on the proper choice of the privacy budget. Changes in the privacy budget translate exponentially into changes in vulnerability ($e^{\varepsilon}$). If $e^{\varepsilon}$ approaches 1, then the vulnerability can be considered as acceptably low as the findings from the data hardly change whether or not a data subject is in the data. Thus, a small privacy budget close to 0 can give the mathematical guarantee that privacy is preserved. If the privacy budget becomes large, however, theoretical privacy presumptions cannot be translated into empirical privacy, or as acknowledged by Dwork et al: "sound reasoning about the semantics of differential privacy [...] becomes unsound when $\varepsilon$ is large" [159]. In such a scenario, the interpretation of the privacy budget may differ across different statistical analyses [160] and is likely to depend on the implementation and introduced noise [161]. However, in practice, large values of epsilon have been observed (e.g., Google 2.64 for COVID-19 daily mobility reports, US Census Bureau used 17.14 for the 2020 Census persons file). Further examples of epsilon values in practice are given in Figure 19. In the case of the US Census, the initial implementation of an epsilon between 0.25 and 8 did not satisfy required utility. Epsilon was then adjusted to 12.2, and in the final production setting to 18.19 given a delta of $10^{-10}$ [162]. The US Census Bureau showed then empirically that a larger epsilon can still provide protection against a privacy attack [13]. The attack was labeled as re-identification attack by the authors while according to our terminology we would better describe it as an undifferentiated attribute disclosure metric. The ability of this metric to measure meaningful privacy has been criticized by further authors [163]. Our point, however, is not the nature of the privacy assessment but that any privacy assessment is very likely to result in different vulnerability for a different dataset with the same value of epsilon. This is due to the unclear (and non-uniform) interpretation of large epsilon values.

Figure 19 also shows that there is substantial variation in choosing epsilon in commercial settings. This is similar in academia: For example, Muralidhar et al. make use of an epsilon of 1.0 [161], Stadler et al. implement an epsilon of 0.1 [63], Li et al. state that 4 is an empirically reasonable value [106], Rosenblatt et al. consider an epsilon < 3.0 as low [153], and Hayes et al. an epsilon < 10 as acceptable [103]. To leverage the theoretical guarantee of differential privacy, a precise definition must be established to determine when an epsilon value is considered as "small" or "close to 0" and thereby can be reliably used to demonstrate privacy. It is, however, unclear whether an epsilon value other than 0 with a clear privacy interpretation exists.

A pragmatic solution to this dilemma may be an empirical privacy evaluation to gain insights into the disclosure vulnerability as in the example above from the US Census Bureau [13]. The seemingly arbitrary choice and wide variation of epsilon give further rise to concerns that the label of differential privacy may be misused to imply a theoretical guarantee of privacy in scenarios where presumptions fail (i.e., with large privacy budgets). Such a concern is, for example, expressed by the developers of differential privacy who question whether currently implemented methodologies are aligned with the original principles or in their words "as we have seen, when [epsilon] is large it can also allow for a form of privacy theatre"[164].

| | Application | Privacy budget epsilon ε | Delta δ | Privacy unit |
|---|---|---|---|---|
| **Google** | COVID-19 community mobility report [165] | 2.64 | | User-day |
| | COVID-19 symptoms search trends [166] | 1.68 | | User-day |
| | COVID-19 vaccination search trends [167] | 2.19 | $10^{-5}$ | User-day |
| | Mobility data [168] | 0.66 | $2.1 \times 10^{-29}$ | User-trip-week |
| | Streaming aggregation of shopping data [169] | 1 | $10^{-9}$ | User-day |
| | Streaming aggregation of trends data [169] | 2 | $10^{-10}$ | User-day |
| **Apple** | QuickType suggestions [170] | 16 | | User-day |
| | Emoji suggestions [170] | 4 | | User-day |
| | Lookup hints [170] | 8 | | User-day |
| | Health types (HealthKit App) [170] | 2 | | User-day |
| | Energy draining domains (Safari) [170] | 8 | | User-day |
| | Crashing domains (Safari) [170] | 8 | | User-day |
| | Auto-play intent detection (Safari) [170] | 16 | | User-day |
| **Facebook** | Movement range maps [171] | 2 | | User-day |
| **Microsoft** | Labor market employers (LinkedIn) [172] | 4.8 | $4 \times 10^{-10}$ | Top-employers-month |
| | Labor market jobs (LinkedIn) [172] | 4.8 | $4 \times 10^{-10}$ | Top-jobs-month |
| | Labor market skills (LinkedIn) [172] | 0.1 | $10^{-10}$ | Top-skills-month |

**Figure 19: Epsilon values used in practice.** Epsilon values and in case of *(ε,δ)*-differential privacy delta values are given for exemplary applications. Privacy unit refers to the unit for which the guaranteed differential privacy is valid. To allow for comparison, privacy units were standardized to privacy budget per user per day (user-day) whenever reasonably achievable. For example, when epsilon was given per contribution and a user could have multiple contributions per days, these values were added up. Note that this is an estimation of the privacy budget per user per day. The actual values may differ as there can be overlapping applications resulting in a higher actual epsilon. A user could also make less contributions per day than the maximum allowed number which would then result in a lower actual epsilon.

### 5.3.2   An Acceptable Privacy-Utility Trade-Off is Sometimes Difficult to Achieve

The use of large privacy budgets is often driven by the observation that substantial utility loss occurs at lower budgets. While this report is focused on privacy, multiple studies document difficulties in achieving an acceptable privacy-utility tradeoff at low epsilon values. This background is helpful in interpreting the performance of differentially private SDG models as well. An exemplar of this challenge was reported by the US Census Bureau whereby the initial implementation of a small privacy budget for the 2020 Census resulted in substantial concerns about its data utility, and was followed by the decision to change epsilon to a larger value [146].

In general, the noise introduced can compromise the utility of the synthetic data. In [173], an $(\varepsilon,\delta)$-differentially private GAN ($\varepsilon$ = 4) is compared against a non-differentially private GAN. Application of differential privacy resulted in inconsistent results in terms of broad utility with higher variability in some but not in all estimates. The public release of a differentially private synthetic version of the Israel's National Live Birth Registry provided sufficient utility for pre-selected queries [151]. The authors implemented further post-processing steps to ensure logical consistency. However, they also note that it is impossible to satisfy both – privacy and utility of all potential statistical queries. Utility in their release is therefore not guaranteed beyond the pre-selected queries. A reasonably good performance in downstream analytical tasks seems to be more challenging. Applying $(\varepsilon,\delta)$-differential privacy ($\varepsilon$ = 10) to a GAN, the authors in [174] show a pronounced decrease in utility compared to a non-differentially private GAN. Similar results were obtained for a differentially private GAN and differentially private Bayesian network in [63]. The authors conclude that a reasonable privacy-utility trade-off cannot be achieved for many use cases even if a high value of epsilon is used. A comprehensive benchmarking of diverse GAN similarly assigned lowest rankings to a differentially-private GAN [91]. Inconsistent results regarding prediction task performance were reported in a study that evaluated a differentially private GAN with an epsilon of 1 or 2 [175]. In [176], the authors observe a similar utility in a differentially-private autoencoder when compared to its non-differentially private counterpart. And in [154], substantial lower accuracy was observed using epsilon values around 0.1 but the same accuracy as with a non-differentially private GAN could be achieved at an epsilon of 10.

Applications outside SDG also face similar parametrization challenges. For example, Fredrikson et al. tried to provide personalized recommendation on Warfarin dosing using differential privacy as a PET [118], and found that they could only prevent attacks with unacceptable utility constraints that would result in a higher exposure of patients to side effects. Similar difficulties with the privacy-utility tradeoff were reported elsewhere [63,94,103,107,108,152,161,177,178].

Note that the results above are in contrast to those in Figure 19, where it is assumed that an acceptable utility-privacy tradeoff *was* achieved with the given epsilon values and privacy units selected. It should also be noted that similar privacy budget values are used in both cases, re-enforcing a key question about how to set that parameter.

### 5.3.3   Setting an Inappropriate Level of Granularity for Differential Privacy Can Give a Wrong Sense of Privacy

Houssiau et al. illustrate the role of assumptions in the choice of the privacy budget in a case where a low vulnerability was stated for differentially private Google Maps data but built upon a wrong assumption [179]. In this example, the low disclosure vulnerability resulted from the assumption that a Google Maps user only contributed with one trip to the aggregated data which according to the authors was unlikely to be true in the real-world. This example fits the caveat raised by Dwork et al. that the resulting privacy is highly dependent on the granularity at which differential privacy is applied [159]. For example, applying differential privacy at a single physician appointment (i.e., event-level) would give a different disclosure vulnerability than applying it at the entire medical history of that patient (i.e., user-level).

These examples illustrate that differential privacy still requires assumptions, such as the granularity (or unit) that requires protection, and wrong assumptions can erode trust in proper data sharing.

### 5.3.4 Implementation Mistakes can Undermine Theoretical Privacy Guarantees in Small Epsilon Values

In differentially private SDG with small epsilon values ($\varepsilon = 0.1$), Stadler et al. found discrepancies between the theoretical privacy guarantee and empirical privacy violations, and showed that these violations were unpredictable [63]. Other authors found discrepancies between empirical disclosure vulnerability and theoretical guarantees in current differentially private SDG implementations [105,180,181]. In [180], the authors audited six popular implementations and identified implementation mistakes such as the leakage of metadata directly from the original data. This shows that the implementation of differentially privacy SDG itself is not trivial and even small mistakes can lead to substantial privacy violations despite a seemingly small privacy budget.

While implementation problems can compromise the performance of any SDG model, the disclosure vulnerability of the generated synthetic datasets from other SDG models can be independently evaluated using the metrics described earlier in this report. For differential privacy, it is assumed that the generative model itself provides the privacy guarantees, which makes the correct implementation particularly crucial.

---

**Summary of Key Points from the Critical Appraisal of Differential Privacy**

- The assumptions of differential privacy fail when the epsilon value is large, which means that disclosure vulnerabilities still need to be assessed.

- There is wide variation in the choice of epsilon values and what constitutes an acceptably low value within academia, industry and government. Considering the exponential nature of the relationship, this means that there is no meaningful consensus on when a value becomes large.

- The implementation of differential privacy can be challenging in SDG. Even with small epsilon values, implementation mistakes have been shown to deteriorate privacy.

---

## 5.4 Recommendations

While differential privacy is a rigorous and appealing concept, it has been shown to be challenging to implement correctly in practice. The major challenge arises from the unclear translation of epsilon values into generally understood privacy concepts. Given the current lack of interpretability of large epsilon values, other privacy metrics are still required and the privacy budget itself may not be sufficient. The implication is that such differentially private SDG is not necessarily more private than a non-differentially private counterpart by definition. Using small epsilon values that are more likely to translate into stronger privacy guarantees, however, can affect utility. From that perspective, an acceptable privacy-utility trade-off may be achieved in differentially private SDG when epsilon is chosen to preserve utility and empirical privacy evaluation of the synthetic data demonstrates a sufficiently low disclosure vulnerability. Complementary approaches as proposed in [182] may be a good way to achieve such a trade-off. Then, as done in the example of the US Census, privacy needs to be evaluated empirically [13]. The same privacy metrics as those used in non-differentially privacy synthetic data should be used to allow for comparisons.

Over time norms on suitable privacy budgets (or ranges) and units may emerge within specific domains and for specific types of datasets, and may be implemented through structures as an Epsilon Registry [159]. These norms would be based on validation against empirical disclosure vulnerability evaluations, and would give more meaning to the privacy guarantees of differential privacy.

# A Consensus Privacy Metrics Framework for Synthetic Data

# Appendix B: Datasets

# 1.    BORN

The BORN collects Ontario's prescribed perinatal, newborn and child registry with the role of facilitating quality care for families across the province. It can be accessed through a data request at https://bornontario.ca/en/data/data.aspx.

# 2.    California

The California dataset contains the patient's hospital 2008 discharge data from California, State Inpatient Databases (SID), Healthcare Cost and Utilization Project (HCUP), Agency for Healthcare Research and Quality [2], and is available for purchase at https://hcup-us.ahrq.gov/tech_assist/centdist.jsp.

# 3.    CCHS

The Canadian Community Health Survey (CCHS) is Canadian population-level information concerning health status, health system utilization and health determinants collected by Statistics Canada through telephone survey. The availability of CCHS data is restricted and requires an access request at https://www150.statcan.gc.ca/n1/pub/82-620-m/2005001/4144189-eng.htm.

# 4.    FAERS

The FDA Adverse Event Reporting System (FAERS) is a database comprising the information on adverse events and medication error reports submitted to FDA and can be downloaded at https://open.fda.gov/data/faers/.

# 5.    Florida

The Florida dataset contains the patient's hospital 2007 discharge data from Florida, State Inpatient Databases (SID), Healthcare Cost and Utilization Project (HCUP), Agency for Healthcare Research and Quality [2], and is available for purchase at https://hcup-us.ahrq.gov/tech_assist/centdist.jsp.

# 6.    New York

The New York dataset contains the patient's hospital 2007 discharge data from New York, State Inpatient Databases (SID), Healthcare Cost and Utilization Project (HCUP), Agency for Healthcare Research and Quality [2], and is available for purchase at https://hcup-us.ahrq.gov/tech_assist/centdist.jsp.

# 7.    Nexoid

The COVID-19 survival dataset Nexoid is a web-based survey data collected by a company called Nexoid in United Kingdom. It is publicly available at https://www.covid19survivalcalculator.com/en/download.

# 8.    Texas

The Texas dataset contains the patient's hospital discharge information for the first quarter of 2012 from Texas in the United States [3], and is publicly available at https://www.dshs.texas.gov/center-health-statistics/chs-data-sets-reports/texas-health-care-information-collection/health-data-researcher-information/texas-inpatient-public-use.

## 9.     Washington

The Washington dataset contains the patient's hospital 2007 discharge data from Washington, State Inpatient Databases (SID), Healthcare Cost and Utilization Project (HCUP), Agency for Healthcare Research and Quality [2], and is available for purchase at https://hcup-us.ahrq.gov/tech_assist/centdist.jsp.

## 10.     Washington 2008

The Washington2008 dataset contains the patient's hospital 2008 discharge data from Washington, State Inpatient Databases (SID), Healthcare Cost and Utilization Project (HCUP), Agency for Healthcare Research and Quality [2], and is available for purchase at https://hcup-us.ahrq.gov/tech_assist/centdist.jsp.

**A Consensus Privacy Metrics Framework for Synthetic Data**

**Appendix C: Methods and Results from Delphi Rounds**

# 1.     Statements and Explanations throughout the Delphi process

The initial question for the consensus process was formulated as follows "How should we evaluate privacy in synthetic data". From this question, the report was initialized which then served to identify relevant statements around privacy metrics in synthetic data. In the first round, there were 15 statements to score. In the second round, statements were revised to address any ambiguity, two new statements were introduced and four statements omitted based on the panelists' feedback. In the third round, two more statements were omitted so that a total of 11 statements were considered in the study.

In the following, the numbering of the statements is aligned with the chronological order of the rounds meaning that it is discontinuous from the second round on due to the omission of certain statements and the introduction of new ones.

## 1.1     Statements of Round 1

### 1.1.1     Overarching Considerations

**S1: Disclosure vulnerability metrics should be based on quasi-identifiers rather than on complete records.**

Explanation: Decades of research and practice on identity disclosure in anonymized data, known re-identification attacks, and respective guidelines are based on the assumption that adversary prior knowledge is represented by the quasi-identifiers. Assuming that an adversary knows all of the variables is not necessarily a worse case assumption for synthetic data.

**S2: When evaluating and SDG model, disclosure vulnerability metrics need to be reported for multiple (e.g. averaged) synthetic datasets.**

Explanation: SDG is a generative process with stochastic variability in its output. When evaluating an SDG model rather than a single synthetic dataset, an aggregate across multiple synthetic datasets from the same model would be appropriate.

**S3: Disclosure vulnerability metrics should not be calculated on a pre-selected subset of "vulnerable" records but for all of the records.**

Explanation: Pre-selection of records mainly involves choosing "vulnerable" records but it remains unclear if these records indeed are the ones with highest disclosure vulnerability or rather vulnerable in a more ethical sense.

### 1.1.2     Record-Level Similarity

**S4: Similarity metrics that are not part of attribute or membership disclosure should not be used to report privacy in synthetic data.**

Explanation: Similarity metrics that are not part of calculating membership or attribute disclosure metrics have no precise interpretation. In contrast to anonymized data, matching a synthetic to training record by itself does not translate into an identity disclosure vulnerability as it does not imply learning correct information. Similarity by itself may or may not translate into disclosure vulnerability. This ambiguity when interpreting such metrics makes clear that they are not proper disclosure vulnerability metrics.

### 1.1.3     Membership Disclosure

**S5: Membership disclosure vulnerability should only be evaluated when the adversary would learn something new for targets drawn from the same population as the training dataset.**

Explanation: Membership disclosure metrics assume that the adversary draws target records from the same population as the training dataset is drawn from. The information they learn from such an attack is membership and in case of interventional studies the information that a certain procedure was applied. A vaccination study is an example where the calculation of membership disclosure vulnerability by current metrics is meaningful.

**S6: Calculating membership disclosure vulnerability is only meaningful when the naïve (i.e., inherent) membership disclosure vulnerability is low.**

Explanation: If the F_naive value is already high then the calculation of an absolute or relative membership disclosure vulnerability of the synthetic data does not provide meaningful guidance in decisions on SDG. The computation does not alter the conclusion that the situation exhibits high membership disclosure and the focus should only be on attribute disclosure.

**S7: In membership disclosure vulnerability, an absolute vulnerability (i.e., F1 value) lower than or equal to 0.5 can be considered as acceptably low.**

Explanation: Membership disclosure vulnerability is a binary classification problem. Accordingly, an accuracy of 0.5 would be a random guess between members and non-members. This threshold applies to both, the naïve (inherent) vulnerability and the absolute vulnerability of the synthetic data

**S8: In membership disclosure vulnerability, a relative vulnerability (i.e., F_rel value) lower than or equal to 0.2 can be considered as acceptably low.**

Explanation: A common way to define thresholds is to rely on precedents. The threshold that have been used in the literature is 0.2 for relative metrics when compared to a naïve baseline (F_rel).

**S9: In membership disclosure vulnerability, a relative vulnerability higher than its threshold should only be considered as unacceptably high when the absolute vulnerability is higher than its threshold.**

Explanation: The relative metric cannot be interpreted uniformly across the entire scale and is probably not always informative by itself.

### 1.1.4 Attribute Disclosure

**S10: Meaningful attribute disclosure vulnerability only applies to individuals that are in the dataset (i.e., members). Penalizing accurate prediction on individuals that have not been part of the dataset (i.e., group privacy) requires a broader ethical framework.**

Explanation: In technical privacy research an increasingly accepted principle is to make sure that being a member in a dataset does not increase the likelihood of an adversary gaining sensitive information about an individual so that everything that can be learned about the individual can also be learned without them being a member of the dataset. From this perspective, it follows that inferences about non-members cannot be attribute disclosure in the sense of a privacy compromise. This would be more consistent with the goals of research where inference about non-members results from publishing aggregate statistics and prognostic results.

**S11: A relative attribute disclosure vulnerability that takes a non-member baseline into account is meaningful.**

Explanation: A baseline derived from the prediction accuracy experienced by non-members distinguishes attribute disclosure vulnerability from knowledge generation. In technical privacy research an increasingly accepted principle is to make sure that being a member in a dataset does not materially increase the likelihood of an adversary gaining sensitive information about an individual so that everything that can be learned about the individual can also be learned without them being a member of the dataset. From this perspective, it follows that inferences about non-members cannot be attribute disclosure in the sense of a privacy compromise.

**S12: In attribute disclosure vulnerability, a distance to the non-member baseline lower than or equal to 0.15 can be considered as acceptably low.**

Explanation: For attribute disclosure vulnerability, thresholds could be based on common interpretations of prediction accuracy. Across different labeling systems, labels such as poor, moderate or good prediction accuracy typically change with differences of 0.1 to 0.2 in AUC. A threshold of 0.15 therefore appears reasonable. The threshold would apply to the difference in AUC between member and non-member accuracy, not to a proportional metric (such as a Kappa-like metrics that are relative to the maximum possible attribute disclosure vulnerability beyond knowledge generation).

**S13: In attribute disclosure vulnerability, an absolute vulnerability (i.e., accuracy of a member) lower than or equal to 0.6 can be considered as acceptably low.**

Explanation: Across different labeling systems, poor accuracy is mainly assigned to an AUC lower than or equal to 0.6. Following this interpretation, a threshold of 0.6 could be established for absolute vulnerability, meaning that only accuracies that exceed poor prediction (i.e., 0.6) can be meaningful disclosure.

**S14: In attribute disclosure vulnerability, a relative vulnerability higher than its threshold is only considered as unacceptably high when the absolute vulnerability is higher than its threshold.**

Explanation: The relative metric cannot be interpreted uniformly across the entire scale and is probably not always informative by itself. Therefore the decision rule for acceptable attribute disclosure needs to also consider the absolute vulnerability value.

### 1.1.5 Differential Privacy

**S15: Non-differentially private SDG models Should be Preferred Over Current Applications of Differential Privacy in SDG.**

Explanation: The value of the privacy budget epsilon has no uniform translation into privacy and can only be interpreted contextually. The consequences of the privacy budget are not the same across different statistical analyses and its value cannot be easily translated into empirical privacy. Differentially private synthetic data would therefore still require a privacy evaluation. Current implementations cannot provide a reasonable privacy-utility trade-off.

## 1.2 Statements of Round 2

### 1.2.1 Overarching Considerations

**S1: Disclosure vulnerability metrics should be based on quasi-identifiers rather than on complete records.**

Explanation: Quasi-identifies represent the background knowledge of an adversary. The primary reasons for this recommendation are:

(a)   Decades of research and practice on identity disclosure in anonymized data, known re-identification attacks, and respective guidelines are based on the assumption that adversary prior knowledge is represented by the quasi-identifiers.

(b)   Assuming that an adversary knows all of the variables is not necessarily a worse case assumption for synthetic data (please see report section 1.5.1 in the report for an explanation of this point). It can be shown that by considering all of the variables we may be underestimating disclosure vulnerability.

(c)     There are generally accepted criteria for deciding what a quasi-identifier is. Quasi-identifiers can vary and are ultimately ascertained by the data controller for a given dataset.

**S2: When evaluating a specific trained SDG model, disclosure vulnerability metrics need to be reported both for individual and multiple synthetic datasets (e.g. averaged across them and variation).**

Explanation: SDG is a generative process with stochastic variability in its output. When evaluating an SDG model rather than a single synthetic dataset, an aggregate (average and standard deviation) across multiple synthetic datasets from the same model would be appropriate. This would then reflect the vulnerability of an SDG model tied to a training dataset. However, if the decision-making scenario is data release, then the disclosure vulnerability for the specific synthetic dataset(s) may be the most relevant. Given that it is not always possible to determine a priori the exact decision-making scenario, it would be prudent to have both types of results.

**S3: Disclosure vulnerability metrics should not be calculated on a pre-selected subset of "vulnerable" records but for all of the records.**

Explanation: Pre-selection of records mainly involves choosing "vulnerable" records but it remains unclear if these records indeed are the ones with highest disclosure vulnerability or rather vulnerable in a more ethical sense. Knowledge on which records are the ones with highest disclosure vulnerability can only be obtained when calculating the vulnerability for all records in the first place.

## 1.2.2    Record-Level Similarity

**S4: Stand-alone similarity metrics (i.e., that are not part of attribute or membership disclosure) should not be used to report privacy in synthetic data.**

Explanation: Stand-alone similarity metrics (in fully synthetic data) that are not part of calculating membership or attribute disclosure metrics have no precise interpretation. Our analysis in Section 2.2.6 shows that these metrics do not convey additional information beyond membership and attribute disclosure metrics.

## 1.2.3    Membership Disclosure

**S5: Membership disclosure vulnerability should only be evaluated when the assumptions of the current metrics hold which is that the adversary would learn something new for targets drawn from the same population as the training dataset.**

Explanation: Current membership disclosure metrics assume that the adversary draws target records from the same population as the training dataset is drawn from. Most narratives, however, make the assumption that targets are drawn from a different population. While the latter may be a reasonable scenario, it is not reflected in current metrics. Therefore, current metrics should only be applied when the assumption of the metric is in line with the adversarial assumptions, otherwise using the current metrics in the wrong context may give wrong vulnerability estimates.

**S16: Because the most commonly used membership disclosure metric is an F1 score, and this is prevalence dependent, it needs to be reported relative to an adversary guessing membership.**

Explanation: It is known that metrics like the F1 score vary with prevalence. The absolute value can therefore not be uniformly interpreted. For example, with a low naïve guess (e.g., 0.01), an F1 score of 0.5 would still be considered as high, reflecting reasonable success given the rarity of members. This naïve baseline can be accounted for by a relative metric (F_rel) or by adjusting threshold according to the baseline.

**S8: As a default threshold value for membership disclosure vulnerability, a relative vulnerability (i.e., F_rel value) less than or equal to 0.2 can be considered as acceptably low.**

Explanation: The relative metric (F_rel) is a good way to account for the naïve baseline in a non uniformly interpretable F1. This threshold has been used in the literature for the relative membership disclosure metric (F_rel). In practice, thresholds that are recommended based on precedents may be adjusted up or down based on the context. Such adjustments need to be considered carefully and be justified. This context is characterized by the sensitivity of the data, potential harm and appropriateness of consent and notice.

### 1.2.4   Attribute Disclosure

**S10: Meaningful attribute disclosure vulnerability only applies to individuals that are in the dataset (i.e., members). Penalizing accurate prediction on individuals that have not been part of the dataset (i.e., group privacy) requires a broader ethical framework.**

Explanation: In technical privacy research an increasingly accepted principle is to make sure that being a member in a dataset does not increase the likelihood of an adversary gaining sensitive information about an individual so that everything that can be learned about the individual can also be learned without them being a member of the dataset. From this perspective, it follows that inferences about non-members cannot be attribute disclosure in the sense of a privacy compromise. This would be more consistent with the goals of research where inference about non-members results from publishing aggregate statistics and prognostic results.

**S11: A relative attribute disclosure vulnerability that takes a non-member baseline into account is meaningful.**

Explanation: A baseline derived from the prediction accuracy experienced by non-members distinguishes attribute disclosure vulnerability from knowledge generation. In technical privacy research an increasingly accepted principle is to make sure that being a member in a dataset does not materially increase the likelihood of an adversary gaining sensitive information about an individual so that everything that can be learned about the individual can also be learned without them being a member of the dataset. From this perspective, it follows that inferences about non-members cannot be attribute disclosure in the sense of a privacy compromise.

**S12: As a default threshold value for attribute disclosure vulnerability, a difference from the non-member baseline less than or equal to 0.15 can be considered as acceptably low.**

Explanation: For attribute disclosure vulnerability, thresholds could be based on common interpretations of prediction accuracy. Across different labeling systems, labels such as poor, moderate or good prediction accuracy typically change with differences of 0.1 to 0.2 in AUC. A threshold of 0.15 therefore appears reasonable. The threshold would apply to the difference in AUC between member and non-member accuracy. In practice, thresholds that are recommended based on precedents may be adjusted up or down based on the context. Such adjustments need to be considered carefully and be justified. This context is characterized by the sensitivity of the data, potential harm and appropriateness of consent and notice.

**S13: As a default threshold value for attribute disclosure vulnerability, an absolute vulnerability (i.e., accuracy of predicting a member sensitive attribute) less than or equal to 0.6 can be considered as acceptably low.**

Explanation: Across different labeling systems, poor accuracy is mainly assigned to an AUC lower than or equal to 0.6. Following this interpretation, a threshold of 0.6 could be established for absolute vulnerability, meaning that only accuracies that exceed poor prediction (i.e., 0.6) can be meaningful disclosure. In practice, thresholds that are recommended based on precedents may be adjusted up or down based on the context. Such adjustments need to be considered carefully and be justified. This context is characterized by the sensitivity of the data, potential harm and appropriateness of consent and notice.

**S14: In attribute disclosure vulnerability, a relative vulnerability higher than its threshold is only considered as unacceptably high when the absolute vulnerability is higher than its threshold.**

Explanation: The relative metric cannot be interpreted uniformly across the entire scale and is probably not always informative by itself. Therefore the decision rule for acceptable attribute disclosure needs to also consider the absolute vulnerability value.

### 1.2.5 Differential Privacy

**S17: The privacy budget epsilon is not an adequate metric to report disclosure vulnerability. Disclosure vulnerability would still need to be evaluated using the same metrics as those applied to non-differentially private synthetic data.**

Explanation: The value of the privacy budget epsilon has no uniform translation into empirical privacy and can only be interpreted contextually. This is in particular true when the privacy budget is relatively large. Differentially private synthetic data would therefore still require a full privacy evaluation.

## 2.  Statements and Explanations of Round 3

### 2.1  Overarching Considerations

**S1: Disclosure vulnerability metrics should be based on quasi-identifiers. These may vary depending on the data context (e.g., can still be all attributes) and are ascertained by the data controller.**

Explanation: Quasi-identifies represent the background knowledge of an adversary. The primary reasons for this recommendation are:

(a)     Decades of research and practice on identity disclosure in anonymized data, known re-identification attacks, and respective guidelines are based on the assumption that adversary prior knowledge is represented by the quasi-identifiers.

(b)     Assuming that an adversary knows all of the variables is not necessarily a worse case assumption for synthetic data (please see section 1.5.1 in the report for a detailed explanation and justification for this point). It can be shown that by considering all the variables we may be underestimating disclosure vulnerability.

(c)     There are generally accepted criteria for deciding what a quasi-identifier is. Quasi-identifiers can vary and are ultimately ascertained by the data controller for a given dataset. Treating all attributes as quasi-identifiers is itself an assumption.

**S2: When evaluating a specific trained SDG model, disclosure vulnerability metrics need to be reported both for individual and multiple synthetic datasets (e.g. averaged across them and variation).**

Explanation: SDG is a generative process with stochastic variability in its output. When evaluating an SDG model rather than a single synthetic dataset, an aggregate (average and standard deviation) across multiple synthetic datasets from the same model would be appropriate. This would then reflect the vulnerability of an SDG model tied to a training dataset. However, if the decision-making scenario is data release, then the disclosure vulnerability for the specific synthetic dataset(s) may be the most relevant. Given that it is not always possible to determine a priori the exact decision-making scenario, it would be prudent to have both types of results.

**S3: Disclosure vulnerability metrics should not be calculated on a pre-selected subset of "vulnerable" records but for all of the records.**

Explanation: Pre-selection of records mainly involves choosing "vulnerable" records but it remains unclear if these records indeed are the ones with highest disclosure vulnerability or rather vulnerable in a more ethical sense. Knowledge on which records are the ones with highest disclosure vulnerability can only be obtained when calculating the vulnerability for all records in the first place.

## 2.2 Record-Level Similarity

**S4: Stand-alone similarity metrics (i.e., that are not part of attribute or membership disclosure) should not be used to report privacy in synthetic data.**

Explanation: Stand-alone similarity metrics (in fully synthetic data) that are not part of calculating membership or attribute disclosure metrics have no precise interpretation. Our analysis in Section 2.2.6 shows that these metrics do not convey additional information beyond membership and attribute disclosure metrics.

## 2.3 Membership Disclosure

**S5: Membership disclosure vulnerability should only be evaluated when the assumptions of the current metrics hold which is that the adversary would learn something new for targets drawn from the same population as the training dataset.**

Explanation: Current membership disclosure metrics assume that the adversary draws target records from the same population as the training dataset is drawn from. Most narratives, however, make the assumption that targets are drawn from a different population. While the latter may be a reasonable scenario, it is not reflected in current metrics. Therefore, current metrics should only be applied when the assumption of the metric is in line with the adversarial assumptions, otherwise using the current metrics in the wrong context may give wrong vulnerability estimates.

**S16: Because the F1 score, which is commonly used in membership disclosure metrics, is prevalence dependent, it needs to be reported relative to an adversary guessing membership.**

Explanation: It is known that metrics like the F1 score are affected by prevalence. The absolute value can therefore not be uniformly interpreted. For example, with a low naïve guess (e.g., 0.01), an F1 score of 0.5 would still be considered as high, reflecting reasonable success given the rarity of members. This naïve baseline can be accounted for by a relative metric (F_rel) or by adjusting threshold according to the baseline.

**S8: As an anchor for membership disclosure vulnerability, a relative F1 score vulnerability (i.e., F_rel value) of 0.2 is suggested.**

Explanation: The relative metric (F_rel) is a good way to account for a naïve baseline to interpret F1. This threshold has been used in the literature for the relative membership disclosure metric (F_rel). In practice, thresholds that are recommended based on precedents may be adjusted up or down based on the context. Such adjustments need to be considered carefully and be justified. This context is characterized by the sensitivity of the data, potential harm and appropriateness of consent and notice.

## 2.4 Attribute Disclosure

**S10: Meaningful attribute disclosure vulnerability only applies to individuals that are in the dataset (i.e., members). Penalizing accurate prediction on individuals that have not been part of the dataset (i.e., group privacy) requires a broader ethical framework.**

Explanation: In technical privacy research an increasingly accepted principle is to make sure that being a member in a dataset does not increase the likelihood of an adversary gaining sensitive information about an individual so that everything that can be learned about the individual can also be learned without them being a member of the dataset. From this perspective, it follows that inferences about non-members cannot be attribute disclosure in the sense of a privacy compromise. This would be more consistent with the goals of research where inference about non-members results from publishing aggregate statistics and prognostic results.

**S11: A relative attribute disclosure vulnerability that takes a non-member baseline into account is meaningful.**

Explanation: A baseline derived from the prediction accuracy experienced by non-members distinguishes attribute disclosure vulnerability from knowledge generation. In technical privacy research an increasingly accepted principle is to make sure that being a member in a dataset does not materially increase the likelihood of an adversary gaining sensitive information about an individual so that everything that can be learned about the individual can also be learned without them being a member of the dataset. From this perspective, it follows that inferences about non-members cannot be attribute disclosure in the sense of a privacy compromise.

**S14: In attribute disclosure vulnerability, a relative vulnerability higher than its threshold is only considered as unacceptably high when the absolute vulnerability is higher than its threshold.**

Explanation: The relative metric cannot be interpreted uniformly across the entire scale and is probably not always informative by itself. Therefore the decision rule for acceptable attribute disclosure needs to also consider the absolute vulnerability value.

## 2.5    Differential Privacy

**S17: The privacy budget epsilon is not an adequate metric to report disclosure vulnerability unless it is set to a value close to 0. Even when differential privacy methods are used, disclosure vulnerability would still need to be evaluated using the same metrics as those applied to non-differentially private synthetic data.**

Explanation: The value of the privacy budget epsilon has no uniform translation into empirical privacy and can only be interpreted contextually. This is in particular true when the privacy budget is relatively large. Differentially private synthetic data would therefore still require a full privacy evaluation.

# 3. Quantitative Results

## 3.1 Summary of Round 1

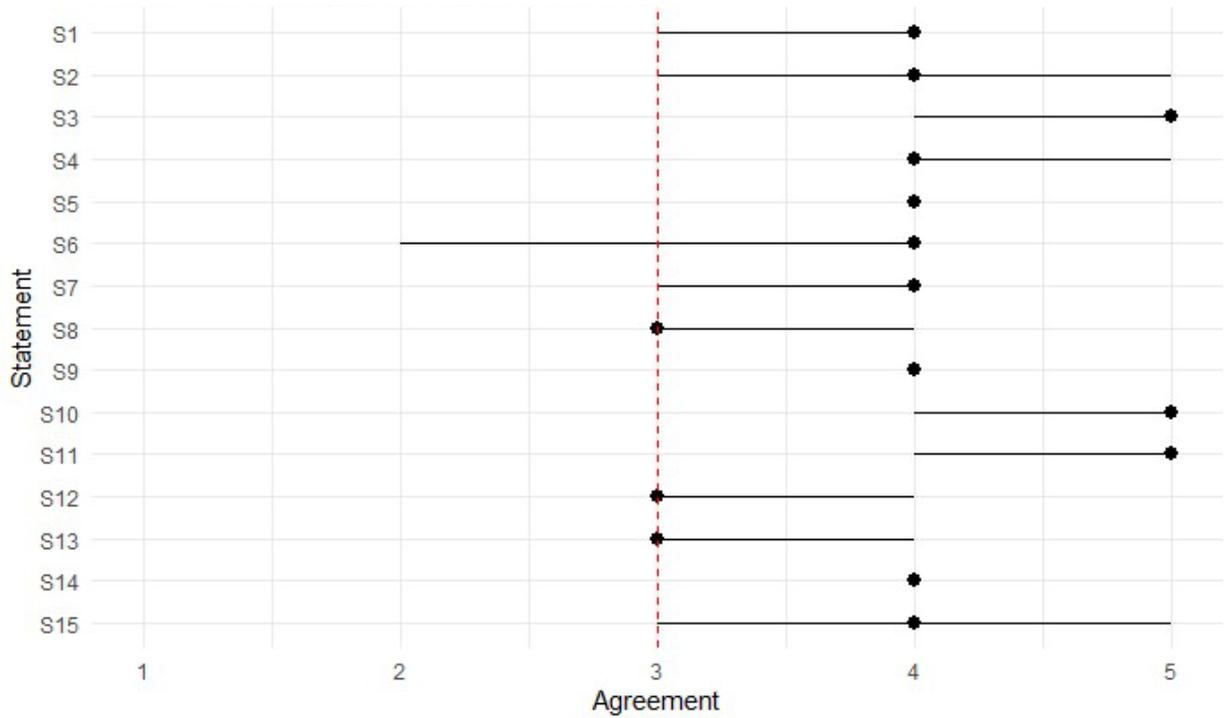

**Figure S1: Quantitative Round Summary of Round 1**. Agreement to each of the 15 statements of the first round is indicated in 5 levels ranging from strongly disagree (1) to strongly agree (5). Median and interquartile range (IQR) are indicated.

| Number | Agreement | | | | |
|--------|-----------------------|---------------|-------------|-------------|---------------------|
|        | Strongly Disagree (1) | Disagree (2)  | Neutral (3) | Agree (4)   | Strongly Agree (5)  |
| S1     | 0/13 (0.0%)           | 3/13 (23.1%)  | 3/13 (23.1%) | 4/13 (30.8%) | 3/13 (23.1%)       |
| S2     | 0/13 (0.0%)           | 2/13 (15.4%)  | 2/13 (15.4%) | 4/13 (30.8%) | 5/13 (38.5%)       |
| S3     | 0/13 (0.0%)           | 0/13 (0.0%)   | 3/13 (23.1%) | 3/13 (23.1%) | 7/13 (53.8%)       |
| S4     | 0/13 (0.0%)           | 0/13 (0.0%)   | 3/13 (23.1%) | 6/13 (46.1%) | 4/13 (30.8%)       |
| S5     | 1/13 (7.7%)           | 2/13 (15.4%)  | 0/13 (0.0%)  | 8/13 (61.5%) | 2/13 (15.4%)       |
| S6     | 0/13 (0.0%)           | 4/13 (30.8%)  | 1/13 (7.7%)  | 6/13 (46.1%) | 2/13 (15.4%)       |
| S7     | 2/13 (15.4%)          | 1/13 (7.7%)   | 3/13 (23.1%) | 5/13 (38.5%) | 2/13 (15.4%)       |
| S8     | 2/13 (15.4%)          | 0/13 (0.0%)   | 6/13 (46.1%) | 4/13 (30.8%) | 1/13 (7.7%)        |
| S9     | 1/13 (7.7%)           | 0/13 (0.0%)   | 2/13 (15.4%) | 8/13 (61.5%) | 2/13 (15.4%)       |
| S10    | 0/13 (0.0%)           | 2/13 (15.4%)  | 0/13 (0.0%)  | 4/13 (30.8%) | 7/13 (53.8%)       |
| S11    | 0/13 (0.0%)           | 0/13 (0.0%)   | 0/13 (0.0%)  | 6/13 (46.1%) | 7/13 (53.8%)       |
| S12    | 1/13 (7.7%)           | 2/13 (15.4%)  | 6/13 (46.1%) | 2/13 (15.4%) | 2/13 (15.4%)       |
| S13    | 1/13 (7.7%)           | 2/13 (15.4%)  | 4/13 (30.8%) | 4/13 (30.8%) | 2/13 (15.4%)       |
| S14    | 1/13 (7.7%)           | 0/13 (0.0%)   | 2/13 (15.4%) | 9/13 (69.2%) | 1/13 (7.7%)        |
| S15    | 0/13 (0.0%)           | 3/13 (23.1%)  | 3/13 (23.1%) | 2/13 (15.4%) | 5/13 (38.5%)       |

**Table S1: Distribution of Responses**. The 15 statements of the first round were rated by the panelists. The agreement is indicated in 5 levels.

## 3.2   Summary of Round 2

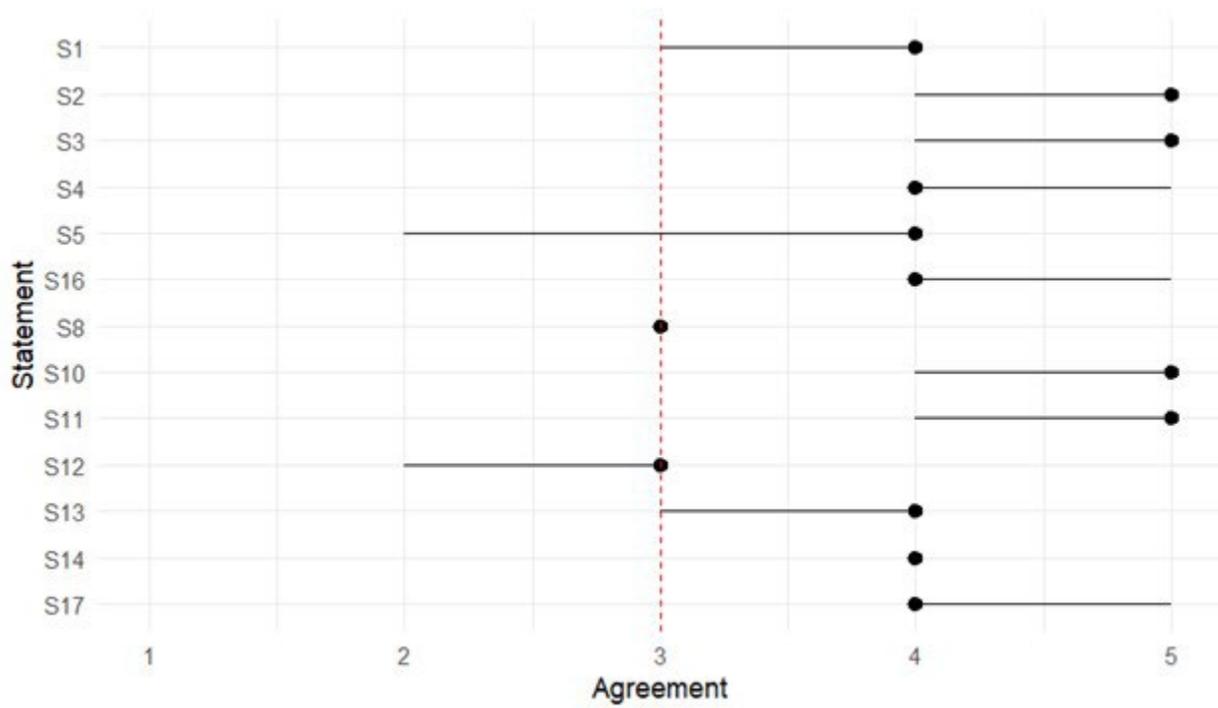

**Figure S2. Quantitative Round Summary of Round 2**. Agreement to each of the 13 statements of the second round is indicated in 5 levels ranging from strongly disagree (1) to strongly agree (5). Median and interquartile range (IQR) are indicated. Note that the numbering of statements is discontinuous due to the omission of certain statements and the introduction of new ones in the first rounds.

| Number | Agreement | | | | |
|--------|-----------------------|----------------|----------------|----------------|-------------------|
|        | Strongly Disagree (1) | Disagree (2)   | Neutral (3)    | Agree (4)      | Strongly Agree (5) |
| S1     | 0/13 (0.0%)           | 3/13 (23.1%)   | 2/13 (15.4%)   | 5/13 (38.5%)   | 3/13 (23.1%)      |
| S2     | 0/13 (0.0%)           | 0/13 (0.0%)    | 1/13 (7.7%)    | 5/13 (38.5%)   | 7/13 (53.8%)      |
| S3     | 0/13 (0.0%)           | 0/13 (0.0%)    | 2/13 (15.4%)   | 2/13 (15.4%)   | 9/13 (69.2%)      |
| S4     | 0/13 (0.0%)           | 0/13 (0.0%)    | 2/13 (15.4%)   | 6/13 (46.1%)   | 5/13 (38.5%)      |
| S5     | 2/13 (15.4%)          | 2/13 (15.4%)   | 0/13 (0.0%)    | 8/13 (61.5%)   | 1/13 (7.7%)       |
| S16    | 1/13 (7.7%)           | 0/13 (0.0%)    | 2/13 (15.4%)   | 4/13 (30.8%)   | 6/13 (46.1%)      |
| S8     | 3/13 (23.1%)          | 0/13 (0.0%)    | 7/13 (53.8%)   | 2/13 (15.4%)   | 1/13 (7.7%)       |
| S10    | 0/13 (0.0%)           | 2/13 (15.4%)   | 0/13 (0.0%)    | 3/13 (23.1%)   | 8/13 (61.5%)      |
| S11    | 0/13 (0.0%)           | 0/13 (0.0%)    | 0/13 (0.0%)    | 5/13 (38.5%)   | 8/13 (61.5%)      |
| S12    | 3/13 (23.1%)          | 1/13 (7.7%)    | 7/13 (53.8%)   | 1/13 (7.7%)    | 1/13 (7.7%)       |
| S13    | 2/13 (15.4%)          | 1/13 (7.7%)    | 3/13 (23.1%)   | 4/13 (30.8%)   | 3/13 (23.1%)      |
| S14    | 1/13 (7.7%)           | 0/13 (0.0%)    | 1/13 (7.7%)    | 9/13 (69.2%)   | 2/13 (15.4%)      |
| S17    | 1/13 (7.7%)           | 1/13 (7.7%)    | 1/13 (7.7%)    | 4/13 (30.8%)   | 6/13 (46.1%)      |

**Table S2: Distribution of Responses**. The 13 statements of the second round were rated by the panelists. The agreement is indicated in 5 levels. Note that the numbering of statements is discontinuous due to the omission of certain statements and the introduction of new ones in the first rounds.

## Summary of Round 3

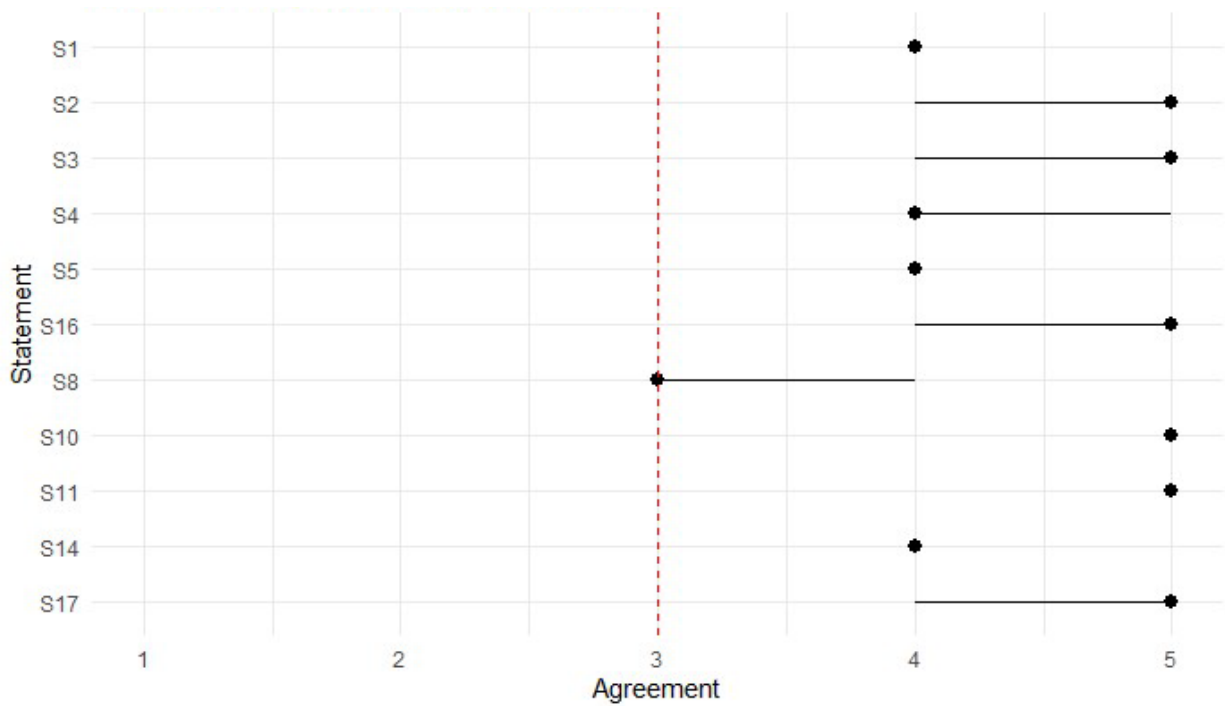

**Figure S3: Quantitative Round Summary of Round 3**. Agreement to each of the 11 statements of the third round is indicated in 5 levels ranging from strongly disagree (1) to strongly agree (5). Median and interquartile range (IQR) are indicated. Note that the numbering of statements is discontinuous due to the omission of certain statements and the introduction of new ones in the first rounds.

| Number | Agreement | | | | |
|---|---|---|---|---|---|
| | Strongly Disagree (1) | Disagree (2) | Neutral (3) | Agree (4) | Strongly Agree (5) |
| S1 | 0/13 (0.0%) | 2/13 (15.4%) | 1/13 (7.7%) | 7/13 (53.8%) | 3/13 (23.1%) |
| S2 | 0/13 (0.0%) | 0/13 (0.0%) | 1/13 (7.7%) | 5/13 (38.5%) | 7/13 (53.8%) |
| S3 | 0/13 (0.0%) | 0/13 (0.0%) | 2/13 (15.4%) | 2/13 (15.4%) | 9/13 (69.2%) |
| S4 | 0/13 (0.0%) | 0/13 (0.0%) | 2/13 (15.4%) | 7/13 (53.8%) | 4/13 (30.8%) |
| S5 | 0/13 (0.0%) | 1/13 (7.7%) | 0/13 (0.0%) | 9/13 (69.2%) | 3/13 (23.1%) |
| S16 | 0/13 (0.0%) | 0/13 (0.0%) | 1/13 (7.7%) | 5/13 (38.5%) | 7/13 (53.8%) |
| S8 | 1/13 (7.7%) | 1/13 (7.7%) | 7/13 (53.8%) | 2/13 (15.4%) | 2/13 (15.4%) |
| S10 | 0/13 (0.0%) | 1/13 (7.7%) | 1/13 (7.7%) | 1/13 (7.7%) | 10/13 (76.9%) |
| S11 | 0/13 (0.0%) | 0/13 (0.0%) | 0/13 (0.0%) | 3/13 (23.1%) | 10/13 (76.9%) |
| S14 | 0/13 (0.0%) | 1/13 (7.7%) | 0/13 (0.0%) | 10/13 (76.9%) | 2/13 (15.4%) |
| S17 | 0/13 (0.0%) | 0/13 (0.0%) | 2/13 (15.4%) | 2/13 (15.4%) | 9/13 (69.2%) |

**Table S3: Distribution of Responses**. The 11 statements of the third round were rated by the panelists. The agreement is indicated in 5 levels. Note that the numbering of statements is discontinuous due to the omission of certain statements and the introduction of new ones in the first rounds.

# 4.    Qualitative Results

Comments were analysed with two objectives: First, to refine the statements and the report in between the rounds and second, to understand the perspective of panelists who indicated an agreement level below 4 in the statements. The comments were coded according to key topics for each statement [1], [2]. Key topics were then classified into the following categories:

1. Misunderstandings and ambiguity: Topics that resulted from misunderstandings and reflected ambiguity in the report and/or statement.
2. Misconceptions: Topics that resulted from misconceptions that have been proved wrong in literature and/or through simulations in the report.
3. Counterarguments: Topics that covered remaining counterarguments to the statement.
4. Unrelated: Topics that were unrelated to the disagreement or uncertainty in the statement.

While comments of categories 1 and 2 were used in between the rounds to adjust statements and report, category 3 informed the discussion presented in this manuscript. In the following, we present the qualitative analysis of statements where an agreement level below 4 was indicated in the final round. Note that key findings of this analysis have been included in the main manuscript where they help to understand the outcomes of the Delphi process.

## 4.1    Disclosure vulnerability metrics should be based on quasi-identifiers. These may vary depending on the data context (e.g., can still be all attributes) and are ascertained by the data controller

Even though there was stable consensus on agreement for this recommendation, we want to acknowledge that there were 2/13 (15.4%) experts who disagreed with this statement. While our simulation provided compelling evidence against the first argument, subjectivity remained as a key topic of disagreement. For example, the diagnosis of obesity or diabetes mellitus can be considered as a QI since its knowability can be quite high (e.g., public awareness, photos, discussions on social media) while the diagnosis of interstitial nephropathy may not be a familiar term to the patient themselves. Consequently, it can be challenging to label the attribute *diagnosis* as either QI or non-QI across all diagnoses. Also, the context of the data may shift over time and an attribute that was once deemed not to be a QI may then be considered one. This contextual interpretation is very likely to result in inter-individual variability. Another key topic argued that any attribute could be a QI depending on the adversary so that entire records should be considered. While it may be the most prudent approach to treat all attributes as QIs, this could, in practice, become computationally problematic as a prudent approach would also require considering all potential combinations of those QIs in matching.

## 4.2    Membership disclosure vulnerability should only be evaluated when the assumptions of the current metrics hold which is that the adversary would learn something new for targets drawn from the same population as the training dataset.

While 12/13 (92.3%) panelists agreed to this statement, there was 1/13 (7.7%) disagreement. Key topics that have been raised as counterarguments are the benefits of measuring membership disclosure in terms of trust, the consideration of membership rather as a flavor of attribute disclosure than a disclosure by itself and the unclear informative value of membership disclosure in practice. The remaining disagreement cannot be clearly attributed to one of these key topics. Given that most common key topics in previous rounds were misunderstandings and ambiguity, it may be a remnant of that.

### 4.3 As an anchor for membership disclosure vulnerability, a relative F1 score vulnerability (i.e., $F_{rel}$ value) of 0.2 is suggested.

The panel's ratings reflected a consensus on uncertainty regarding this statement. In the qualitative analysis, two key topics could be identified: First, any threshold, anchor or default value for disclosure vulnerability is considered as inadequate. One reason that was given for this perspective is the contextual interpretation of privacy vulnerability as described above. Another reason is that vulnerability can be reported very differently, for example, based on an average value as well as a maximum value. A concern is that when giving an anchor value people may rely on this value without considering relevant aspects that may trigger adjustments to the value. And second, there is not enough evidence for the very anchor value proposed. Implications of this result are discussed below.

### 4.4 Meaningful attribute disclosure vulnerability only applies to individuals that are in the dataset (i.e., members). Penalizing accurate prediction on individuals that have not been part of the dataset (i.e., group privacy) requires a broader ethical framework.

While achieving broad agreement among the panelists (11/13, 84.6%), this statement also resulted in 1/13 (7.7%) rating on disagreement and 1/13 (7.7%) on uncertainty. There were no counterarguments in the key topics identified regarding this statement. However, there are opinions in the literature that would support a different conclusion. A legal analysis, for example, describes that inferring information about non-members could indeed be interpreted as a privacy violation [3]. At the same time, however, the authors acknowledge that in the United States (US) this would probably raise First Amendment challenges. The First Amendment in the US protects freedom of speech and inference constitutes speech. Similarly, it has been proposed that the concept of privacy can be extended to groups or to collective privacy in the context of big data [4], [5]. The remaining disagreement may be in line with these opinions, may oppose to the idea of a broader ethical framework or result from misunderstanding and/or ambiguity in the statement.

### 4.5 In attribute disclosure vulnerability, a relative vulnerability higher than its threshold is only considered as unacceptably high when the absolute vulnerability is higher than its threshold.

Among our experts, there was 1/13 (7.7%) who disagreed with this statement. Counterarguments in key topics related to this statement did not question the relevance of the absolute scale but rather the application of a threshold in this context or the use of two metrics instead of a proportional metric as in [6]. The discussion on thresholds in general is presented below in the main manuscript. A proportional metric as in [6] is defined relative to the maximum possible attribute disclosure vulnerability beyond the non-member baseline:

$$R = \frac{r_{train} - r_{control}}{1 - r_{control}} \tag{1}$$

where $r_{train}$ is the prediction accuracy for members and $r_{control}$ the one for non-members. The idea is that this metric accounts for both absolute and relative scale. It treats differences at the upper end of the scale as more severe than those at the lower end. At the same time, however, it penalizes high non-member baselines and thereby penalizes potentially relevant population-level information (i.e., knowledge generation). Assuming, for example, a non-member baseline of 0.9 accuracy. This is commonly seen as a very good or excellent model [7]. The high accuracy on unseen data (i.e., non-members) suggests that relevant and externally valid population-level information have been derived from the data. At this end of the scale, however, a slightly higher member accuracy of 0.95 would result in a high proportional vulnerability estimate of 0.5. Using a threshold of 0.2 which has been used for such ratios elsewhere [8], [9], [10], this would be considered as an unacceptably high vulnerability. A proportional metric thus comes with the challenge that potentially highly valuable insights would be considered as disclosure even though the actual difference is negligible.